\documentclass{pas}

\usepackage{multirow}
\usepackage{hyperref}

\begin{document}

\lefttitle{Publications of the Astronomical Society of Australia}
\righttitle{Geethika Santhosh \textit{et al.}}

\jnlPage{1}{18}

\articletitt{Research Paper}

\title{Star formation outside galaxies undergoing gravitational and hydrodynamic
interactions: Dust attenuation and the star formation rate}

\author{Geethika Santhosh$^1$}
\author{Rakhi R$^1$}
\affil{$^1$Department of Physics, N.S.S. College, Pandalam (Affiliated to University of Kerala), Kerala 689501, India}

\author{Koshy George$^2$}
\affil{$^2$University Observatory, LMU Faculty of Physics, Scheinerstrasse 1, 81679 Munich, Germany}

\author{Bianca M. Poggianti$^3$}
\affil{$^3$INAF-Astronomical Observatory of Padova,  vicolo dell'Osservatorio 5 35122 Padova, Italy}

\author{Smitha Subramanian$^4$}
\affil{$^4$Indian Institute of Astrophysics, Bangalore 560034, India}

\author{ Kavila Indulekha$^5$}
\affil{$^5$School of Pure and Applied Physics, Mahatma Gandhi University, Kottayam, Kerala 686560, India}

\corresp{Geethika Santhosh, Email: geethikagks@gmail.com}



\begin{abstract}
Galaxies undergo perturbations, either gravitational or hydrodynamic in origin, which can generate extragalactic structures such as rings and tails, where in situ star formation may take place. 
We selected a sample consisting of JO201 and JW100, undergoing ram-pressure stripping, and NGC 5291 and NGC 7252, formed through gravitational interactions, to investigate how different perturbation mechanisms influence dust content and star formation in extragalactic features.
In 
both cases, star formation can be observed outside the main disks of the galaxies. 
We present new results of dust attenuation for JO201 and JW100, while for NGC 5291 and NGC 7252 we use results from our previous study, based on high-resolution observations obtained with the Ultraviolet Imaging Telescope onboard AstroSat. Dust attenuation is determined from the ultraviolet continuum slope ($\beta$) calculated using the FUV–NUV colour, and the star formation rates of the star-forming knots are corrected accordingly.
It is seen that dust attenuation and dust-corrected SFR densities of the knots in the ram-pressure stripped tails of JO201
and JW100 are comparable to those in the collisional ring of the NGC 5291 system and the tidal tails of the
NGC 7252 system.
We conclude that, though the formation scenarios of the tails of JO201 and JW100, the NGC 5291 ring, and the NGC 7252 tails 
are different, their dust content and star formation activity are notably similar. 
\end{abstract}

\begin{keywords}
galaxies: interactions, galaxies: star formation, ultraviolet: galaxies, galaxies: evolution,  galaxies: formation
\end{keywords}

\maketitle

\section{Introduction}
\label{section1}

Galaxies experience perturbations
which can broadly be classified into two types: gravitational and hydrodynamic.
The nature, frequency, strength, speed, and effects of such perturbations are indeed environment dependent \citep{2004cgpc.symp..277M,2006MNRAS.367.1029S,2006PASP..118..517B,2010MNRAS.407.1514E}. 
Gravitational perturbations can occur in both high- and low-density environments, while hydrodynamic perturbations are mostly confined to dense environments since low-density environments lack a hot, ionised plasma like an intracluster or intragroup medium.
Such perturbations, in addition to altering the morphology, mass distribution, colour, star formation rate (SFR) and active galactic nucleus (AGN) activity \citep{1980ApJ...236..351D,2008ApJ...674..784P,2012A&A...539A..45L,2012A&A...539A..46A,2015ApJ...806..219C,2017Natur.548..304P,2022A&ARv..30....3B,2025MNRAS.538..327W}, can create peculiar structures outside the disk of galaxies, such as rings (collisional rings, polar rings, accretion rings) \citep{1999IAUS..186...97A,2003A&A...401..817B,1987ApJ...320..454S}, tails (tidal tails, ram-pressure stripped tails) \citep{2013LNP...861..327D,2019MNRAS.482.4466P}, bridges, plumes, loops, streams and shells \citep{2024MNRAS.532..883S}. In many cases, substructures such as small star-forming knots and even dwarf galaxy-sized objects are formed in situ in these structures, offering unique opportunities to study and understand the origin and evolution of galaxies \citep{2007IAUS..237..323D,2024ApJ...969...24L}. 

The environment has a significant impact on how galaxies form and evolve
\citep{ 1980ApJ...236..351D,2003MNRAS.346..601G,2009MNRAS.399..966S,2010ApJ...723..197K,2024ApJ...969...24L}. Galaxies in high-density environments are susceptible to multiple perturbing processes that drive
rapid evolution \citep{2006PASP..118..517B}. In rich environments like galaxy clusters, early-type galaxies and quiescent systems are the dominant populations \citep{1993ApJ...407..489W,2010ApJ...721..193P}. The minority population of spirals observed in clusters has a lower atomic and molecular gas content, and therefore they have reduced star formation rates (SFR) compared to their field counterparts \citep{2002AJ....124.2440S,2003ApJ...584..210G,2005A&A...429..439G,2014A&A...564A..67B,2014A&A...570A..69B}.
The reason for this is that spiral galaxies in dense environments are susceptible to a variety of physical processes, such as ram-pressure stripping \citep{1972ApJ...176....1G}, thermal evaporation \citep{1977Natur.266..501C}, starvation or strangulation \citep{1980ApJ...237..692L}, viscous stripping \citep{1982MNRAS.198.1007N}, galaxy harassment, \citep{1996Natur.379..613M} and tidal interactions \citep{1981ApJ...243...32F}. These processes, particularly hydrodynamic interactions such as ram-pressure stripping, viscous stripping, and thermal evaporation, are capable of predominantly removing cold gas from the disk of a galaxy, causing the rate of star formation to decrease and eventually stop completely \citep{2006PASP..118..517B, 2013ApJ...775..126H,2020ApJ...892..146V,2020ApJ...905...31L}. 

Ram-pressure stripping is often regarded as the dominant hydrodynamic perturbing mechanism that effectively strips spiral galaxies in dense clusters of their cold gas \citep{2006PASP..118..517B,2007ApJ...671.1434T,2013A&A...553A..90G,2014A&A...570A..69B,2017ApJ...844...48P,2017ApJ...846...27G,2019MNRAS.487.4580R,2020MNRAS.494.5029D,2022ApJ...927...39L,2022MNRAS.516.2683D,2022A&ARv..30....3B}. Ram pressure stripping can give rise to distinctive morphological features in galaxies, most notably the interstellar medium (ISM) being drawn out into extended tails.
The gas tails of these stripped galaxies have been observed in atomic, molecular, ionised, and hot gas phases due to different physical processes acting on the stripped gas \citep{2006ApJ...637L..81S,2014MNRAS.445.4335F, 2017ApJ...839..114J,2018MNRAS.480.2508M, 2021A&A...650A.111R, 2021A&A...652A.153R}. 

The most interesting process of ram-pressure stripping is the occurrence of in situ star formation in the stripped gas tails of galaxies 
\citep{2006AJ....131.1974O,2007MNRAS.376..157C,2010MNRAS.408.1417S,2012ApJ...750L..23O,2014ApJ...781L..40E,2014MNRAS.442..196R,2014MNRAS.445.4335F,2016AJ....151...78P,2017A&A...606A..83C,2017ApJ...844...48P,2017ApJ...846...27G,2017ApJ...844...49B,2018MNRAS.479.4126G,2018ApJ...866L..25V,2018A&A...615A.114B,2019MNRAS.482.4466P,2019ApJ...887..155P,2019MNRAS.485.1157B,2020ApJ...899...13G,2023ApJ...949...72G,2023MNRAS.519.2426G,2025A&A...700A..38G}.
The triggering of star formation in the stripped gaseous tails is rather puzzling considering the hostile environment (ICM), with high temperatures of $\sim$10$^7$ - 10$^8$ K 
\citep{2015A&A...582A...6V, 2022A&ARv..30....3B}.
Studies indicate that magnetic fields play a crucial role in this phenomenon \citep{2021Galax...9..116M, 2021NatAs...5..159M}. The ICM magnetic fields are instrumental in protecting the stripped tails from the harsh environment of the ICM. As the galaxy moves through the magnetised plasma, a magnetic draping layer forms around the stripped tails. Such a layer helps shield the gas tails from evaporation, thereby enabling the cooling and condensation of gas within the stripped tails, leading to the formation of molecular clumps \citep{2021Galax...9..116M}.
Galaxies that have undergone ram-pressure stripping, with enhanced star formation along their disks and in their stripped tails, are referred to as `jellyfish' galaxies due to their appearance, which resembles jellyfish with tentacles.

Dust and gas can be well mixed in galaxy disks; under ram pressure,
dust can be stripped from the galaxy disks along with the gas, forming tails of gas and dust.
\citep{2005ASPC..331..281C,2014ApJ...796...89S,2014AJ....147...63A,2016AJ....152...32A,2020A&A...633L...7L,2020A&A...644A.161L,2022MNRAS.509.3938L,2022A&ARv..30....3B}. 
\citet{2016AJ....152...32A} observed structures such as linear dust filaments and ridges in some ram-pressure stripped galaxies in the Virgo cluster. They propose that galactic magnetic fields can bind together the multi-phase, multi-density components of the ISM. This magnetic binding could explain the large, coherent dust structures observed in these galaxies \citep{2016AJ....152...32A}.
Using far-infrared (FIR) data from the Herschel Space Telescope, \citet{2020A&A...644A.161L} detected diffuse dust, in the HI and H$\alpha$ tails of three galaxies (NGC 4330, NGC 4522, and NGC 4654) in the Virgo cluster, undergoing ram-pressure stripping. The estimated values, of the gas-to-dust ratio in the tails, were
typical of those observed in the outer disk of spirals, confirming ram-pressure stripping of the dust component along with the gas \citep{2020A&A...644A.161L,2022A&ARv..30....3B}.
Besides,  \citet{2020MNRAS.492.4599B} observed that cluster galaxies in general have lower dust-to-stellar mass ratios compared to similar galaxies in the field. 
Direct ram-pressure stripping of molecular gas has been observed in some galaxies in nearby clusters \citep{2018MNRAS.480.2508M,2019MNRAS.483.2251Z,2019ApJ...883..145J,2019asrc.confE..80C, 2020ApJ...889....9M,2020ApJ...897L..30M, 2020ApJ...901...95C,2023ApJ...955..153M}. Since dust is primarily associated with molecular gas in galaxies, the stripping of molecular gas further strengthens the possibility of stripping of dust via ram-pressure.
Dust found in the tails of ram-pressure stripped galaxies can originate in two different ways: dust that is stripped along with the gas from the
galaxy disk, and dust that forms in situ within the star-forming clumps born in the stripped gas \citep{2022A&ARv..30....3B}.
Previous studies of dust extinction in jellyfish tails have found that the
dust distribution is inhomogeneous. The  distribution is clumpy, characterised by high-extinction star-forming knots ($A_v$ $\gtrsim$ 1 mag), and low-extinction inter-knot regions ($A_v$ $\lesssim$ 0.5 mag) \citep{2017ApJ...844...48P,2017ApJ...846...27G}.

As mentioned earlier, gravitational perturbations can also create unique structures outside galaxy disks. Signatures of gravitational perturbations can be observed in the form of peculiar extended features such as tidal tails, bridges, collisional rings, ripples, shells and warps. Tidal tails and collisional rings are two of the most conspicuous structures formed by galaxy-galaxy gravitational interactions. 
Tidal tails are formed when tidal forces pull out gas, dust and stars from the disks of interacting or merging galaxies. 
Collisional ring structures around galaxies do not have a direct tidal origin \citep{2013LNP...861..327D}. 
They are formed when a disk galaxy experiences a high-speed nearly head-on collision with an intruder galaxy. This high-speed collision triggers radial density waves that move outwards from the center of the disk galaxy, forming a ring around the galaxy \citep{1976ApJ...209..382L, 1999IAUS..186...97A,1996FCPh...16..111A}.
Both tidal tails and collisional rings have been observed to host star-forming knots and dwarf galaxy-sized objects that are formed by the compression of gas 
removed from the parent galaxies during galaxy-galaxy interactions.
Tidal tails and collisional rings contain dust, as gas, dust, and stars can be removed from a galaxy during gravitational interactions.

Though dust constitutes only a small fraction of the total mass of the ISM
of a galaxy, it is a vital constituent as it plays a major role in galaxy growth by facilitating star formation.
Dust grains act as substrates for the formation of H$_2$ molecules, shield these molecules from photodissociation and also act as an ISM cooling agent \citep{1971ApJ...163..155H,2017MNRAS.465..885S}. Therefore, dust and molecular gas within galaxies are particularly closely linked with each other. Dust severely interferes with the detectability of star formation in the ultraviolet to optical regime. Dust in the debris of gravitational and hydrodynamic interactions\textemdash{}collisional rings, tidal tails, and ram-pressure stripped tails\textemdash{}can significantly obscure the emission from young stars. Emission from young stars is used as tracer for star formation activity, and in the quantification of star formation activity, such obscuration will lead to underestimation. 
Hence, analysing the dust content alongside of star formation is essential for a complete understanding of galactic evolution.

Both ram-pressure stripping and gravitational tidal interactions are outside-in processes where the peripheral HI gas is removed first \citep{2006PASP..118..517B}. However, while hydrodynamic processes predominantly affect the diffuse ISM components, gravitational perturbations affect all the components of a galaxy, including stars, gas, dust and dark matter.  Differential stripping can result in a gradient in dust content and star formation along the stripped tails. Besides, dust cannot survive long in hostile environments such as galaxy clusters due to its destruction by thermal sputtering \citep{1979ApJ...231..438D,2019MNRAS.487.4870V,2024A&A...682A.162W}.
Hence, we may expect low dust content in the interaction debris of cluster galaxies undergoing perturbations \citep{2024A&A...682A.162W}.
Comparing the dust content and star formation activity in exotic extended structures like collisional rings and tidal tails (gravitational origin) and ram-pressure stripped tails (hydrodynamic origin) can provide crucial insights into the characteristic similarities and/or differences among the diverse perturbing mechanisms through which the galaxies evolve, and this study aims to explore this in detail.

The ultraviolet (UV) continuum serves as a direct probe of recent star formation as it is dominated by photospheric emission from hot, young, massive O, B, and A stars with ages $\leq$ 200 Myr \citep{2012ARA&A..50..531K}. While UV emission effectively traces ongoing star formation in galaxies, it is also very
sensitive to dust attenuation \citep{2012ARA&A..50..531K, 1998ARA&A..36..189K}. 
Ultraviolet dust obscuration can be quantified using the UV spectral slope ($\beta$) method  \citep{1999ApJ...521...64M}, which uses the
the slope of the ultraviolet continuum as an indicator of dust attenuation in galaxies \citep{1989ApJ...345..245C,1999ApJ...521...64M}. 
This method is most commonly employed when infrared (IR) data are unavailable, such as in high-redshift studies, or when the available IR data do not offer sufficient depth or resolution to meet the scientific objectives.

The present study aims to compare the dust attenuation, estimated from UV slope $\beta$ derived from FUV-NUV colour, and dust-corrected far-ultraviolet (FUV) star formation rate (SFR) of the resolved star-forming knots outside galaxies undergoing  gravitational and hydrodynamic interactions using a sample of four targets: the jellyfish galaxies JO201 and JW100 (hydrodynamic interactions) \citep{2016AJ....151...78P, 2017ApJ...844...49B, 2019ApJ...887..155P}, and the galaxy systems NGC 5291  and NGC 7252 (gravitational interactions) \citep{1972ApJ...178..623T,2009AJ....137.4561B}. 
The galaxies in our sample are characterised by extended structures---the ram-pressure stripped tails of JO201 and JW100, the collisional ring of the NGC 5291 system and the tidal tails of the NGC 7252 system\textemdash{}where substructures like star-forming clumps or even dwarf-like objects are observed. 

The main objective of the study is to compare the dust content and star formation in the JO201 (z$\sim$0.056) and JW100 (z$\sim$0.055) jellyfish tails with that in the collisional ring of the NGC 5291 system (z$\sim$0.015) and tidal tails of the NGC 7252 post-merger system (z$\sim$0.016).
The study utilises high-resolution FUV and NUV imaging data from the Ultraviolet Imaging Telescope (UVIT) onboard AstroSat. For JO201 and JW100, we present results from our new analysis based on the UV slope method, while for NGC 5291 and NGC 7252, we use results from our recent UVIT study \citep{2025PASA...42...73S}.
  
This paper is organised as follows: Section \ref{section2} is a brief overview of the selected sample galaxies and galaxy systems. Section \ref{section3} describes data reduction, source extraction, and identification of star-forming knots in JO201 and JW100. The results of the analysis of JO201 and JW100 are presented in Section \ref{section4}.  Section \ref{section5} presents the detailed comparison of dust attenuation and star formation in the debris of gravitational and hydrodynamic interactions. Conclusions are presented in Section \ref{section6}.

Throughout this paper, we adopt the standard  $\Lambda$CDM cosmology with the following cosmological parameters, H$_0$ = 70 km$^{-1}$ s$^{-1}$ Mpc$^{-1}$, $\Omega_M$ = 0.3, $\Omega_\Lambda$ = 0.7.

\section{Sample selection}
\label{section2}

\subsection{Galaxies JO201 \& JW100}

\begin{figure}
    \includegraphics[width=0.47\textwidth]{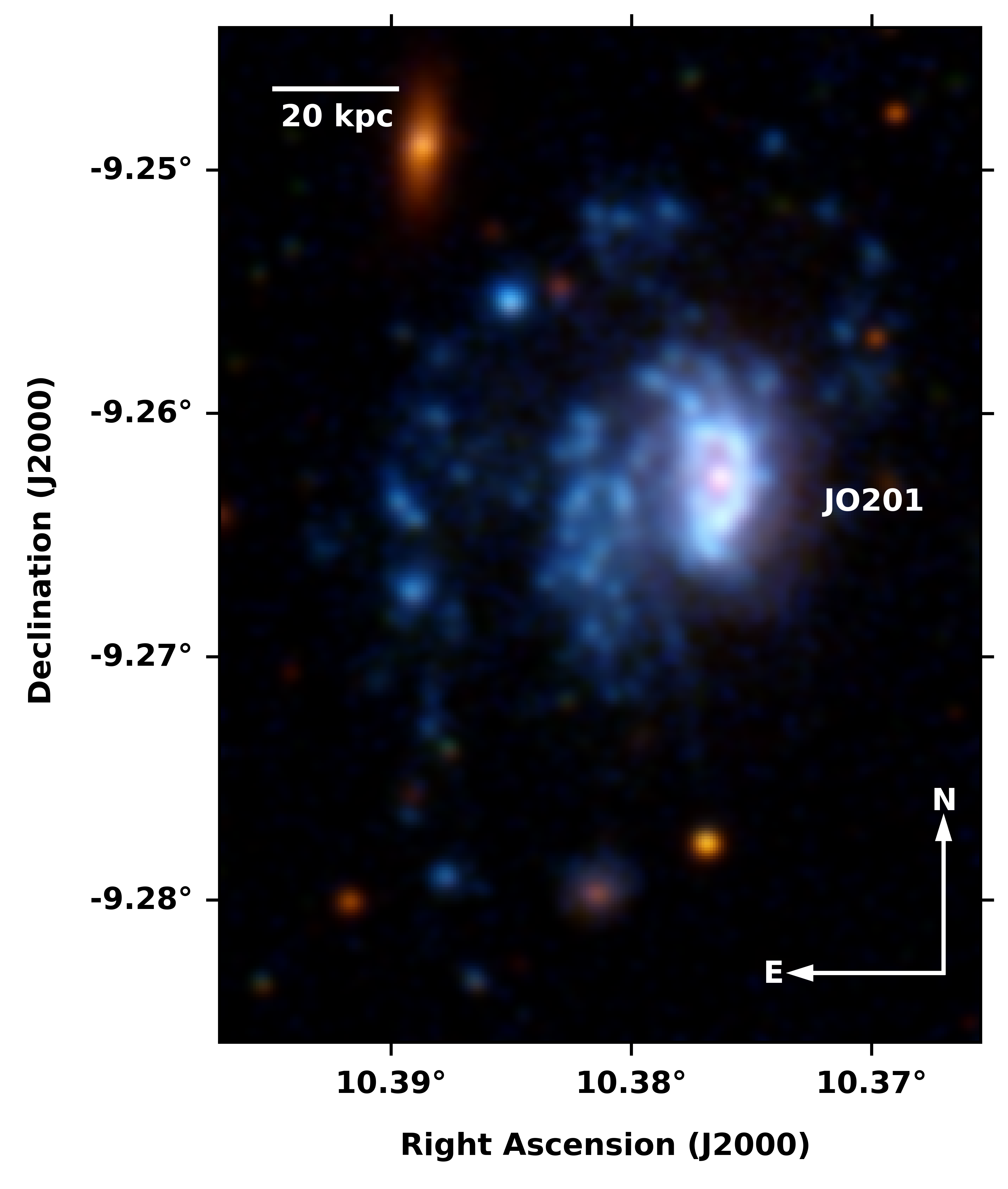} 
    \vfill
    \vspace{1.5em}
    \includegraphics[width=0.47\textwidth]{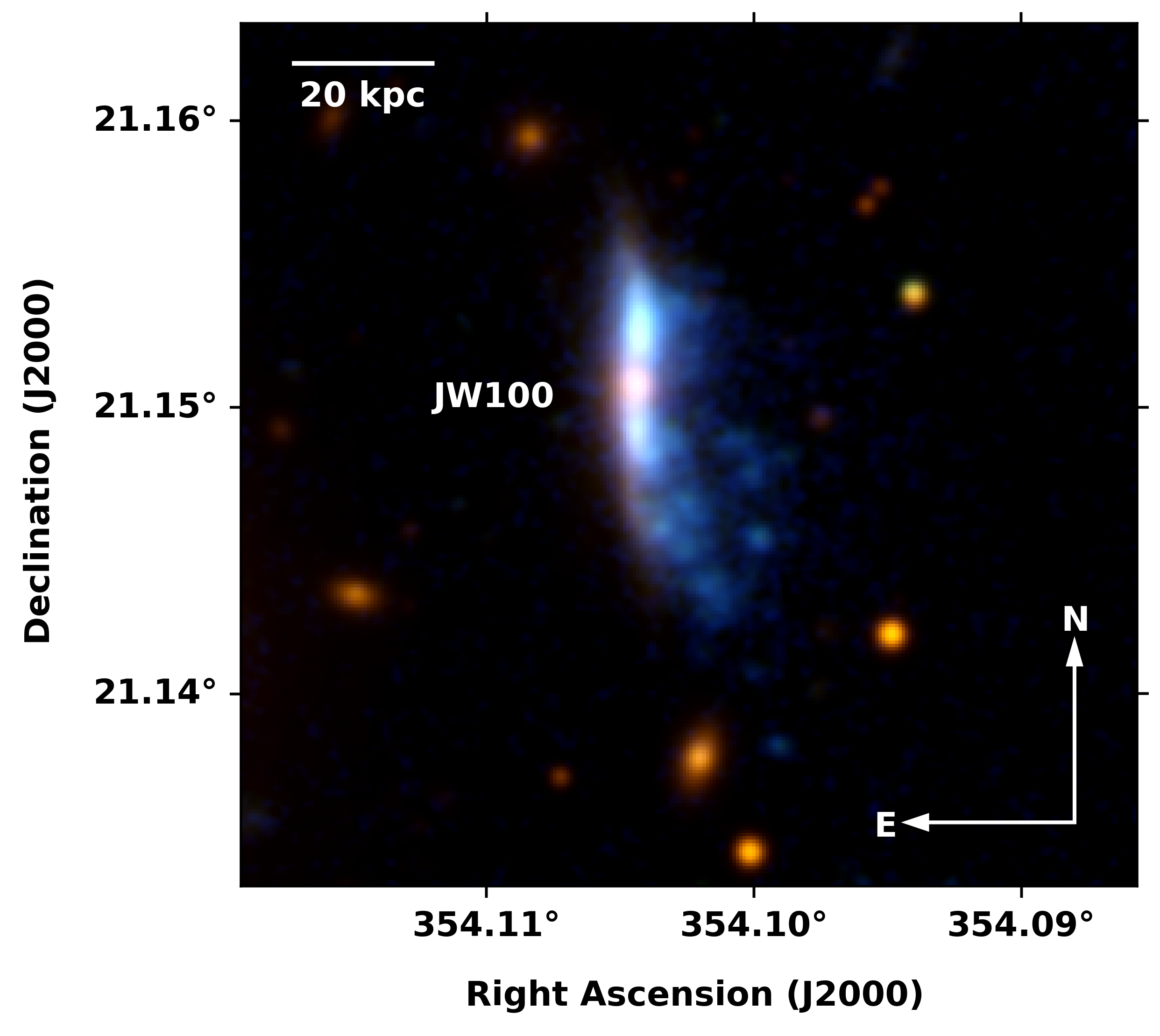} 
    \caption{Colour composite images of JO201 (top) and JW100 (bottom) made using FUV (blue), NUV (green) and DECaLS r-band images (red).}
    \label{fig:rgb-jo201-jw100}
\end{figure}

The galaxies, JO201 and JW100 (Figure \ref{fig:rgb-jo201-jw100}) are experiencing extreme ram-pressure stripping and have intense star formation in their disk and stripped tails \citep{2019MNRAS.485.1157B,2019MNRAS.482.4466P,2019ApJ...887..155P,2018MNRAS.479.4126G,2020ApJ...899...13G,2023ApJ...949...72G}.
The almost face-on galaxy JO201 belongs to the Abell 85 galaxy cluster (cluster redshift z = 0.0559) and lies at a small projected radial distance of 360 kpc from the brightest cluster galaxy (BCG). The total stellar mass of the galaxy is $\sim$3.55$\times$10$^{10}$ $M_{\odot}$. 
The galaxy has a high velocity of 3364 km s$^{-1}$ relative to the cluster and is falling into the cluster for the first time, from behind. JO201 is moving towards the observer along the line of sight; due to a slight inclination of the velocity vector with respect to the line of sight,
the projected jellyfish tails are pointing towards the east \citep{2017ApJ...844...49B}. The almost edge-on galaxy JW100 is a member of the Abell 2626 galaxy cluster (cluster redshift z = 0.0548) and has a stellar mass of 3.2$\times$10$^{11}$ $M_{\odot}$ \citep{2017Natur.548..304P}. JW100 lies at a projected distance of only 83 kpc from the BCG and has a line-of-sight velocity of 1807 km s$^{-1}$ relative to the cluster \citep{2019ApJ...887..155P}.

\citet{2018MNRAS.479.4126G} conducted a detailed study on star formation in the disk and tail of JO201 and estimated dust-corrected SFRs in FUV and H$\alpha$. \citet{2018ApJ...866L..25V} and \citet{2019MNRAS.482.4466P} estimated the dust-corrected current SFR in JO201 and JW100 from H$\alpha$ luminosities. \citet{2019ApJ...887..155P} presented a preliminary UVIT analysis of star formation in JW100, identifying sites of recent star formation using UVIT NUV imaging data. More recently, a high-resolution, pixel-by-pixel study on star formation in JO201 and JW100 is presented by \citet{2024ApJ...976...90T} using UVIT FUV and NUV images. In these studies, the dust correction was performed using the Balmer decrement method. 
In the present study, we use the $\beta$ slope method  to correct the FUV luminosities for dust attenuation. Since our main goal is to compare the dust attenuation and star formation in the JO201 and JW100 jellyfish tails with those in the NGC 5291 ring and NGC 7252 tails\textemdash{}where the $\beta$ slope method was employed \citep{2025PASA...42...73S}, we adopt the same method here for the jellyfish galaxies to ensure methodological consistency.

\subsection{Galaxy systems NGC 5291 \& NGC 7252}

\begin{figure}
    \includegraphics[width=0.48\textwidth]{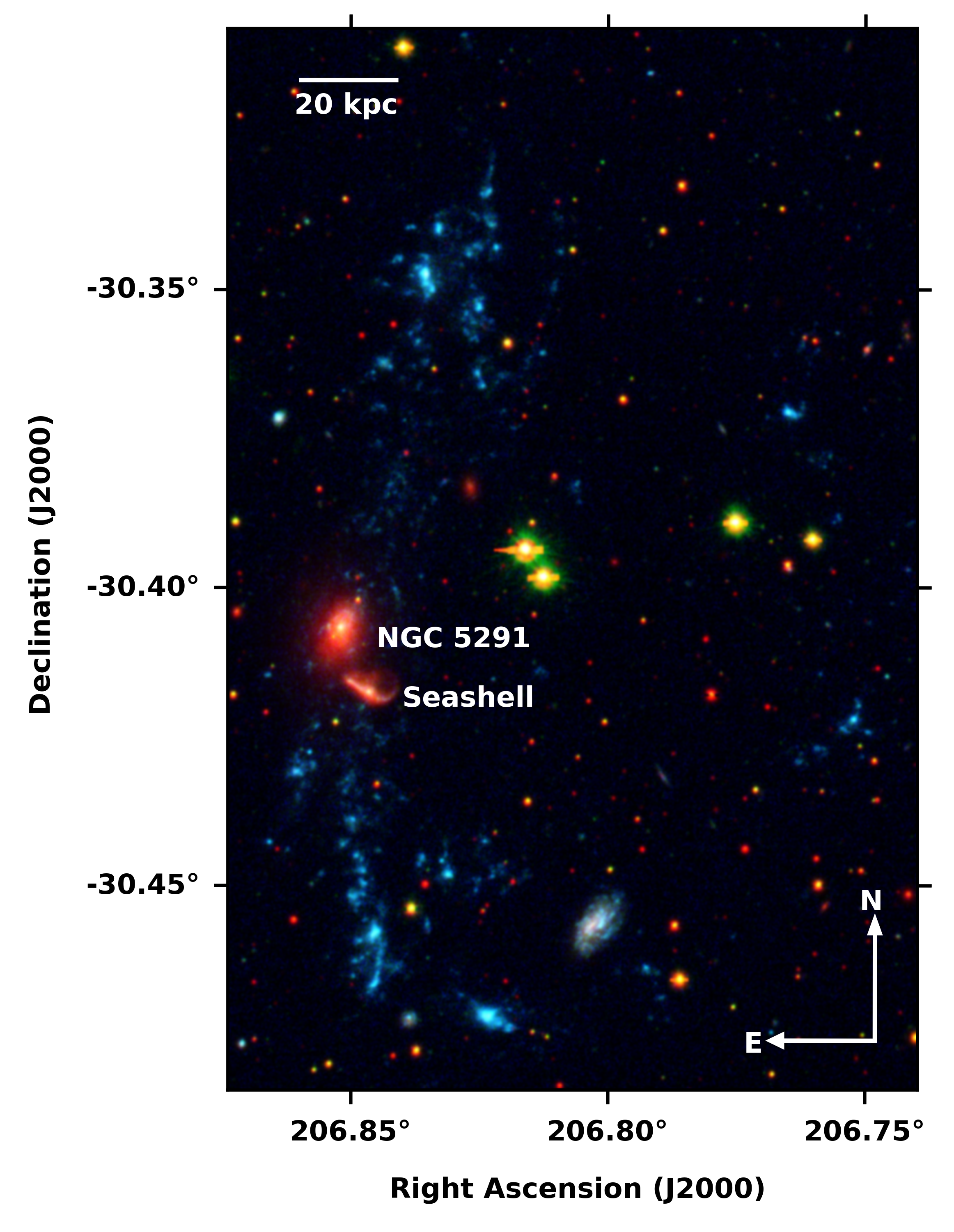}  
    \vfill
    \vspace{1.5em}
    \includegraphics[width=0.48\textwidth]{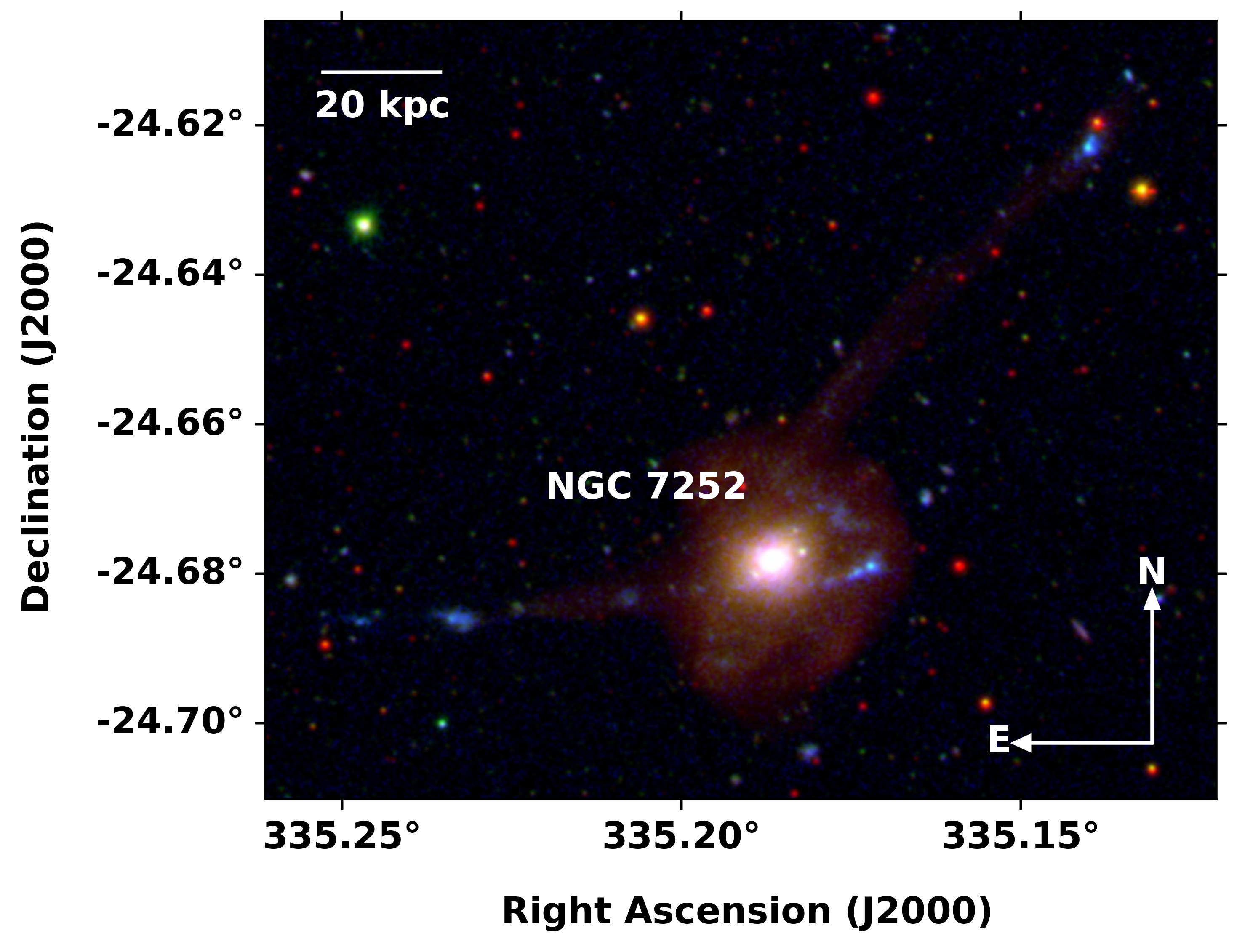}
    \caption{Colour composite images of the NGC 5291 (top) and NGC 7252 (bottom) systems made using FUV (blue), NUV (green) and DECaLS r-band images (red).}
    \label{fig:rgb-5291-7252}
\end{figure}

The collisional ring system, NGC 5291 and the post-merger system, NGC 7252 (Figure \ref{fig:rgb-5291-7252}) are two galaxy systems formed from the gravitational interaction (collision or merger) between galaxies. The NGC 5291 system, located on the edge of the Abell 3574 galaxy cluster, consists of two galaxies—NGC 5291 and the Seashell—along with a massive HI-dominated ring structure that surrounds the galaxies
and hosts numerous young star-forming knots and candidate tidal dwarf galaxies (TDGs). The redshift of the NGC 5291 galaxy z = 0.0146.
The numerical model of \citet{2007Sci...316.1166B} predicts that a violent, high-speed, nearly head-on collision of the NGC 5291 galaxy with a massive elliptical galaxy, IC 4329,  resulted in the formation of the the NGC 5291 ring system. IC 4329 is currently outside the field of view at a projected distance of $\sim 430$ kpc from the NGC 5291 galaxy. According to the model, NGC 5291 was initially a disc galaxy that transformed into an early-type galaxy after the collision. The model suggests that the Seashell galaxy\textemdash{}only weakly interacting with the NGC 5291 galaxy\textemdash{}is an unlikely progenitor for the ring and is probably an interloper \citep{2007Sci...316.1166B}.

The NGC 7252 post-merger system, located in a field environment, is of purely tidal origin and comprises the merger remnant and two prominent tidal tails, both of which are believed to have formed from the interaction and subsequent merger of two massive, gas-rich spiral galaxies.
Star-forming knots and two bona fide TDGs are located along the tidal tails of the NGC 7252 system. The NGC 7252 galaxy has a redshift z = 0.0159. 
 \citet{2007A&A...467...93B} presented a multi-wavelength study of the star-forming knots in the NGC 5291 collisional system, and a more detailed study on star formation in the collisional debris using a sample of interacting galaxies including the NGC 5291 and NGC 7252 systems is later published in \citet{2009AJ....137.4561B}. The bona fide TDGs in the NGC 5291 system were studied in detail in \citet{2019A&A...628A..60F}. High-resolution ultraviolet studies on star formation in NGC 7252 and NGC 5291 systems using UVIT data were presented in \citet{2018A&A...614A.130G} and \citet{2023MNRAS.522.1196R}, respectively. \citet{2025PASA...42...73S} presented a detailed comparison of dust attenuation and star formation in the NGC 5291 and NGC 7252 interacting galaxy systems, where the dust attenuation is estimated from the $\beta$ slope derived from UVIT FUV-NUV colour. 

\begin{table*}
    \setlength{\tabcolsep}{4.35pt}
    \caption{Log of UVIT observations}
    \label{tab:uvit-observations-jellyfish}
    {\tablefont\begin{tabular}{@{\extracolsep{\fill}}lcccccccccc}
    \toprule
Target,   & RA & Dec & z & Luminosity  & Linear distance & Channel & Filter   & 
$\lambda$\textsubscript{mean} &  $\Delta \lambda$   & Integration  \\
{Observation ID} & {} & {} & {} & distance & corresponding  & {} & {} & {} & {} & {time} \\
{\& Date} & {} & {} & {} & (Mpc) & to 1$''$ (kpc) & {} & {} & {\AA} & {\AA} & {(s)} \\
\hline
JO201 galaxy &00$^h$41$^m$30.2951$^s$ & -09$^{\circ}$15$'$46.002$''$ & 0.0559 & 249.6 & 1.085 &FUV & F148W & 1481 & 500 & 18043\\
G06\_019T01\_9000000796 & & & & & & NUV & N242W & 2418 & 785 & 15045\\
15-Nov-2016 & & & & & & & & & &\\

\hline
JW100 galaxy & 23$^h$36$^m$25.0459$^s$ &  +21$^{\circ}$09$'$02.998$''$ & 0.0548 & 244.5  & 1.065 & FUV & F148W & 1481 & 500 & 11899\\
G07\_002T02\_9000001432 & & & & & & NUV & N242W & 2418 & 785 & 11939\\
04-Aug-2017 & & & & & &  &  &  &  & \\
\hline
NGC 5291 system &13$^h$47$^m$24.5087$^s$ & -30$^{\circ}$24$'$25.603$''$ & 0.0146 &  62 & 0.30 & FUV & F148W & 1481 & 500 & 8242\\
 G07\_003T01\_9000001290 & & & & & & NUV & N242W & 2418 & 785 & 8079\\
13-Jun-2017 & & & & & &  & &  &  & \\
\hline
NGC 7252 system & 22$^h$20$^m$44.7679$^s$ &  -24$^{\circ}$40$'$41.942$''$ & 0.0159 & 68  & 0.32 & FUV & F148W & 1481 & 500 & 8138\\
G07\_003T02\_9000001564 & & & & & & NUV & N242W & 2418 & 785 & 7915\\
27-Sep-2017 & & & & & &  &  &  &  & 
\botrule
\end{tabular}}

\begin{tabnote}
{$\lambda_{mean}$ and $\Delta\lambda$ respectively are the effective wavelength and bandwidth of the filters. Details on the UVIT filters can be found in \citet{2017AJ....154..128T}.
For JO201 and JW100, the redshifts given  are the cluster redshifts \citep{2019MNRAS.482.4466P}. Distances to JO201 and JW100 are estimated from the cluster redshifts ({\url{https://www.astro.ucla.edu/~wright/CosmoCalc.html}}).}
\end{tabnote}
\end{table*}

\begin{figure*}
    
    \includegraphics[width=0.95\textwidth]{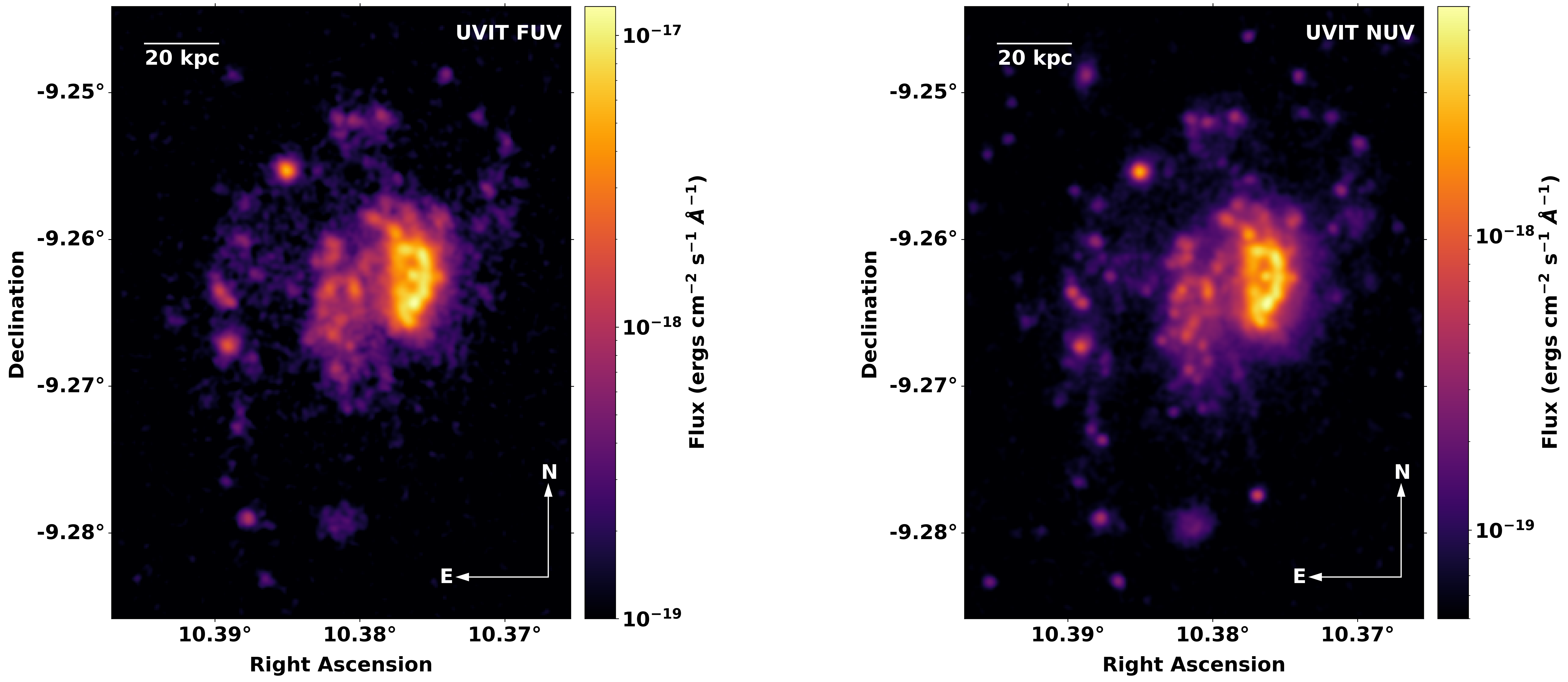}
    \centering
    \includegraphics[width=0.95\textwidth]{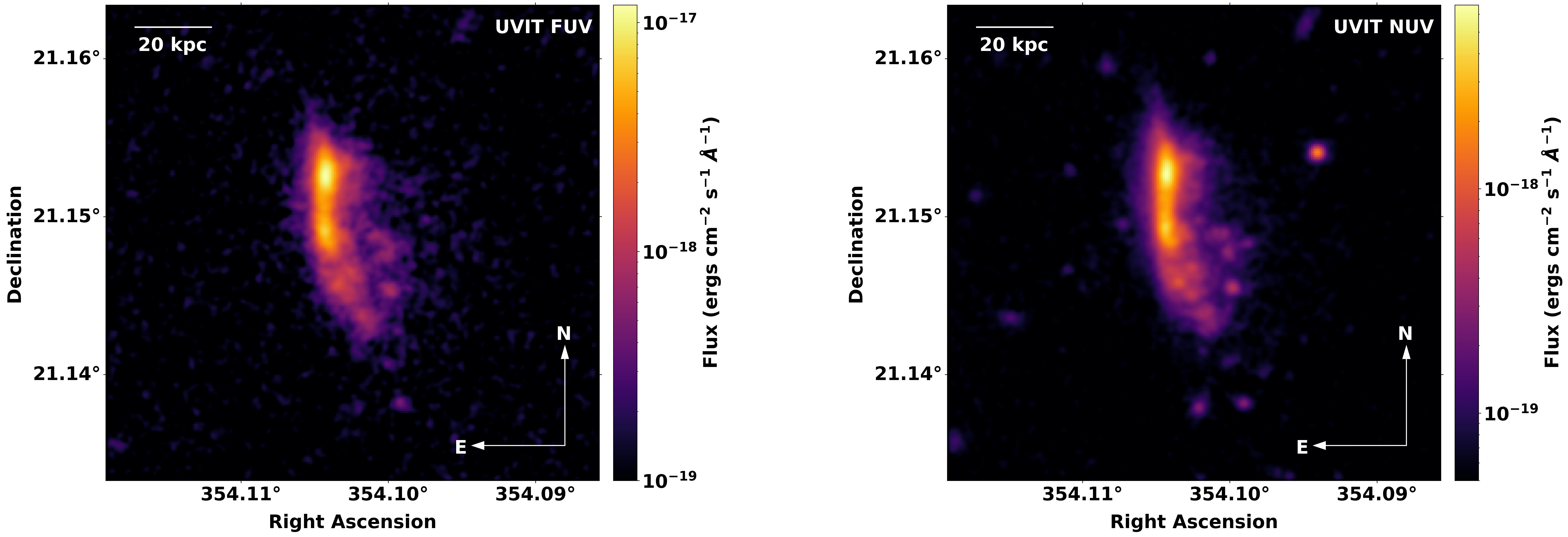}
    
    \caption{UVIT images of JO201 (top) and JW100 (bottom). The contrast level is adjusted to highlight the features.}
    \label{fig:uvit-fuv-nuv}
\end{figure*}

\section{Observations and data analysis} 
\label{section3}
In this section, we present the details of observation and data analysis for the jellyfish galaxies JO201 and JW100. The analysis of the NGC 5291 and NGC 7252 galaxy systems is detailed in \citet{2023MNRAS.522.1196R} and  \citet{2025PASA...42...73S}.

\subsection{Ultraviolet imaging}

To study the dust attenuation and recent star formation activity in the two jellyfish galaxies: JO201 in Abell 85 and JW100 in Abell 2626, the FUV and NUV imaging data from the UVIT  instrument \citep{2006AdSpR..38.2989A} are used (PI: Koshy George).
UVIT consists of two co-aligned telescopes, one for FUV  (1300 - 1800 \AA) and the other for NUV  (2000 - 3000 \AA) and VIS  (3200 - 5500 \AA) ranges. The telescope observes simultaneously in the FUV, NUV and VIS channels with a circular field of view of $\sim$28$^{\prime}$ diameter and,  spatial resolution of $\sim$1.2$^{\prime\prime}$ in NUV and $\sim$1.4$^{\prime\prime}$ in FUV. Multiple filters are provided for observation for both FUV and NUV channels \citep{2016SPIE.9905E..1FS,2017AJ....154..128T,2020AJ....159..158T}. Details of the UVIT observations of the galaxies are given in Table \ref{tab:uvit-observations-jellyfish}.

The Level 1 (L1) data of the targets are reduced to science-ready Level 2 (L2) images using CCDLAB which is a UVIT data reduction software \citep{2017PASP..129k5002P,2021JApA...42...30P}. The L1 data comprises of multi-orbit data sets. CCDLAB extracts and digests the LI data and then applies various corrections to the data sets, including fixed-pattern noise correction, distortion correction, and spacecraft drift correction. CCDLAB then aligns orbit-wise images to a common frame and  merges them to create a single master image on which astrometric corrections are performed. The final science-ready images (L2 images) are in counts. The UVIT FUV and NUV images of the jellyfish galaxies are shown in Figure \ref{fig:uvit-fuv-nuv}. 

The galaxy systems, NGC 5291 and NGC 7252, were also observed with AstroSat/UVIT (PI: Koshy George). The same FUV and NUV filters were used for observations, as in the cases of JO201 and JW100.  Data reduction was carried out using the CCDLAB pipeline , following the same procedure described for the jellyfish galaxies. The details of UVIT observations, data reduction and the final integration times are given in \citet{2023MNRAS.522.1196R} and \citet{2025PASA...42...73S}, and summarised in Table \ref{tab:uvit-observations-jellyfish}.

\subsection{Optical imaging}

We use r-band images to perform a detailed colour analysis of the star-forming  knots in JO201 and JW100, which will be discussed in Section \ref{section4}. The optical r-band images of the two galaxies are obtained from the DECaLS\footnote{\url{https://www.legacysurvey.org/decamls/}} (Dark Energy Camera Legacy Survey) database \citep{2019AJ....157..168D}.  DECaLS makes use of the Dark Energy Camera (DECam; \cite{2015AJ....150..150F})  mounted on the Victor M. Blanco 4m telescope at the Cerro Tololo Inter-American Observatory. The r-band (effective wavelength = 6382.6 {\AA}; \cite{2011ApJ...737..103S}) coadded images of the target galaxies downloaded from the survey database are science-ready images that can be used directly for photometry. 

A similar analysis was also performed for the NGC 5291 and NGC 7252 galaxy systems using DECaLS r-band imaging data, as presented in \citet{2025PASA...42...73S}.

\subsection{Source extraction and identification of star-forming knots}

To identify the star-forming (SF) knots belonging to JO201 and JW100, particularly those outside the galaxy disks, we use the UVIT FUV images of the targets. For each galaxy, sources are extracted from the FUV image, and the FUV source segmentation map is forced on the NUV and r-band images for extracting NUV and r-band fluxes. The star-forming knots located outside the disks of the jellyfish galaxies are identified using the segmentation maps, by overlaying stellar disk contours to distinguish between knots within the disk and those in the tails. The procedure is detailed in the following subsections.

\subsubsection{AGN flux removal}

Both jellyfish galaxies, JO201 and JW100, host AGN. Before source extraction, flux from the AGN-dominated regions of the two galaxies is first removed by using the optical emission line ratio diagnostic diagrams given in \citet{2019MNRAS.482.4466P}. To remove the AGN flux from the UV and optical images, the area corresponding to the central AGN is outlined using an optimal polygon, and the pixels within the polygonal region are replaced with the per-pixel sky background values. The AGN-subtracted images thus obtained are then used for source extraction and the identification of SF knots.

\subsubsection{Source extraction}

\begin{figure}
    \includegraphics[width=0.475\textwidth]{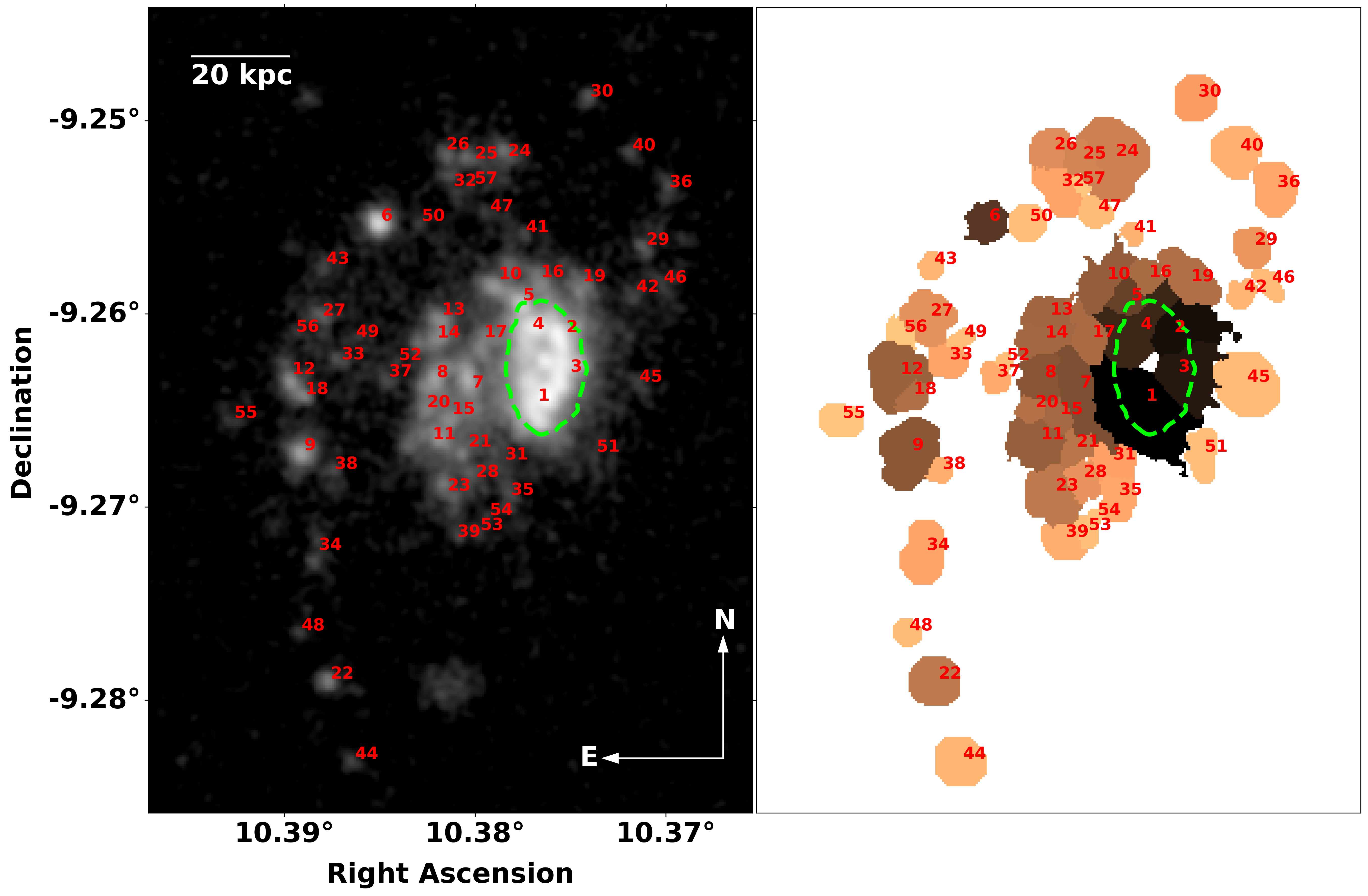}

    \includegraphics[width=0.475\textwidth]{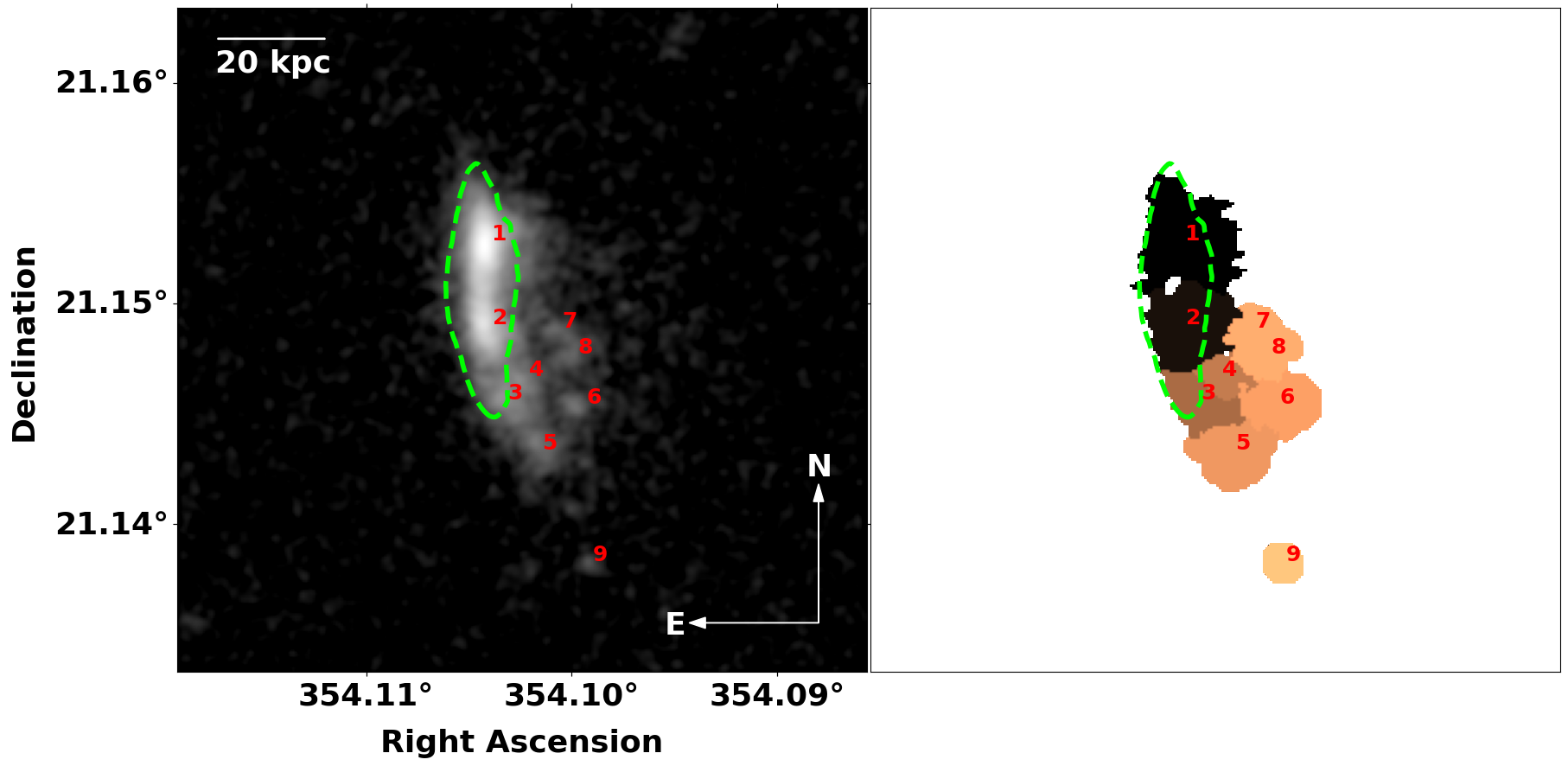}
    
    \caption{FUV images of JO201 (top left) and JW100 (bottom left) with SF knots marked in red. The segmentation maps of JO201 (top right) and JW100 (bottom right) are shown in the right. The colours of the segments indicate the relative brightness of SF knots with brighter SF knots corresponding to darker shades of brown. z-band isophotes corresponding to 22 mag/arcsec$^2$ shown (green-dashed contours).}
    \label{fig:segmentation}
\end{figure}

Source extraction is performed on ultraviolet (FUV and NUV bands) and optical (r-band) images of target galaxies using the ProFound source extraction package \citep{2018MNRAS.476.3137R}. The sources are identified from the FUV image first (the coarser resolution image). The FUV images are in counts. The counts image, along with the per-pixel sky and sky standard deviation values, is provided to the ProFound function. 

Some of the ProFound function parameters\textemdash{}\textit{skycut}, \textit{sigma}, \textit{pixcut}, \textit{ext} and \textit{tolerance} (see the documentation\footnote{\url{https://www.rdocumentation.org/packages/ProFound/versions/1.14.1/topics/ProFound}} and \citet{2018MNRAS.476.3137R} for details on these parameters)\textemdash{}are also modified from their default values to achieve optimal source extraction and source segmentation. 
ProFound identifies sources from the FUV image and generates the source segmentation map along with the statistics including right ascension (RA), declination (Dec), flux counts, and area for each extracted source (segment). Finally, forced photometry is performed on the  NUV and  r-band images using the FUV segmentation map. 

To derive the FUV and NUV fluxes and magnitudes of the identified sources, the calibration equations provided in \citet{2017AJ....154..128T} are used along with the zero-point magnitudes and unit conversion factors of the UVIT filters \citep[Table 4]{2017AJ....154..128T}. The r-band magnitudes are calculated using the conversion equation, $AB\,mag=22.5-2.5\log \, (flux_{nanomaggy})$, given in the header file of the DECaLS coadded images.

\subsubsection{Star-forming knots in JO201 and JW100}

The identified star-forming  knots and the corresponding ProFound segmentation map showing the extent of each knot are shown in Figure \ref{fig:segmentation}. The DECaLS z-band images of the jellyfish galaxies are used to generate isophotes corresponding to a surface brightness level of $\sim$22 mag/arcsec$^2$. This contour is defined to trace the stellar disk of the jellyfish galaxies and to distinguish SF knots in the disk from the tail.  

For JO201, a total of 57 SF knots (5 in the disk\footnote{Segments whose centroids lie  within or really close to the green contour are considered as disk SF knots in the present analysis.} and 52 in the tails) are identified from the FUV image with  size\footnote{Size is the approximate elliptical major axis length of each segment} ranging from 4.0 to 22.3 kpc. For JW100,  9 SF knots (3 in the disk and 6 in the tails) with  size ranging from 8.3 to 21.5 kpc are identified.  All identified SF knots have  signal-to-noise ratio (SNR) greater than 4.

\subsection{Galactic extinction correction}

The observed UV (FUV and NUV) and r-band fluxes are corrected for 
foreground Galactic extinction using the \citet{1989ApJ...345..245C}  and \citet{1994ApJ...422..158O} extinction laws, respectively, 
assuming a visual extinction to reddening ratio, $R_v$[$=A_v/E(B-V)$] = 3.1. The E(B-V) values are taken from the reddening map of \citet{2011ApJ...737..103S}. 
For JO201, the estimated values for Galactic extinction in the FUV, NUV, and r bands are: $A_{FUV}$(Galactic) = 0.261 mag, $A_{NUV}$(Galactic) = 0.242 mag and $A_{r}$(Galactic) = 0.081 mag. For JW100, $A_{FUV}$(Galactic) = 0.460 mag, $A_{NUV}$(Galactic) = 0.427 mag, and $A_{r}$(Galactic) = 0.143 mag.

\section{Results}
\label{section4}

This section presents the new results (dust attenuation estimated from slope $\beta$) for JO201 and JW100. 
It is to be noted that the $\beta$ slope is also sensitive to the age distribution of stars in galaxies, i.e., the star formation history (SFH) \citep{2011ApJ...741..124H, 2012A&A...539A.145B}. The presence of old stars could redden the intrinsic UV slope.
This sensitivity could potentially bias our measurements of dust attenuation. 
This is particularly significant in the disks of the galaxies, where a mix of old and young stellar populations is present. However, the knots in the debris are characterised by relatively young ages as described in Section \ref{section-beta-slope}. To minimise the bias, we restrict our analysis to the SF knots in the stripped tails of JO201 and JW100, where the stellar populations have a relatively narrow age range \citep{2024A&A...682A.162W}.

The corresponding results for the NGC 5291 and NGC 7252 galaxy systems are presented in \citet{2025PASA...42...73S} and also summarised in Table \ref{tab:comparison_gravi_hydro}.

\subsection{NUV-r colours}

\begin{figure}
    \centering
    \includegraphics[width=0.9\linewidth]{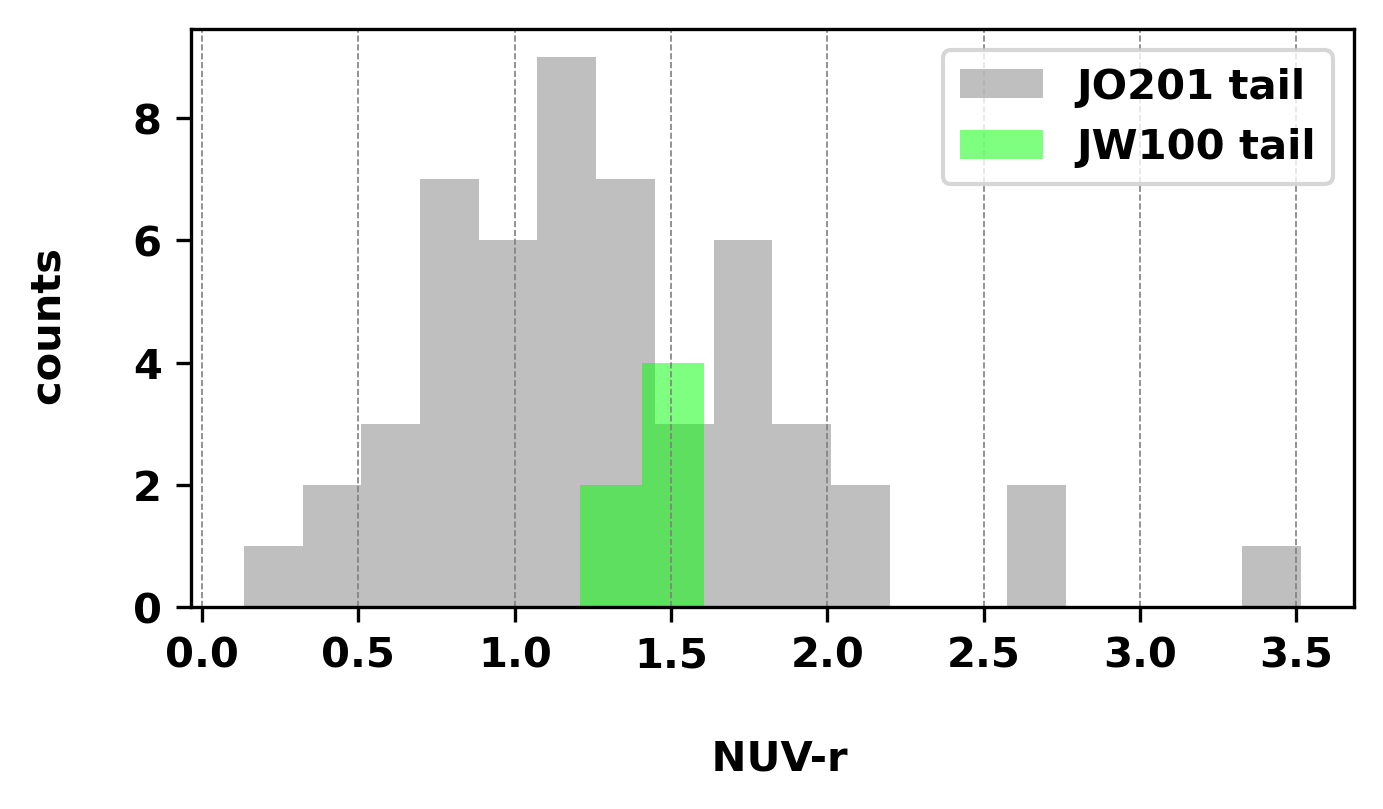}
    \caption{Distribution of NUV-r colours of the SF knots in the tails of JO201 and JW100}
    \label{fig:jellyfish_nuvr_hist}
\end{figure}

To confirm that the identified sources are indeed knots with recent star formation and to check for any contaminants such as background ellipticals exhibiting UV upturn phenomenon due to the presence of evolved population of stars on the horizontal branch \citep{1993ApJ...419..596D, 1995ApJ...442..105D}, we analyse their NUV-r colours.
The NUV-r colours of the SF knots in the tails of the jellyfish galaxies JO201 and JW100 are estimated from the Galactic extinction-corrected NUV and r-band magnitudes and are presented in Figure \ref{fig:jellyfish_nuvr_hist}. For the 
SF knots in the tails of JO201, the NUV-r colours range from  0.13 to 3.51. For JW100, the NUV-r colours of the SF knots in the tails range from 1.21 to 1.61. The mean and median NUV-r colours in the  tails of JO201  are 1.30 and 1.22, respectively. For JW100, the mean and median NUV-r colours in the tails are 1.44 and 1.47, respectively.
All the identified knots have NUV-r colours $<$ 5.4, indicating recent star formation \citep{2007ApJS..173..512S, 2007ApJS..173..619K}.

\begin{figure}
    \centering
    \includegraphics[width=0.9\linewidth]{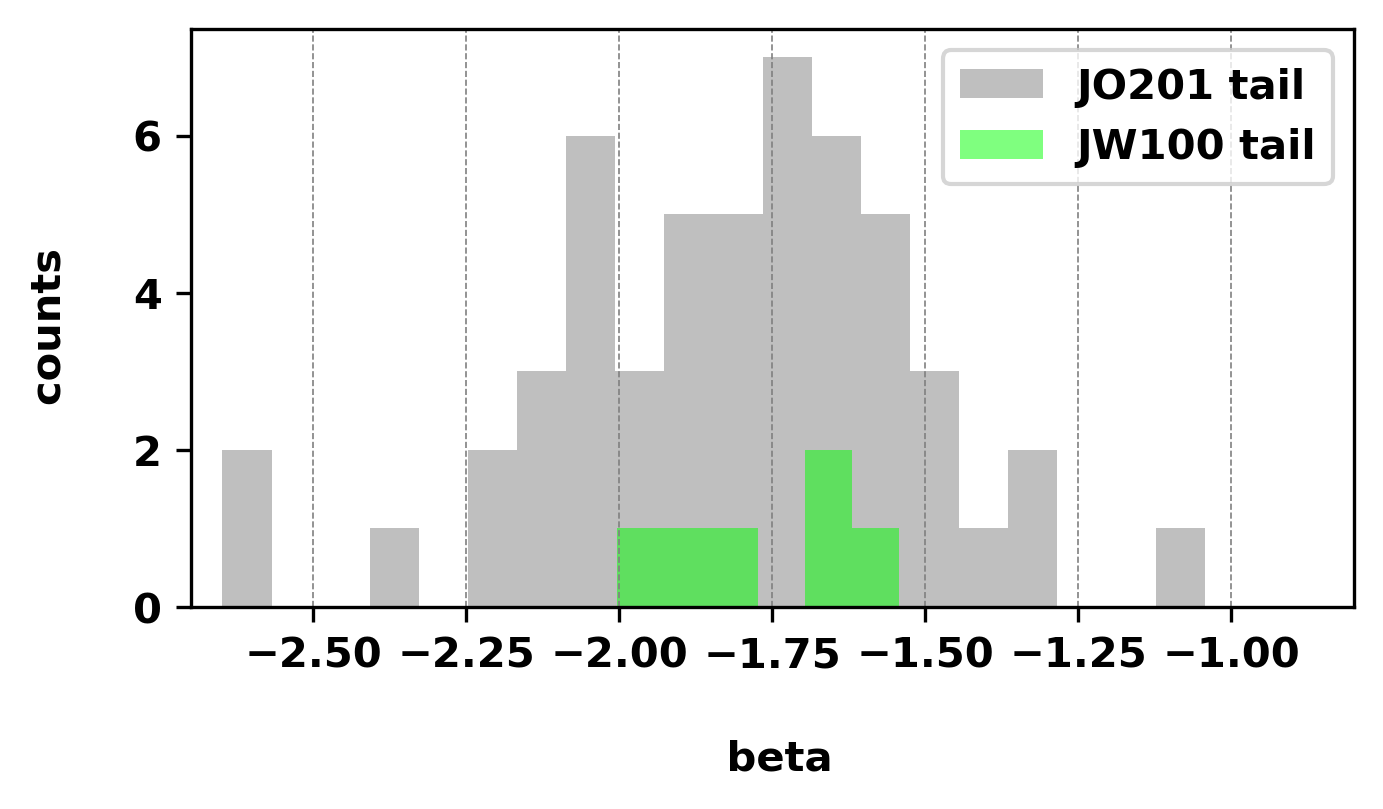}
    \caption{The distribution of $\beta$ of the SF knots in the tails of the jellyfish galaxies JO201 and JW100}
    \label{fig:jellyfish-beta-hist}
\end{figure}

\begin{figure*}
    \includegraphics[width=\linewidth]{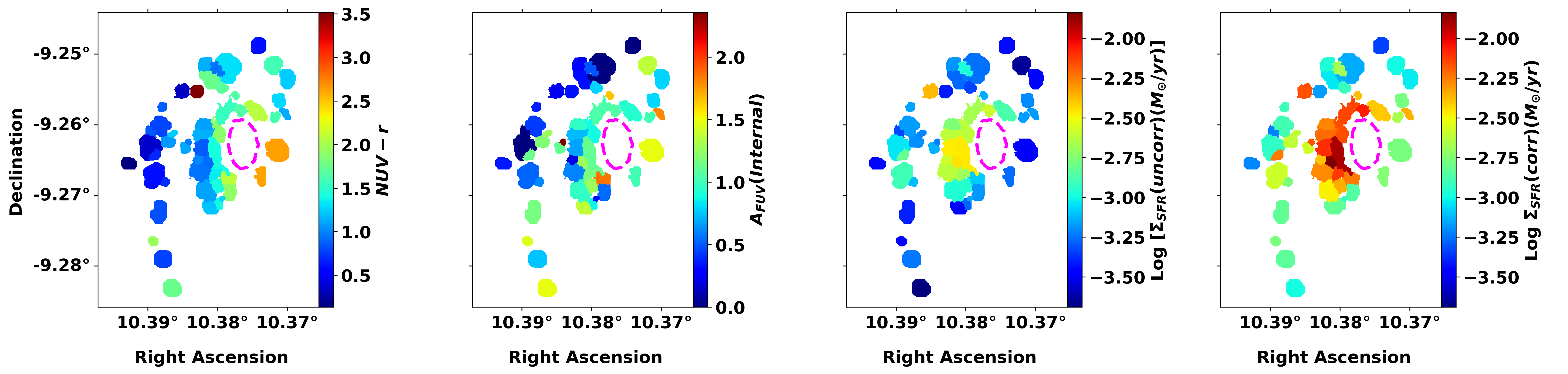}

    \vspace{3em}
    
    \includegraphics[width=\linewidth]{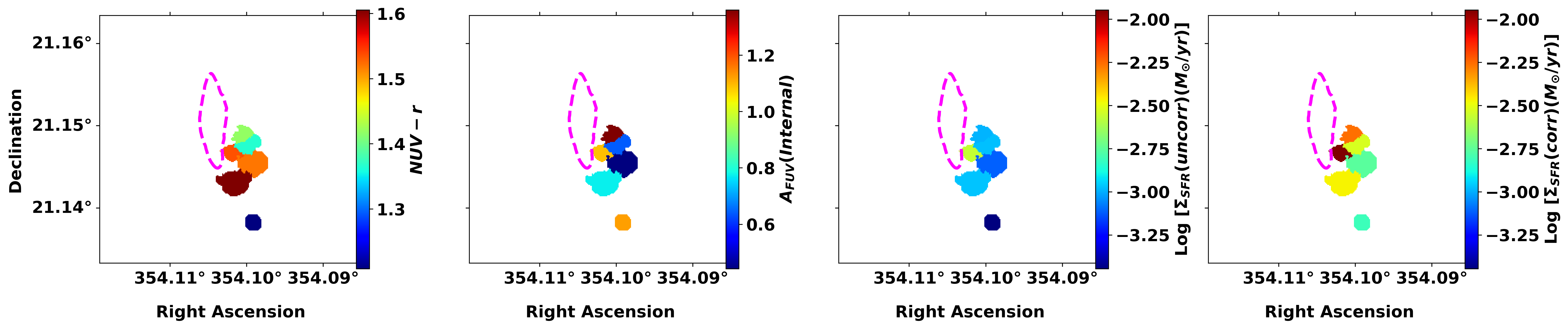}
    \caption{Estimated parameters of JO201 (top) and JW100 (bottom). The magenta contours trace the disks of the galaxies.}
    \label{fig:jo201-jw100-spatial-map}
\end{figure*}

\subsection{Slope of the UV continuum \& internal attenuation}
\label{section-beta-slope}

In the present study, we use the slope of the UV continuum, $\beta$, to estimate the internal attenuation in the resolved SF knots of the two jellyfish galaxies.  
The UV continuum slope $\beta$ is an indicator of the dust attenuation in actively SF galaxies where the UV spectrum is dominated by the photospheric emission of young stars.

For the UVIT bands,  $\beta$ is estimated using the relation given in \citet{2023MNRAS.522.1196R}:
\begin{equation}
   \hspace{1.5cm} \beta_{UVIT}=1.88(m_{FUV}-m_{NUV})-2.0
    \label{eq:beta}
\end{equation}  
where $m_{FUV}$ and $m_{NUV}$ are the FUV and NUV magnitudes corrected for Galactic extinction. 
The internal attenuation is then estimated using the Meurer relation \citep[hereafter M99]{1999ApJ...521...64M} (for starburst case): 
\begin{equation}
\hspace{1.5cm} A_{FUV}\,\,(Internal)=4.43+1.99\;\beta
    \label{eq:afuv}
\end{equation}
where $\beta$ is determined using Equation \ref{eq:beta}.

The $\beta$ values of the SF knots in JO201 and JW100 are shown in Figure \ref{fig:jellyfish-beta-hist}. For the SF knots in the tails of JO201, the $\beta$ values range from -2.65 to -1.04. M99 relation assumes that, for dust-free SF regions, the UV continuum has an intrinsic slope $\beta_i$ = -2.23.  Three SF knots in the JO201 tails have $\beta$ values: -2.65, -2.59 and -2.39. M99 relation cannot be used for these knots as it yields negative attenuation values. Since the $\beta$ values of these SF knots are close to the $\beta$ of dust-free SF regions, they are assumed to be dust-free SF knots in the present study. For JW100, the $\beta$ values of the SF knots in the tails range from -2.00 to -1.54.

For the SF knots in the tails of the jellyfish galaxy JO201, the estimated internal attenuation values range from 0 to 2.35 with a mean value of 0.85 and a median of 0.87. For JW100, the internal attenuation values of the SF knots in the tails range from  0.44 to 1.36 with mean and median values of 0.91 and 0.93, respectively.

From the internal attenuation map (Figure \ref{fig:jo201-jw100-spatial-map}), we observe that the dust distribution is not uniform in JO201 and JW100. Knots with high internal attenuation values are seen on tails of the jellyfish galaxies. (The high attenuation values in the tails of the jellyfish galaxies could also be due to foreground/background contaminants.) 

It is to be noted that while the UV continuum is dominated by emission from massive stars with ages $< 100$ Myr, stars with ages upto 1 Gyr can contribute to the UV emission, especially NUV  \citep{2011ApJ...741..124H}. Old stars can thus redden the FUV-NUV colour and hence the intrinsic UV slope.  The knots in the debris, formed in situ from the gas removed during interactions, are characterised by relatively young ages. For JO201 and JW100, the knots have ages $\lesssim 100$ Myr \citep{2024A&A...682A.162W}. 
The ages of the knots in the NGC 5291 ring range from $\sim$ 5 Myr to a few $10\times10^6$ years \citep{2007A&A...467...93B}.  While the exact ages of the identified knots in the NGC 7252 tails is not known, the dynamical model of \citet{2010MNRAS.407...43C} predict that the stellar populations have ages between $\sim 10 - 600$ Myr.
Though the SF regions are dominated by young stars, the presence of an underlying old stellar component, such as those removed from the progenitors, can also contaminate the flux measured within a segment. In the jellyfish tails, which originate from hydrodynamic interactions, there cannot be the presence of such an old component.
\citet{2012MNRAS.424.1522W} predict that for SF galaxies with a constant SFR over a duration of 10 Myr - 1 Gyr, the change in the FUV-NUV colour is small, $\sim 0.12$. Using Equation \ref{eq:beta}, this corresponds to a change in intrinsic slope, $\delta \beta_i \sim 0.23$, which corresponds to an uncertainty in attenuation, $\delta A_{FUV} \sim 0.5$ mag. Even for the limiting case of an instantaneous burst, for relatively young ages, the variation in the FUV-NUV colour is only $\sim 0.3$ over a duration of 10 - 100 Myr \citep{2012MNRAS.424.1522W}.  
Considering the relatively young ages of the knots  and their locations outside the progenitor disks, the reddening of $\beta_i$ due to older stars can be considered modest over the SF timescale. However, it should also be noted that these predictions are for SF galaxies while the environment in collisional ring, tidal tails and ram-pressure stripped tails can be rather complex.

\subsection{Star formation rate}

For the estimation of SFR, we follow the relation given in \citet{2006ApJS..164...38I}  which is based on the assumption of a constant SFR over a timescale of $10^8$ years, with a Salpeter initial mass function (IMF) \citep{1955ApJ...121..161S} for stars in the mass range 0.1 to 100 M$_\odot$. The relation is: 

\begin{equation}
SFR_{FUV}[M_\odot/yr] = \frac{L_{FUV}[erg/sec]}{3.83 \times 10^{33}}\times 10^{-9.51}
        \label{eq:sfr}
\end{equation}
where, $L_{FUV}$ is the FUV luminosity of the source.

\begin{table}
    \setlength{\tabcolsep}{3.9pt}
    \caption{SFR and  $f_{obscured}$ in the tails of the jellyfish galaxies: JO201 and JW100}
    \label{tab:jellyfish-sfr}
    {\tablefont\begin{tabular}{@{\extracolsep{\fill}}lcccc}
    \toprule
    Region   & $SFR_{FUV}(uncorr)$  & $SFR_{FUV}(Galactic)$ & $SFR_{FUV}(corr)$  & $f_{obscured}$\\

    {}   &  {$(M_\odot/yr)$}  & {$(M_\odot/yr)$} & {$(M_\odot/yr)$} & {}\\
    
    \hline
    JO201 tail     & 3.7 & 4.6 & 10.3 & 0.55\\

    \hline
    JW100 tail     & 0.54 & 0.83 & 1.9 & 0.56
   \botrule
    \end{tabular}}
\end{table}
 
The SFR values in the tails of the jellyfish galaxies are given in Table \ref{tab:jellyfish-sfr}. $SFR_{FUV}(uncorr)$ is the SFR uncorrected for both foreground Galactic extinction and internal attenuation, $SFR_{FUV}(Galactic)$ is the SFR corrected for Galactic extinction but uncorrected for internal attenuation and, $SFR_{FUV}(corr)$ is the SFR corrected for both Galactic extinction and internal attenuation. The fraction of internal dust-obscured star formation, $f_{obscured}$ ($=\frac{SFR_{FUV}(corr)-SFR_{FUV}(Galactic)}{SFR_{FUV}(corr)}$) is also presented in Table \ref{tab:jellyfish-sfr}.

In the JO201 tail, the integrated SFR before internal attenuation correction is 4.6 $M_{\odot}/yr$, while the integrated SFR after correction is 10.3 $M_{\odot}/yr$. In the JW100 tails, the total SFR before internal attenuation correction is 0.83 $M_{\odot}/yr$, while the total SFR after correction is 1.9 $M_{\odot}/yr$ (Table \ref{tab:jellyfish-sfr}).
For both galaxies, the integrated SFR in the tails has increased by a factor\footnote{Increase factor = $\frac{SFR_{FUV}(corr)}{SFR_{FUV}(Galactic)}$} of $\sim$2 after accounting for internal dust attenuation.

The integrated SFRs derived from dust-corrected H$\alpha$  luminosities, assuming a Chabrier IMF \citep{2003PASP..115..763C}, in the jellyfish tails are: JO201 $\sim 1$ $M_{\odot}/yr$ and JW100 $\sim 0.8$ $M_{\odot}/yr$ \citep{2019MNRAS.482.4466P}. However, our estimates are based on the assumption of a Salpeter IMF. On converting from Chabrier to Salpeter IMF\footnote{The SFR increases by a constant factor of $\sim 1.6$ on converting from  Chabrier to Salpeter IMF \citep{2014ARA&A..52..415M}.}, the $SFR_{H\alpha}$ in the tails become: JO201 $\sim 1.6$ $M_{\odot}/yr$ and JW100 $\sim 1.3$ $M_{\odot}/yr$. For JW100,  our corrected $SFR_{FUV}$ is comparable to $SFR_{H\alpha}$. Whereas, for JO201, both our corrected and uncorrected $SFR_{FUV}$ values are higher than $SFR_{H\alpha}$. We note that, for JO201, the definition of disk differs between \citet{2019MNRAS.482.4466P} and our study. The disk considered in our study is smaller and consistent with the inner disk defined in \citet{2023ApJ...949...72G}. If we consider a larger disk, roughly similar to the one defined in \citet{2019MNRAS.482.4466P}, then the corrected $SFR_{FUV}$ in the JO201 tail decreases from 10.3 $M_{\odot}/yr$ to $\sim 6 - 7 $ $M_{\odot}/yr$. Therefore, a significant factor contributing to the discrepancy in the SFR values is how the galaxy boundary is defined. Besides, the discrepancy could also be due to differences in the choice of apertures, i.e., the sizes of the knots located outside the galactic disks.

\begin{figure}
    \centering
    \includegraphics[width=0.9\linewidth]{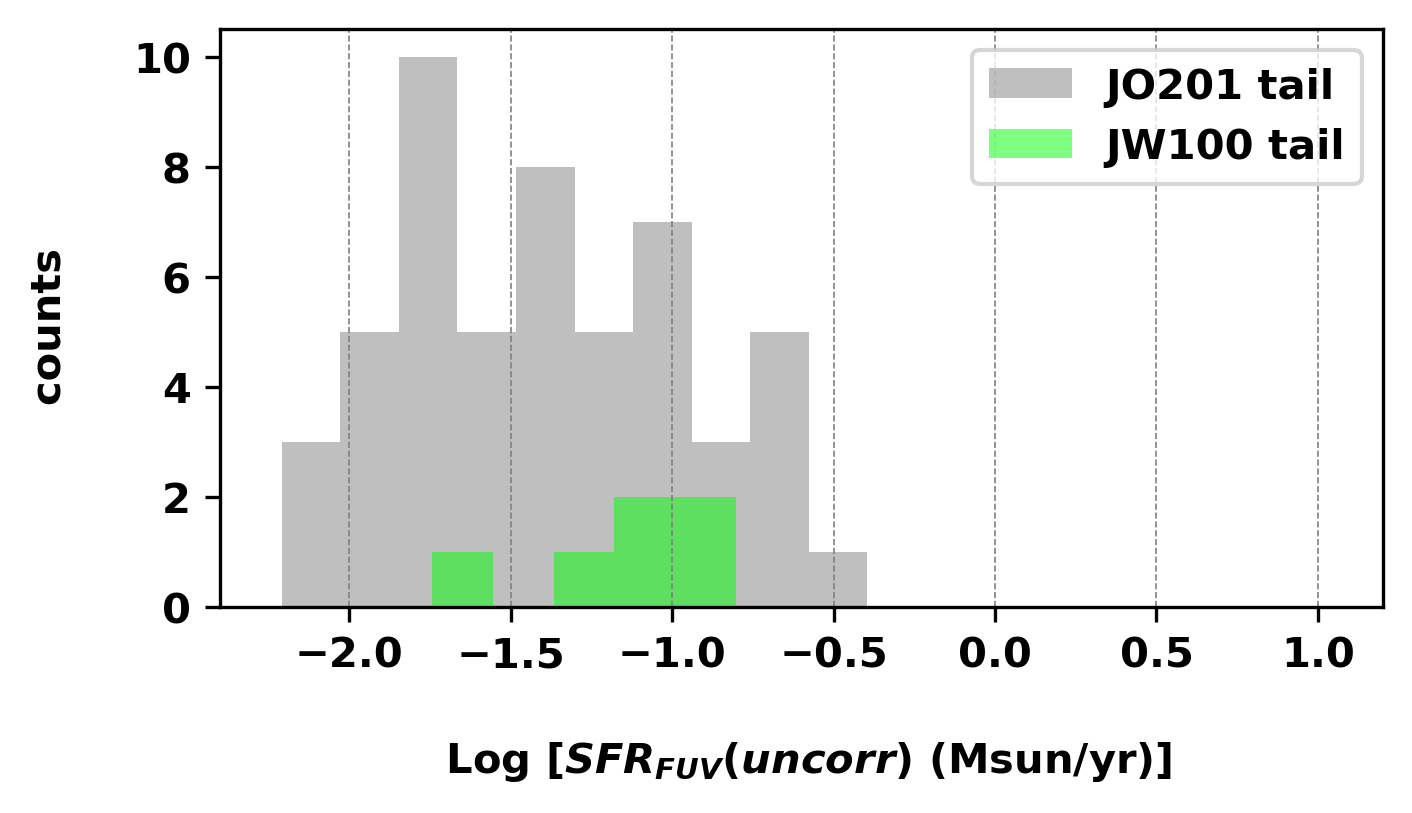}
    \includegraphics[width=0.9\linewidth]{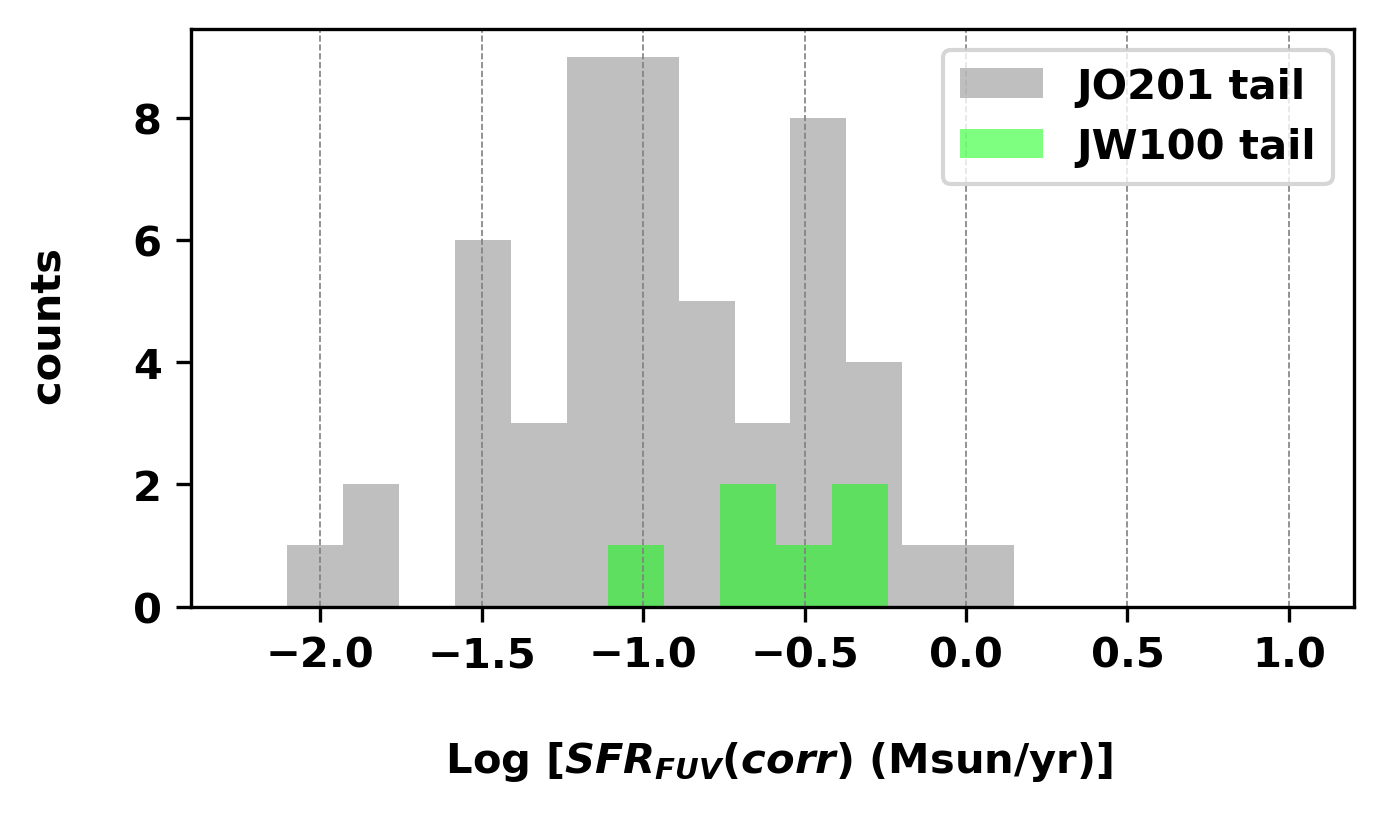}
    \caption{The distribution of $SFR_{FUV}(uncorr)$ (top) and $SFR_{FUV}(corr)$ (bottom) of the SF knots in the disk and tail of JO201 and JW100.}
    \label{fig:jellyfish-sfr-hist}
\end{figure}

The distribution of SFR of the SF knots in JO201 and JW100 is shown in Figure \ref{fig:jellyfish-sfr-hist}.  
The SFR surface density, $\Sigma_{SFR_{FUV}}$, of the resolved SF knots in JO201 and JW100 is shown in Figure \ref{fig:jo201-jw100-spatial-map} along with the NUV-r colour and internal attenuation. Table \ref{tab:jellyfish-sfr-sigmasfr-statistics} gives the range, mean and median values of $SFR_{FUV}(corr)$ and $\Sigma_{SFR_{FUV}}(corr)$ for the identified knots in the 
tails of JO201 and JW100 galaxies.
Overall, we observe a significant change in the SFR and SFR densities after correcting the FUV luminosities for the effect of internal dust.
The average corrected  SFR and SFR surface densities in the 
tails of JW100 are observed to be higher than those in the 
tails of JO201. 

\begin{table}
    \setlength{\tabcolsep}{12.5pt}
    \caption{$SFR_{FUV}(corr)$ and $\Sigma_{SFR_{FUV}}(corr)$ statistics of the resolved knots in the tails of JO201 and JW100}
    \label{tab:jellyfish-sfr-sigmasfr-statistics}
    {\tablefont\begin{tabular}{@{\extracolsep{\fill}}lccc}
    \toprule
    Region   & Statistic & $SFR_{FUV}(corr)$  & $\Sigma_{SFR_{FUV}}(corr)$\\

    {}   & {} &  $(M_\odot/yr)$ & $(M_\odot/yr/kpc^2)$\\
    
    \hline
    JO201 tail     & Range & (0.0079, 1.41) & (0.00046, 0.014) \\
                    & Mean & 0.20 & 0.0034 \\
                    & Median & 0.10 & 0.0020 \\

    \hline
    JW100 tail     & Range  & (0.078, 0.57) & (0.0015, 0.011) \\
                    & Mean & 0.31 & 0.0044\\
                    & Median &  0.27 & 0.0032
    \botrule
    \end{tabular}}
\end{table}

\subsection{Ram-pressure stripping of dust}

JW100 is an almost edge-on galaxy that is undergoing strong ram-pressure stripping, enabling a clear view of the ISM stripping. 
As mentioned in Section \ref{section1}, dust can also get stripped due to ram-pressure \citep{2005ASPC..331..281C,2014AJ....147...63A,2016AJ....152...32A,2020A&A...633L...7L,2020A&A...644A.161L,2022A&ARv..30....3B,2022MNRAS.509.3938L,2025A&A...701A..40G}, and since stripping of dust is not as efficient as the stripping of the diffuse HI component of the galaxy, we expect dust to be stripped to a smaller extent compared to gas \citep{2022A&ARv..30....3B}. Over the last two decades, studies have observed and confirmed the process of ram-pressure stripping of dust from the disk of galaxies, mostly by studying galaxies in the Virgo cluster \citep{2005ASPC..331..281C,2014AJ....147...63A,2020A&A...644A.161L}.

\citet{2023ApJ...945...54G} observed multiple dust lanes distributed irregularly in the nuclear regions of JW100 (see Figure 5 of \citet{2023ApJ...945...54G}).
In our study, it is observed that some knots in the tails of JW100\textemdash{knots 4 and 7 in Figure \ref{fig:segmentation} that lie very close to the galaxy's disk in projection} have high internal dust attenuation values.
The high dust attenuation observed in knots 4 and 7 of the tails may result from the ram-pressure stripping of dust from the disk of JW100. 
Additionally, the NUV–r colours of knots 4 and 7 closely resemble those of other star-forming knots in the tail.   
Our observations, along with previous studies by \citet{2023ApJ...945...54G}, suggest that strong ram-pressure is likely removing dust from the disk of JW100 and redistributing it into the tail regions.

The jellyfish galaxy, J0201 is also undergoing strong ram-pressure stripping similar to JW100. Unlike JW100, JO201 is an almost face-on galaxy (observer's view), and we are observing the one-sided tails due to the slight inclination of the velocity vector 
with respect to the line of sight.  
We observe that the SF knots located at the outermost parts of the JO201 tail generally exhibit the lowest attenuation values.

\section{Discussion}
\label{section5}

Gravitational and hydrodynamic interactions can remove and redistribute components of the interstellar medium (ISM), triggering star formation in the debris removed from and thus located outside galaxy disks.
The ISM of a galaxy is made up of gas and dust. Although dust contributes only $\sim$1\% to the total mass budget of the ISM, it is an essential part of the ISM \citep{2020A&A...633A.100C}. Dust can obscure more than half of the stellar radiation in the UV-optical range \citep{2001PASP..113.1449C}, limiting our ability to understand the true SFR of galaxies. 
However, the degree to which dust in the debris of hydrodynamic and gravitational interactions masks star formation activity is still unknown.
The main objective of this study is to compare the dust attenuation and star formation activity in the debris of gravitational (collisional ring of NGC 5291 and tidal tails of NGC 7252)  and hydrodynamic (ram-pressure stripped tails of JO201 and JW100)  interactions. The study makes use of high-resolution AstroSat/UVIT data. The UV continuum slope ($\beta$) method is employed to estimate the ultraviolet dust attenuation, and the SFR is derived from the dust-corrected FUV luminosities.

We present a detailed comparative analysis of the star-forming knots in the collisional ring of the NGC 5291 system, the tidal tails of the NGC 7252 system, and the ram-pressure stripped tails of JO201 and JW100.
For the NGC 5291 and NGC 7252 systems, we utilise data and results from our previous study \citep{2025PASA...42...73S}.
The estimated parameters of the SF knots  in the debris of gravitational \citep{2025PASA...42...73S} and hydrodynamic interactions (present study) are detailed in Table \ref{tab:comparison_gravi_hydro} and the results are discussed below.

\begin{table*}
\centering
\caption{Estimated parameters of the SF knots in the debris of gravitational versus hydrodynamic interactions}
\label{tab:comparison_gravi_hydro}
{\tablefont\begin{tabular}{@{\extracolsep{\fill}}ll|cc|cc}
\toprule

& & \multicolumn{2}{c|}{\textbf{Gravitational interaction}} & \multicolumn{2}{c}{\textbf{Hydrodynamic interaction}} \\

& & \multicolumn{2}{c|}{\textbf{\citet{2025PASA...42...73S}}} & \multicolumn{2}{c}{\textbf{Present study}} \\

\textbf{Parameter} & \textbf{Statistic} & \textbf{NGC 5291 ring} & \textbf{NGC 7252 tails} & \textbf{JO201 tails} & \textbf{JW100 tails} \\
\hline
$\beta$   
  & Range & (-2.26, -1.79) & (-1.66, -1.06) & (-2.65, -1.04) & (-2.00, -1.54) \\
  & Mean & -2.01 & -1.42 & -1.82 & -1.77 \\
  & Median & -2.00 & -1.40 & -1.78 & -1.76 \\
  & SD & 0.12 & 0.21 & 0.30 & 0.16 \\
\hline

$A_{FUV}(Internal)$
  & Range & (0, 0.87) & (1.12, 2.33) & (0, 2.35) & (0.44, 1.36) \\
$(mag)$ & Mean & 0.42 & 1.61 & 0.85 & 0.91 \\
  & Median & 0.45 & 1.64 & 0.87 & 0.93 \\
  & SD & 0.24 & 0.42 & 0.53 & 0.31 \\
\hline

$SFR_{FUV}(uncorr)$  
  & Range & (0.0014, 0.015) & (0.0025, 0.037) & (0.0062, 0.40) & (0.018, 0.16) \\
$(M_\odot/yr)$  & Mean & 0.018 & 0.018 & 0.070 & 0.091 \\
  & Median & 0.0088 & 0.019 & 0.038 & 0.086 \\
  & SD & 0.029 & 0.013 & 0.079 & 0.047 \\
\hline

$SFR_{FUV}(Galactic)$
  & Range & (0.0021, 0.23) & (0.0031, 0.045) & (0.0079, 0.51) & (0.028, 0.24) \\
$(M_\odot/yr)$  & Mean & 0.027 & 0.022 & 0.089 & 0.14 \\
  & Median & 0.013 & 0.023 & 0.049 & 0.13 \\
  & SD & 0.044 & 0.016 & 0.10 & 0.071 \\
\hline

$SFR_{FUV}(corr)$
  & Range & (0.0021, 0.40) & (0.017, 0.20) & (0.0079, 1.41) & (0.078, 0.57) \\
$(M_\odot/yr)$  & Mean & 0.042 & 0.091 & 0.20 & 0.31 \\
  & Median & 0.019 & 0.10 & 0.10 & 0.27 \\
  & SD & 0.068 & 0.068 & 0.23 & 0.17 \\
\hline

$\Sigma_{SFR_{FUV}}(uncorr)$
  & Range & (0.0006, 0.0080) & (0.00037, 0.0014) & (0.0002, 0.0043) & (0.00036, 0.0027) \\
$(M_\odot/yr/kpc^2)$  & Mean & 0.0015 & 0.00082 & 0.0011 & 0.0012 \\
  & Median & 0.0011 & 0.00063 & 0.00067 & 0.0010 \\
  & SD & 0.0012 & 0.00045 & 0.00098 & 0.00074 \\
\hline

$\Sigma_{SFR_{FUV}}(Galactic)$
  & Range & (0.00091, 0.012) & (0.00045, 0.0018) & (0.00026, 0.0055) & (0.00055, 0.0041) \\
$(M_\odot/yr/kpc^2)$  & Mean & 0.0023 & 0.001 & 0.0015 & 0.0018 \\
  & Median & 0.0017 & 0.00077 & 0.00086 & 0.0016 \\
  & SD & 0.0018 & 0.00055 & 0.0012 & 0.0011 \\
\hline

$\Sigma_{SFR_{FUV}}(corr)$
  & Range & (0.00097, 0.014) & (0.0015, 0.0080) & (0.00046, 0.014) & (0.0015, 0.011) \\
$(M_\odot/yr/kpc^2)$  & Mean & 0.0034 & 0.0043 & 0.0034 & 0.0044 \\
  & Median & 0.0027 & 0.0040 & 0.0020 & 0.0032 \\
  & SD & 0.0026 & 0.0021 & 0.0030 & 0.0034 
\botrule
\end{tabular}}
\end{table*}

\subsection{Dust in collisional ring, tidal tails and ram-pressure stripped tails}

\begin{figure*}
    \centering
    \includegraphics[width=0.43\linewidth]{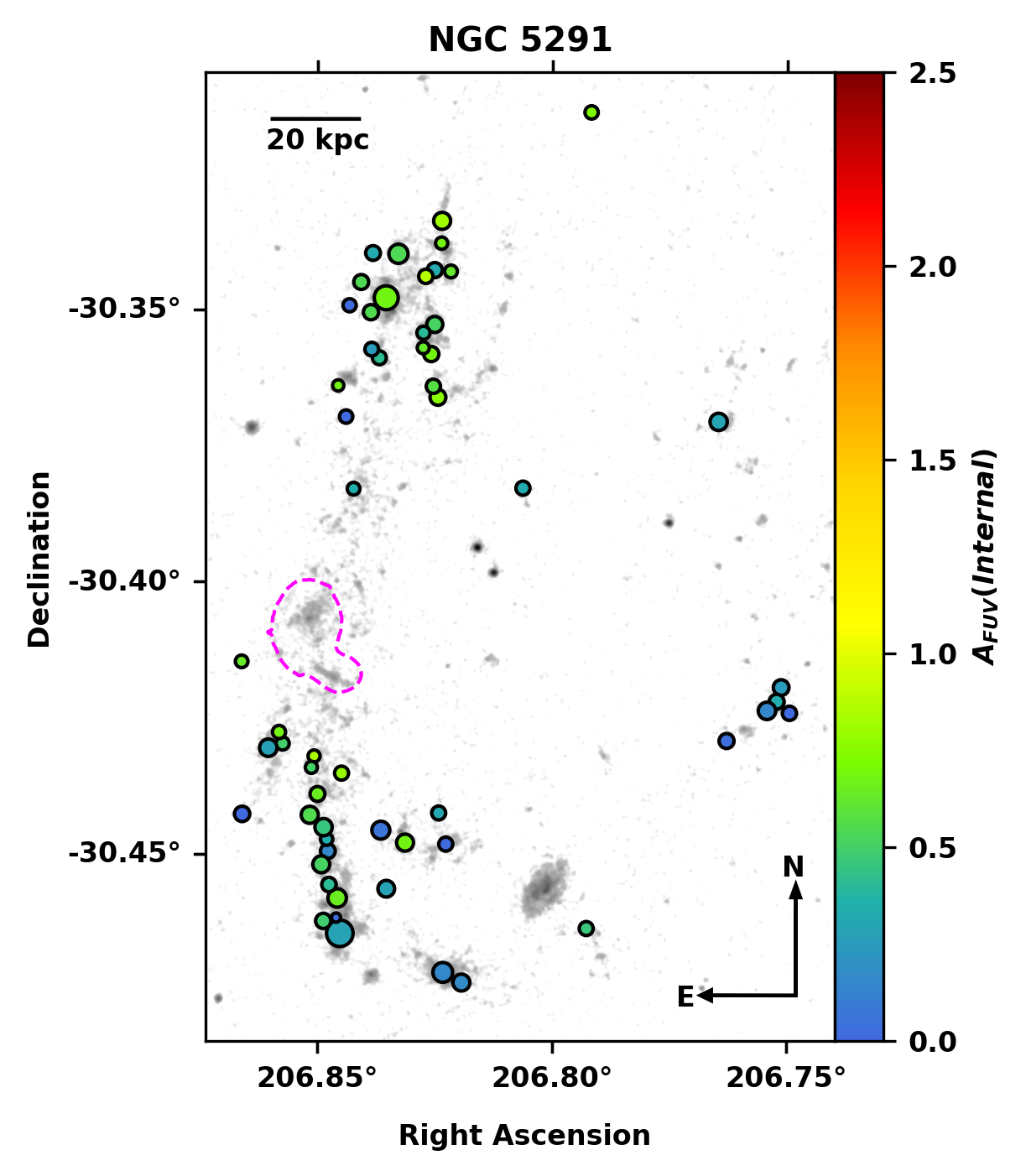}
    \hfill
    \includegraphics[width=0.48\linewidth]{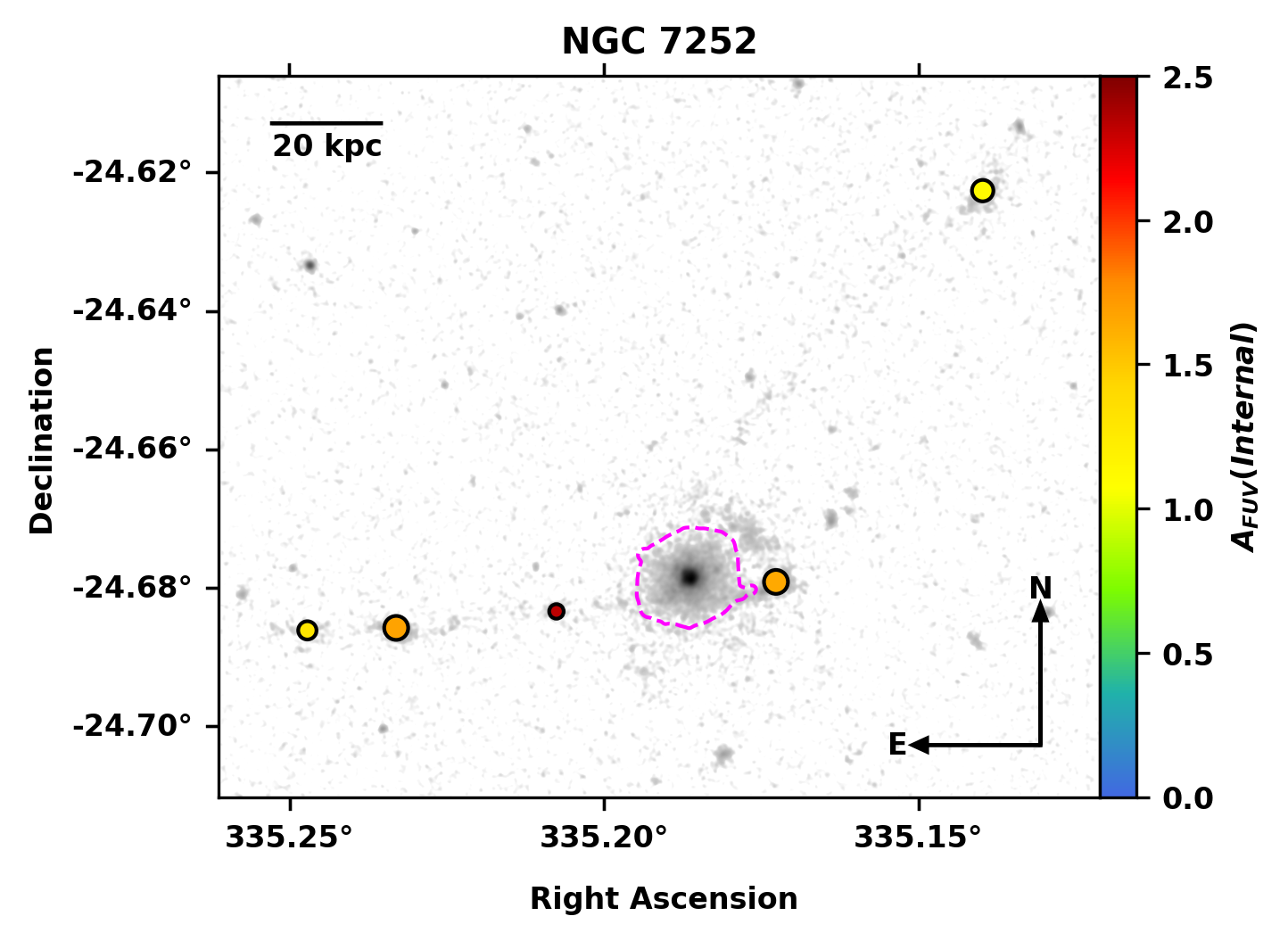}

    \vfill
    
    \includegraphics[width=0.43\linewidth]{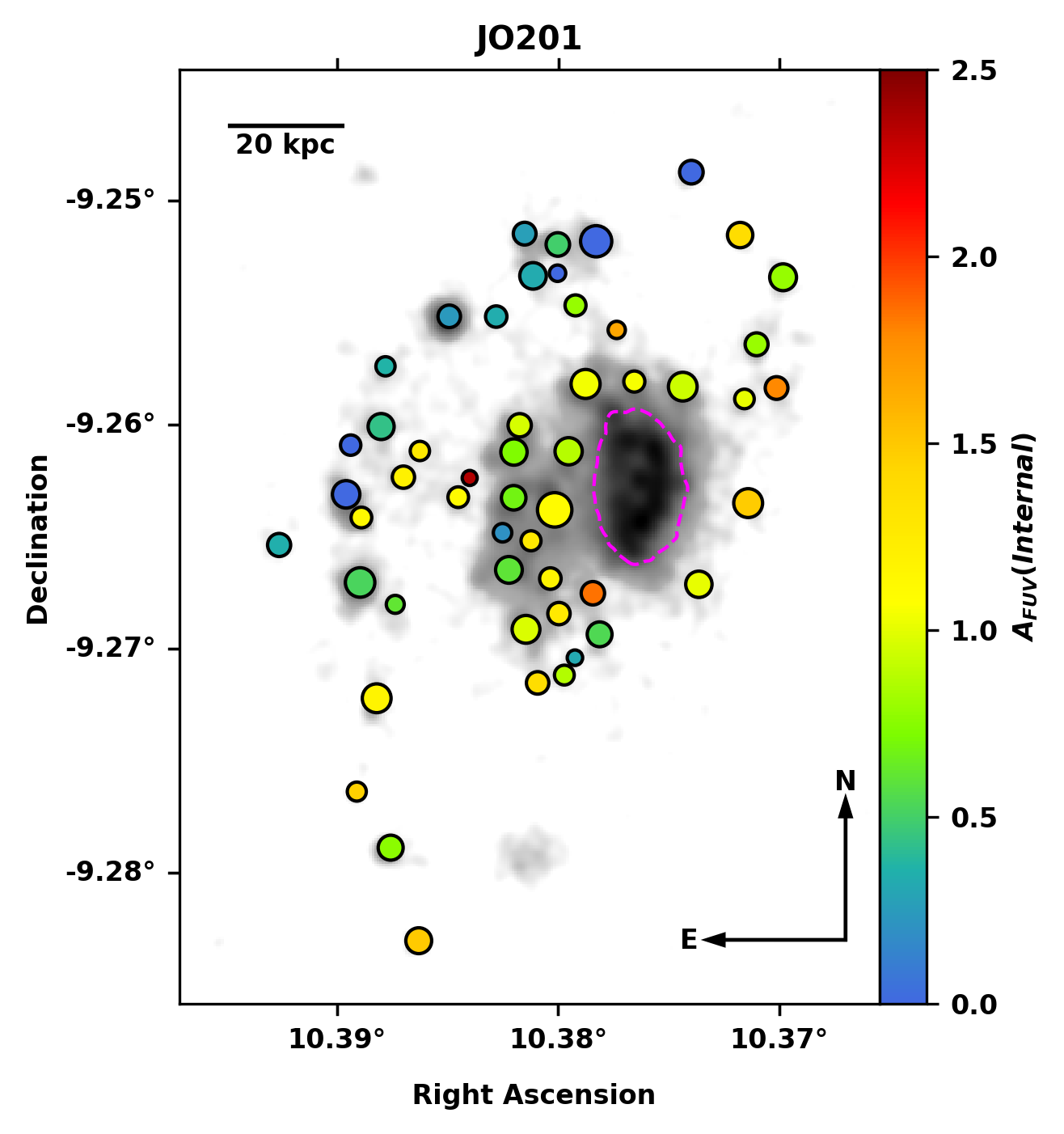}
    \hfill
    \includegraphics[width=0.48\linewidth]{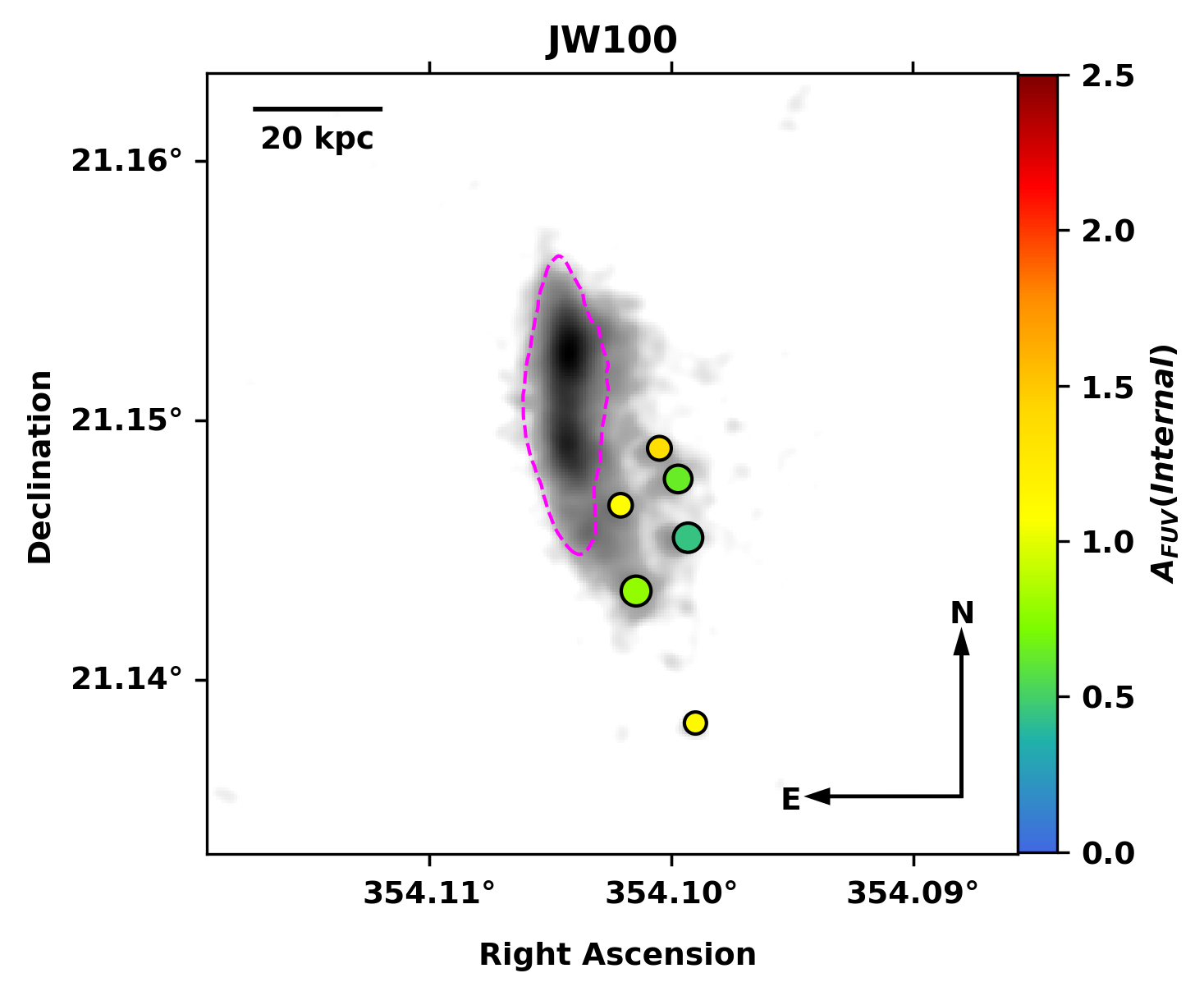}
   
    \caption{Distribution of internal attenuation, $A_{FUV}(Internal)$, plotted over the UVIT FUV images. The magenta contours trace the disks of the galaxies.}
    \label{fig:afuv_internal-all}
\end{figure*}

Figure \ref{fig:afuv_internal-all} gives the internal attenuation values ($A_{FUV}(Internal)$) corresponding to the location of the SF knots in the collisional ring of the NGC 5291 system, tidal tails of the NGC 7252 system (Top panel)  and, ram-pressure stripped tails of JO201 and JW100 galaxies (Bottom panel). The attenuation values for the SF knots in the NGC 5291 ring and NGC 7252 tails have been computed in exactly the same way  as those for the JO201 and JW100 tails. It is observed that the SF knots in the NGC 5291 ring show the least spread in attenuation (0 - 0.87 mag) compared to those in the NGC 7252 (1.12 - 2.33 mag), JO201 (0 - 2.35 mag)  and JW100 tails (0.44 - 1.36 mag).

\begin{figure*}
\centering
    \includegraphics[width=0.5\linewidth]{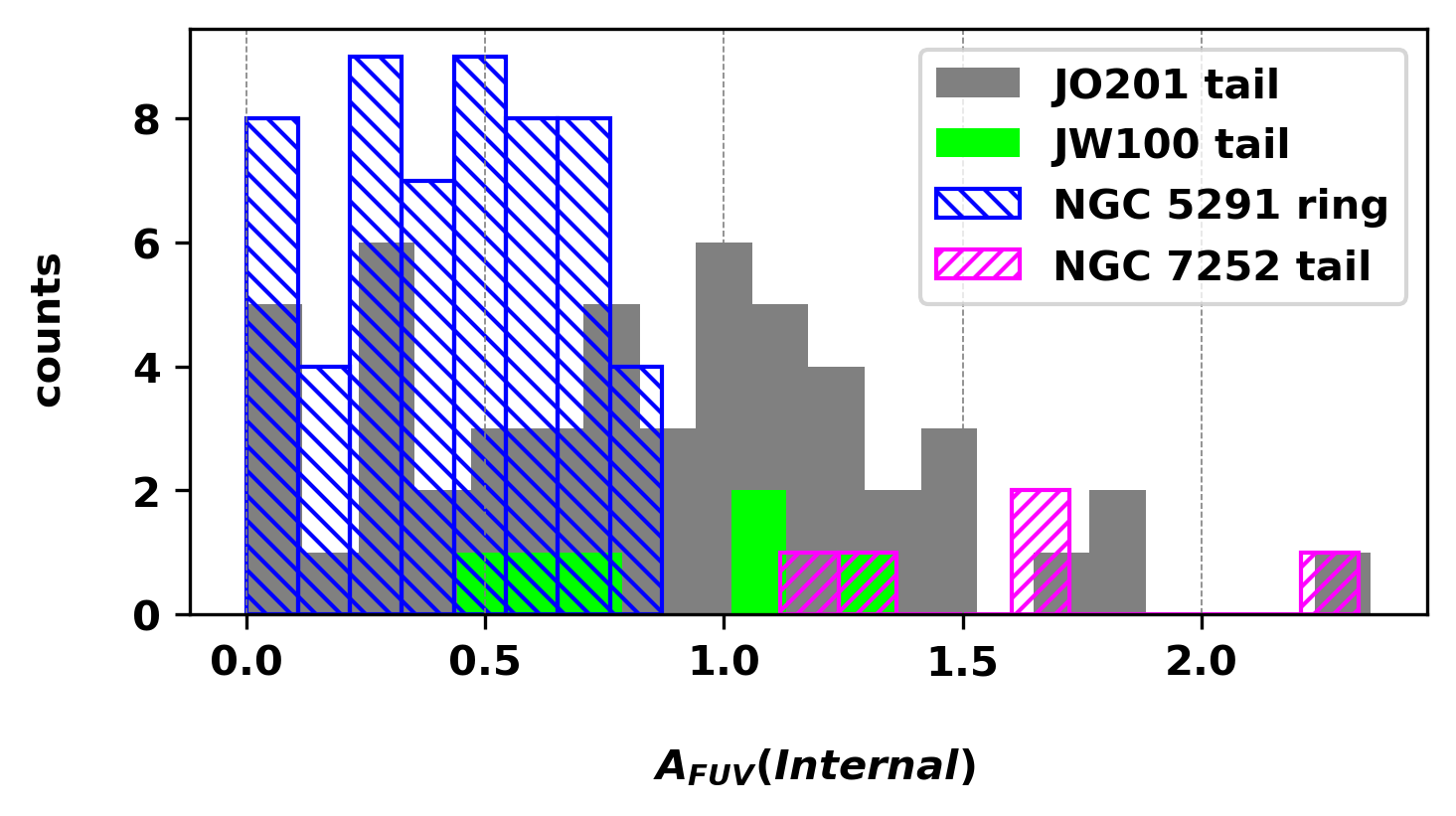}
    \centering {\caption{Comparison of internal attenuation, $A_{FUV}(Internal)$, of the SF knots.}}
    \label{fig:afuv_internal-comp-hist-all}
\end{figure*}

Figure \ref{fig:afuv_internal-comp-hist-all} gives the comparison of $A_{FUV}(Internal)$ (derived from the UV spectral slope) for the SF knots in the ring and tails of the galaxies.
The attenuation values of the knots in the NGC 5291 ring and the NGC 7252 tails are observed to be within the range of attenuation values observed for the SF knots in the ram-pressure stripped tails. All the knots in the NGC 5291 collisional ring have attenuation values that are less than or equal to the median attenuation in the tails of JO201 (0.87 mag) and JW100 (0.93 mag) galaxies. In contrast, the attenuation values of the knots in the NGC 7252 tidal tails are all greater than the median attenuation in the jellyfish tails (see Figure \ref{fig:afuv_internal-comp-hist-all}). The jellyfish tails thus exhibit two types of star-forming knots:
a) knots with little or no dust, resembling those in the NGC 5291 ring, and
b) dusty knots, similar to those found in the NGC 7252 tidal tails. While star-forming knots with little or no dust are found farther from the galaxy disk, dusty knots are mainly concentrated closer to the disk.

\subsection{Star formation in collisional ring, tidal tails and ram-pressure stripped tails}

\begin{figure*}
    \centering
    \includegraphics[width=0.43\linewidth]{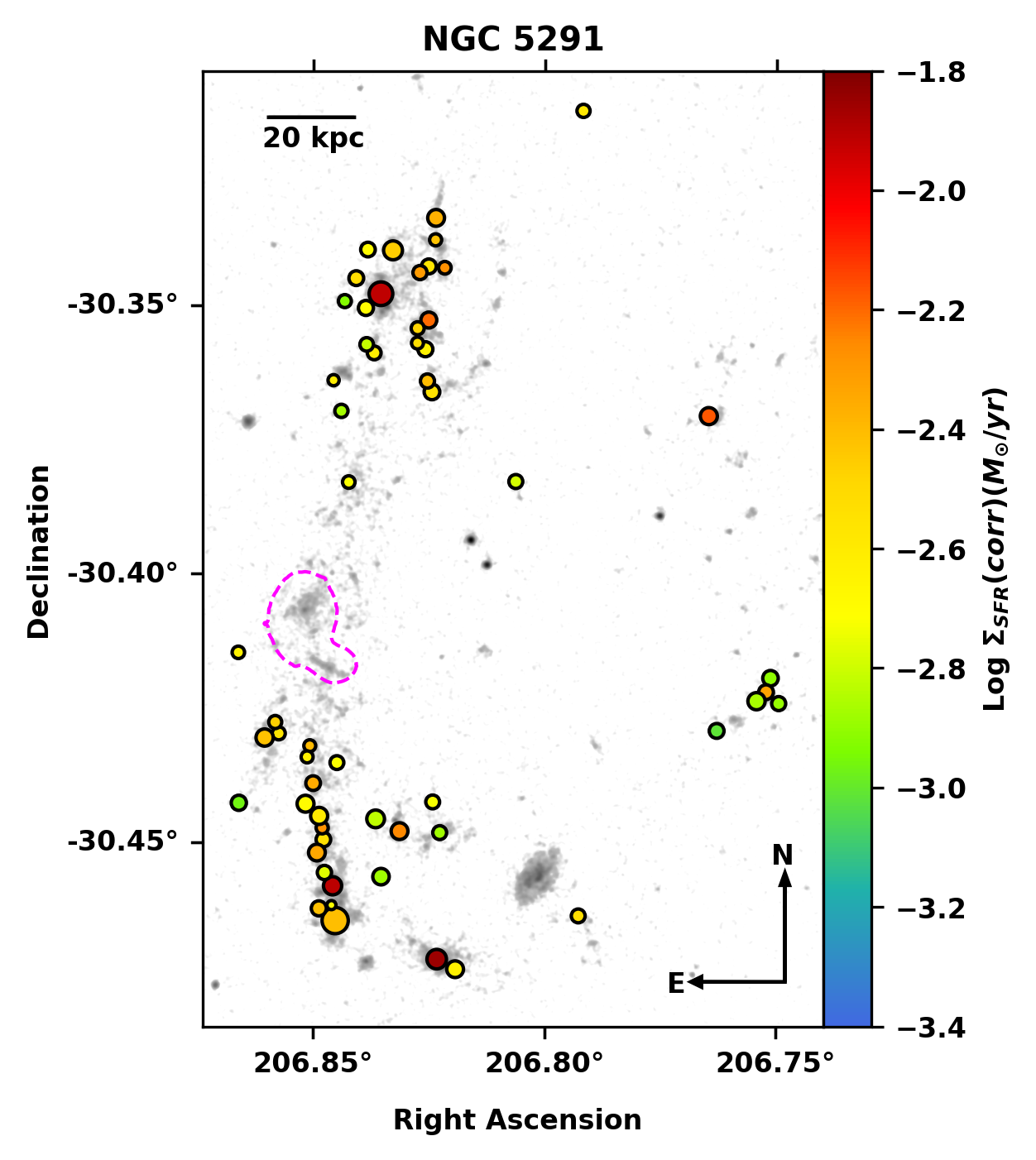}
    \hfill
    \includegraphics[width=0.48\linewidth]{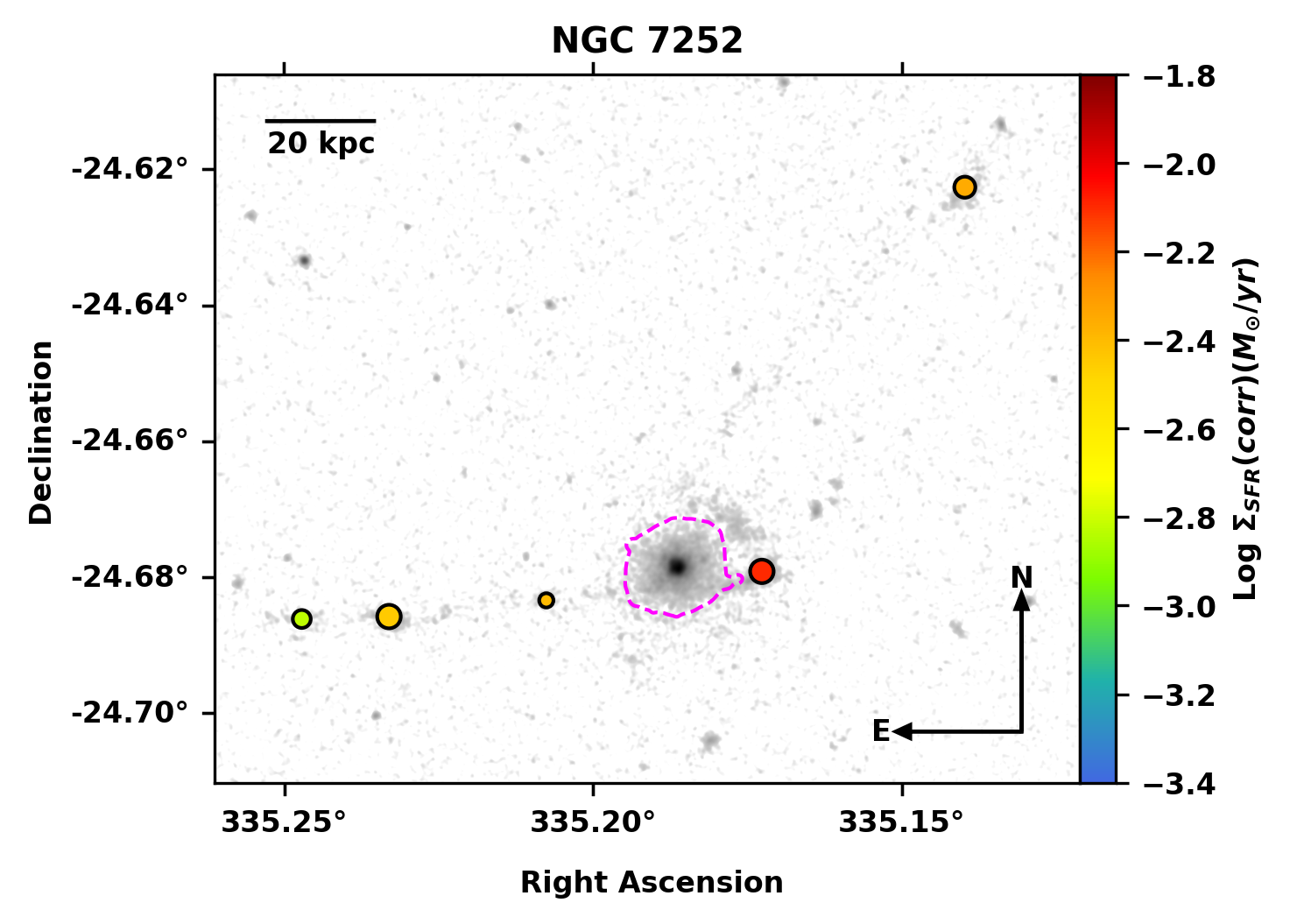}
    \vfill
    \vspace{2em}
    \includegraphics[width=0.43\linewidth]{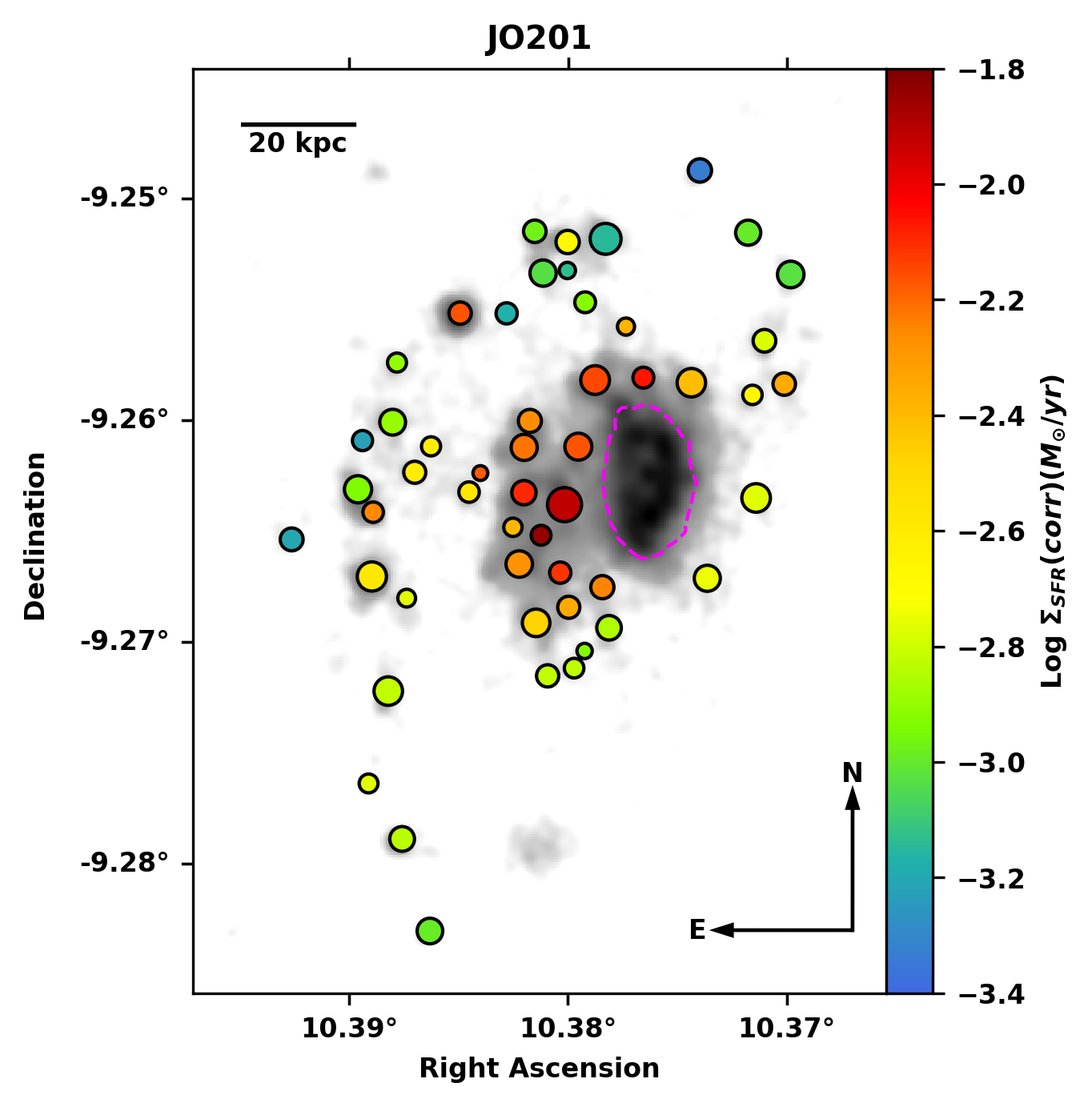}
    \hfill
    \includegraphics[width=0.48\linewidth]{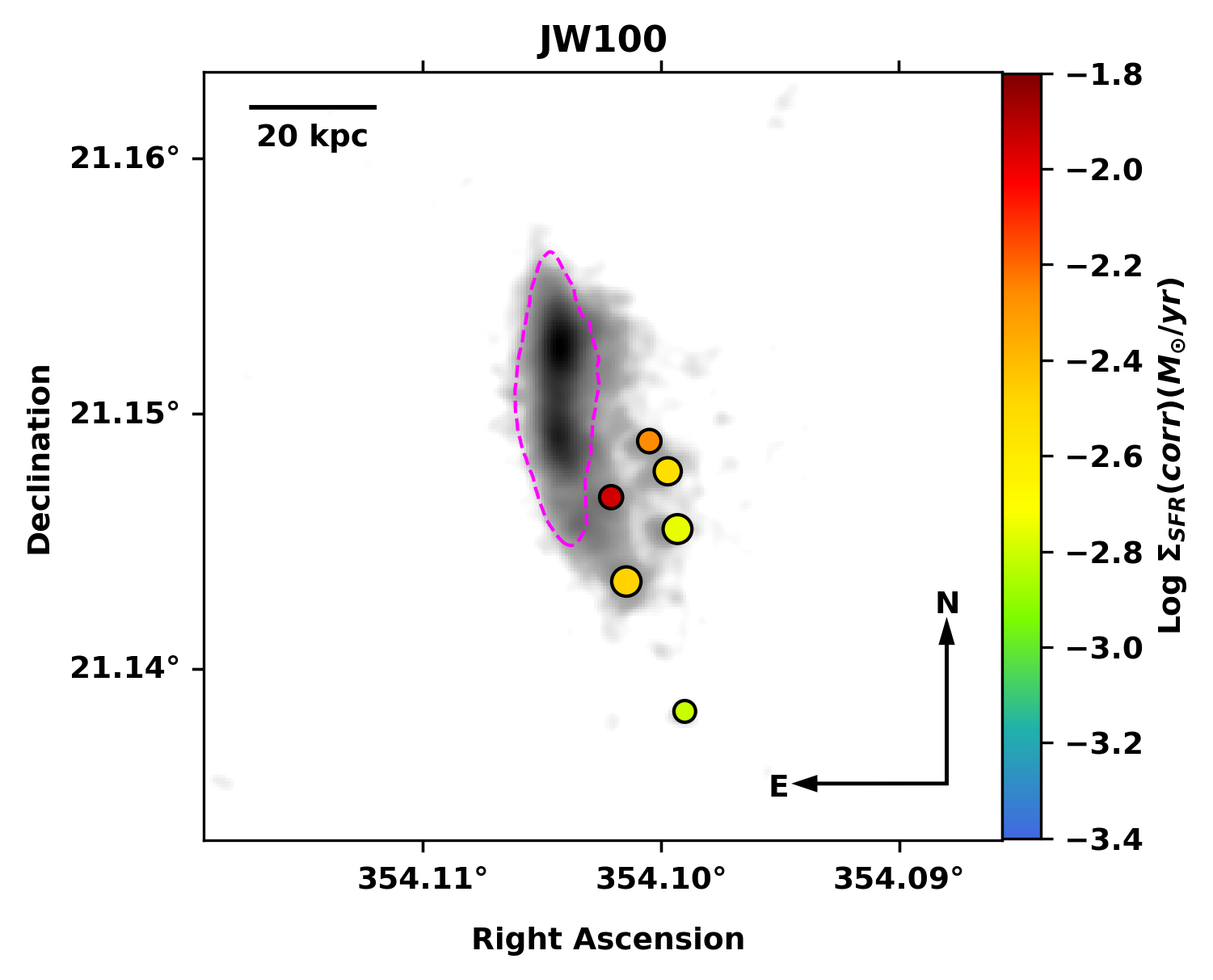}
    \caption{Distribution of SFR density, $\Sigma_{SFR_{FUV}}(corr)$, plotted over the UVIT FUV images. The magenta contours trace the disks of the galaxy.}
    \label{fig:sfr_corr_density-all}
\end{figure*}

Figure \ref{fig:sfr_corr_density-all} shows the spatial distribution of corrected SFR density, $\Sigma_{SFR_{FUV}}(corr)$, in the SF knots along the NGC 5291 ring, NGC 7252 tails (Top panel), and JO201 and JW100 tails (Bottom panel). 

A spatial gradient in the SFR density is observed for the SF knots in the stripped tails of the galaxies 
JO201 and JW100: $\Sigma_{SFR_{FUV}}(corr)$ is generally higher at the locations closer to the disk and decreases gradually with increasing distance from the disk. 
This can be attributed to the outside-in nature of ram-pressure stripping. In this process, the loosely bound, diffuse HI gas is stripped first and displaced to larger distances, while the denser, more gravitationally bound gas remains closer to the galaxy for a longer time. Therefore, high density gas will be concentrated at the
vicinity of disk.  Since the gas surface densities are higher,
the SFR densities could also be higher near the disk. We also see a subtle trend in attenuation in the jellyfish galaxies, where knots of high attenuation are generally concentrated near the disk, while knots further out show lower attenuation. This could be partly due to the lower efficiency of dust stripping via ram pressure, which limits how far dust can be displaced from the disk \citep{2022A&ARv..30....3B, 2024A&A...682A.162W}.  In addition, in a hostile environment like a cluster, dust may also be destroyed through processes such as sputtering due to ISM-ICM interactions \citep{1979ApJ...231..438D}.
Dust forms in dense gas, such as in molecular clouds, and is protected from sputtering losses by the higher density of the cloud; lower-density gas will suffer more sputtering losses. Molecular clouds, being more massive and presumably moving as a bulk, will suffer less acceleration and can be shifted out of a galaxy only at smaller velocities; hence, they will travel shorter distances compared to low-density gas. In contrast, lower-density gas can be stripped away at higher speeds and thus can move to greater distances from the jellyfish galaxy, and is more susceptible to dust loss due to sputtering.

\begin{figure*}
    \centering
    \includegraphics[width=0.5\linewidth]{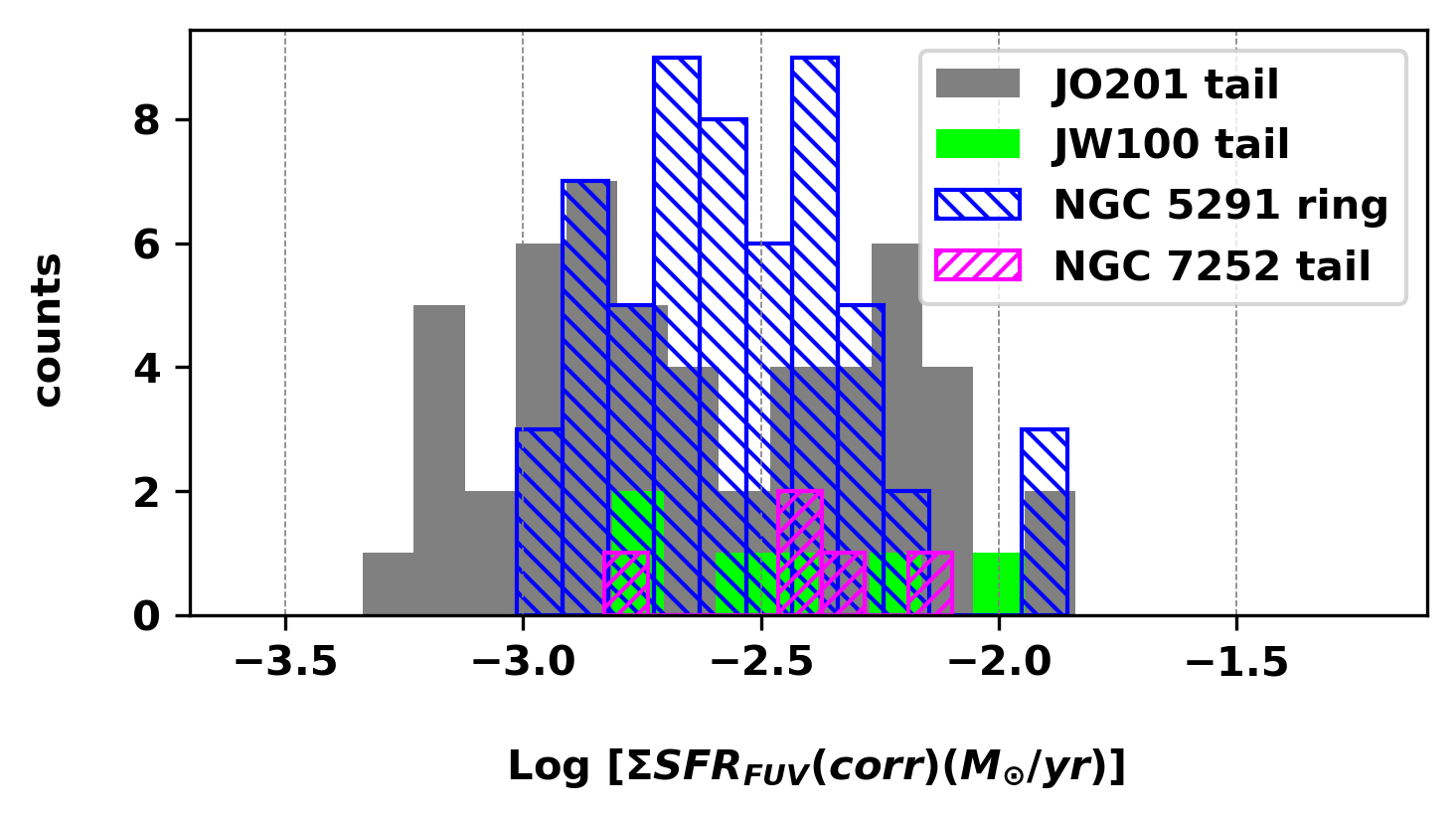}
    \caption{Comparison of corrected SFR density, $\Sigma_{SFR_{FUV}}(corr)$, of the SF knots.}
    \label{fig:sfr_density-comp-hist-all}
\end{figure*}

The comparison of $\Sigma_{SFR_{FUV}}(corr)$ for the SF knots is given in Figure \ref{fig:sfr_density-comp-hist-all}.
The estimated dust-corrected SFR density values of the knots in the NGC 5291 ring and 
in the NGC 7252 tails are comparable and lie within the range of SFR density values of the SF knots in the jellyfish tails.

Similar to ram-pressure stripping, tidal stripping is also an outside-in process, where the peripheral HI is stripped first \citep{2006PASP..118..517B}. NGC 7252, being a late-stage merger, is subject to a range of complex dynamical and astrophysical processes. Gas, stars, and dust may have been significantly redistributed through tidal interactions, extending even to the outermost regions of the tidal tails. These tails could exhibit enhanced dust content—not only due to tidally stripped dust from the progenitor galaxies but also due to dust produced within star-forming knots that have formed in situ. This is in contrast to systems in earlier stages of merging.   Furthermore, observations indicate that gas is falling back from the tidal tails into the remnant body of the NGC 7252  system \citep{1994AJ....107...67H,1995AJ....110..140H},  potentially contributing to a radial gradient in star formation activity \citep{2018A&A...614A.130G}. Taken together, these factors likely contribute to the observed gradient in dust attenuation and star formation rate density along the tidal tails.
In contrast, the collisional ring of NGC 5291 does not result from tidal interactions; but rather from a high-velocity, head-on collision. Consequently, this system does not undergo differential stripping, which may account for the lack of a clear gradient in star formation or dust attenuation. The ring is predominantly HI-rich, and the embedded star-forming knots are extremely young, exhibiting little to no dust content. The dust attenuation levels in these knots are comparable to those observed in the outermost parts of jellyfish galaxy tails, which are likewise associated with young stellar populations.  Moreover, the NGC 5291 system is located near the periphery of the Abell 3574 galaxy cluster, where interactions with the intracluster medium\textemdash{}though not dominant\textemdash{}may have contributed to dust destruction within the star-forming knots.

The average $\Sigma_{SFR_{FUV}}(corr)$ values in NGC 5291 ring (mean: 0.0034 $M_\odot/yr/kpc^2$) and JO201 tails (mean: 0.0034 $M_\odot/yr/kpc^2$) are similar; the same trend is also observed for the NGC 7252 tails (mean: 0.0043 $M_\odot/yr/kpc^2$) and JW100 tails (mean: 0.0044 $M_\odot/yr/kpc^2$).

Despite the similarities in SFR densities, the gas surface densities (and gas content) in the NGC 5291 ring, the NGC 7252 tails and the jellyfish tails are different \citep{2020A&A...640A..22R, 2020ApJ...897L..30M, 2020ApJ...889....9M, 2023MNRAS.526.1940K, 2025PASA...42...73S}. The molecular gas surface density of the SF knots in the JW100 tails ranges from approximately 20 to 63 $ M_{\odot}pc^{-2}$ \citep{2020ApJ...889....9M}, which is significantly higher than the densities found in the NGC 5291 ring and NGC 7252 tails, which range from about 1.6 to 2.6 $M_{\odot}pc^{-2}$ \citep{2023MNRAS.526.1940K, 2025PASA...42...73S}.  This indicates that in the ram-pressure stripped tails, the efficiency of conversion of molecular gas into stars is relatively low. This finding is consistent with previous observations indicating that jellyfish tails exhibit low star formation efficiencies \citep{2020ApJ...889....9M}.

\begin{table}
    \setlength{\tabcolsep}{5.1pt}
    \caption{SFR and  $f_{obscured}$ in the collisional ring, tidal tails and ram-pressure stripped tails.}
    \label{tab:sfr-fobs-all}
    {\tablefont\begin{tabular}{@{\extracolsep{\fill}}l|cc|cc}
    \toprule
    \textbf{Parameter}   & \textbf{NGC 5291}  & \textbf{NGC 7252} & \textbf{JO201} & \textbf{JW100} \\
    \textbf{}   & \textbf{ring}  & \textbf{tails} & \textbf{tails} & \textbf{tails} \\
    
    \hline
    $SFR_{FUV}(uncorr)$ $(M_\odot/yr)$     & 1.0 & 0.092 & 3.7 & 0.54 \\
    $SFR_{FUV}(Galactic)$ $(M_\odot/yr)$    & 1.6 & 0.11 & 4.6 & 0.83 \\
    $SFR_{FUV}(corr)$  $(M_\odot/yr)$         & 2.4 & 0.46 & 10.3 & 1.9 \\
    
    $f_{obscured}$     & 0.34 & 0.76 & 0.55 & 0.56 
    
    \botrule
    \end{tabular}}
\end{table}

Table \ref{tab:sfr-fobs-all} gives the total SFR in the NGC 5291 ring, NGC 7252 tails, and JO201 and JW100 tails. The fraction of internal dust obscured star formation, $f_{obsccured}$, in the interaction debris outside the galaxies is given as follows: NGC 5291 ring = 0.34, NGC 7252 tails = 0.76, JO201 tails = 0.55, and JW100 tails = 0.56. This indicates that along these structures, 30 percent or more of the star formation activity is obscured by dust.

\section{Summary}
\label{section6}

In this paper, we investigated the  dust content and star formation in two cluster galaxies undergoing strong hydrodynamic effects\textemdash{}the jellyfish galaxies JO201 in Abell 85 and JW100 in Abell 2626 undergoing intense ram-pressure stripping\textemdash{}using ultraviolet imaging observations from AstroSat/UVIT. For the resolved star-forming knots in these galaxies, we used the UV spectral slope ($\beta$) derived from the FUV–NUV colour, as an indicator of dust attenuation. The dust-corrected FUV luminosities were then used to estimate the SFRs. We then compared the  dust attenuation and star formation activity in the jellyfish galaxy tails (formed by hydrodynamic interaction) with those in the collisional ring of the NGC 5291 system and the tidal tails of the NGC 7252 post-merger system (both formed by gravitational interactions), to investigate any variations in the dust content and star formation activity along these structures formed by gravitational or hydrodynamic perturbations.

Our key findings are summarised below: 
\begin{itemize}
        \item [-] We identified a total of 52 star-forming knots in the JO201 tails and 6 in the JW100 tails from the UVIT FUV images.

        \item[-] The fraction of internal dust-obscured star formation, $f_{obscured}$, in the  
        tails of JO201 is 0.55, while the same in JW100 is 0.56.

        \item [-] On comparing the SF knots in the JO201 and JW100 tails with those in the NGC 5291 collisional ring and NGC 7252 tidal tails, we find that the attenuation of the SF knots in the jellyfish tails span a range of values which encompasses those of the SF knots in the NGC 5291 ring and NGC 7252 tails.

        \item [-] The median internal attenuation values in the tails of JO201 and JW100 are 0.87 mag and 0.93 mag, respectively. All knots in the NGC 5291 ring show attenuation values that are less than or equal to the median values, whereas the knots in the NGC 7252 tails exhibit attenuation values that exceed these medians. The jellyfish tails therefore have a) knots with attenuation values similar to those found in the NGC 5291 ring, located in the outermost parts of the jellyfish tails and b) dusty knots similar to those present in the NGC 7252 tidal tails, mainly concentrated closer to the jellyfish disk.

        \item [-] The dust-corrected SFR densities of the SF knots in the jellyfish galaxy tails are comparable to those found in the NGC 5291 ring and NGC 7252 tails. 
        Additionally, the SFR densities of the SF knots in the NGC 5291 ring and NGC 7252 tails fall within the range of SFR densities observed in the jellyfish tails.

        \item[-] The fraction of internal dust obscured star formation in the NGC 5291 ring, NGC 7252 tails, and jellyfish tails ranges from 0.34 to 0.76, indicating that more than 30 percent of the star formation in the structures is obscured by dust.
\end{itemize}

The present study uses AstroSat/UVIT ultraviolet imaging to compare star formation and dust attenuation in galaxies perturbed by  gravitational interactions (NGC 5291, NGC 7252) and  hydrodynamic interactions (jellyfish galaxies JO201, JW100). 
We see that both gravitational and hydrodynamic interactions can induce intense star formation in extended structures like tails and rings. This study enhances our understanding of how interactions trigger star formation outside galaxy disks\textemdash{crucial for models of galactic growth and morphological transformation.}


\paragraph{Acknowledgements}

The authors GS and RR acknowledge the support of Indian Space Research Organisation (ISRO) under AstroSat archival data utilisation program  (No. DS\_2B-13013(2)/9/2020-Sec.2). This publication uses data from the AstroSat mission of ISRO, archived at the Indian Space Science Data Centre (ISSDC). RR acknowledges visiting associateship of IUCAA, Pune.

\paragraph{Data Availability Statement}

The Astrosat UVIT imaging data underlying this article are available in ISSDC Astrobrowse archive \url{https://astrobrowse.issdc.gov.in/astro\_archive/archive/Home.jsp}

\bibliography{reference}

\begin{thebibliography}{}

\bibitem[{Abramson} et~al., 2016]{2016AJ....152...32A}
{Abramson}, A., {Kenney}, J., {Crowl}, H., \& {Tal}, T. 2016, {HST Imaging of Dust Structures and Stars in the Ram Pressure Stripped Virgo Spirals NGC 4402 and NGC 4522: Stripped from the Outside In with Dense Cloud Decoupling}.
\newblock {\em \aj}, 152(2), 32.

\bibitem[{Abramson} and {Kenney}, 2014]{2014AJ....147...63A}
{Abramson}, A. \& {Kenney}, J. D.~P. 2014, {Hubble Space Telescope Imaging of Decoupled Dust Clouds in the Ram Pressure Stripped Virgo Spirals NGC 4402 and NGC 4522}.
\newblock {\em \aj}, 147(3), 63.

\bibitem[{Agrawal}, 2006]{2006AdSpR..38.2989A}
{Agrawal}, P.~C. 2006, {A broad spectral band Indian Astronomy satellite {\textquoteleft}Astrosat{\textquoteright}}.
\newblock {\em Advances in Space Research}, 38(12), 2989--2994.

\bibitem[{Alonso} et~al., 2012]{2012A&A...539A..46A}
{Alonso}, S., {Mesa}, V., {Padilla}, N., \& {Lambas}, D.~G. 2012, {Galaxy interactions. II. High density environments}.
\newblock {\em \aap}, 539, A46.

\bibitem[{Appleton}, 1999]{1999IAUS..186...97A}
{Appleton}, P.~N.
\newblock {Collisional Ring Galaxies}.
\newblock In {Barnes}, J.~E. \& {Sanders}, D.~B., editors, {\em Galaxy Interactions at Low and High Redshift} 1999,, volume 186 of {\em IAU Symposium}, ~97.

\bibitem[{Appleton} and {Struck-Marcell}, 1996]{1996FCPh...16..111A}
{Appleton}, P.~N. \& {Struck-Marcell}, C. 1996, {Collisional Ring Galaxies}.
\newblock {\em \fcp}, 16, 111--220.

\bibitem[{Bellhouse} et~al., 2017]{2017ApJ...844...49B}
{Bellhouse}, C., {Jaff{\'e}}, Y.~L., {Hau}, G.~K.~T., {McGee}, S.~L., {Poggianti}, B.~M., {Moretti}, A., {Gullieuszik}, M., {Bettoni}, D., {Fasano}, G., {D'Onofrio}, M., {Fritz}, J., {Omizzolo}, A., {Sheen}, Y.~K., \& {Vulcani}, B. 2017, {GASP. II. A MUSE View of Extreme Ram-Pressure Stripping along the Line of Sight: Kinematics of the Jellyfish Galaxy JO201}.
\newblock {\em \apj}, 844(1), 49.

\bibitem[{Bellhouse} et~al., 2019]{2019MNRAS.485.1157B}
{Bellhouse}, C., {Jaff{\'e}}, Y.~L., {McGee}, S.~L., {Poggianti}, B.~M., {Smith}, R., {Tonnesen}, S., {Fritz}, J., {Hau}, G.~K.~T., {Gullieuszik}, M., {Vulcani}, B., {Fasano}, G., {Moretti}, A., {George}, K., {Bettoni}, D., {D'Onofrio}, M., {Omizzolo}, A., \& {Sheen}, Y.~K. 2019, {GASP. XV. A MUSE view of extreme ram-pressure stripping along the line of sight: physical properties of the jellyfish galaxy JO201}.
\newblock {\em \mnras}, 485(1), 1157--1170.

\bibitem[{Bianconi} et~al., 2020]{2020MNRAS.492.4599B}
{Bianconi}, M., {Smith}, G.~P., {Haines}, C.~P., {McGee}, S.~L., {Finoguenov}, A., \& {Egami}, E. 2020, {LoCuSS: exploring the connection between local environment, star formation, and dust mass in Abell 1758}.
\newblock {\em \mnras}, 492(4), 4599--4612.

\bibitem[{Boquien} et~al., 2012]{2012A&A...539A.145B}
{Boquien}, M., {Buat}, V., {Boselli}, A., {Baes}, M., {Bendo}, G.~J., {Ciesla}, L., {Cooray}, A., {Cortese}, L., {Eales}, S., {Gavazzi}, G., {Gomez}, H.~L., {Lebouteiller}, V., {Pappalardo}, C., {Pohlen}, M., {Smith}, M.~W.~L., \& {Spinoglio}, L. 2012, {The IRX-{\ensuremath{\beta}} relation on subgalactic scales in star-forming galaxies of the Herschel Reference Survey}.
\newblock {\em \aap}, 539, A145.

\bibitem[{Boquien} et~al., 2007]{2007A&A...467...93B}
{Boquien}, M., {Duc}, P.~A., {Braine}, J., {Brinks}, E., {Lisenfeld}, U., \& {Charmandaris}, V. 2007, {Polychromatic view of intergalactic star formation in NGC 5291}.
\newblock {\em \aap}, 467(1), 93--106.

\bibitem[{Boquien} et~al., 2009]{2009AJ....137.4561B}
{Boquien}, M., {Duc}, P.~A., {Wu}, Y., {Charmandaris}, V., {Lisenfeld}, U., {Braine}, J., {Brinks}, E., {Iglesias-P{\'a}ramo}, J., \& {Xu}, C.~K. 2009, {Collisional Debris as Laboratories to Study Star Formation}.
\newblock {\em \aj}, 137(6), 4561--4576.

\bibitem[{Boselli} et~al., 2014a]{2014A&A...564A..67B}
{Boselli}, A., {Cortese}, L., {Boquien}, M., {Boissier}, S., {Catinella}, B., {Gavazzi}, G., {Lagos}, C., \& {Saintonge}, A. 2014,a {Cold gas properties of the Herschel Reference Survey. III. Molecular gas stripping in cluster galaxies}.
\newblock {\em \aap}, 564a, A67.

\bibitem[{Boselli} et~al., 2018]{2018A&A...615A.114B}
{Boselli}, A., {Fossati}, M., {Cuillandre}, J.~C., {Boissier}, S., {Boquien}, M., {Buat}, V., {Burgarella}, D., {Consolandi}, G., {Cortese}, L., {C{\^o}t{\'e}}, P., {C{\^o}t{\'e}}, S., {Durrell}, P., {Ferrarese}, L., {Fumagalli}, M., {Gavazzi}, G., {Gwyn}, S., {Hensler}, G., {Koribalski}, B., {Roediger}, J., {Roehlly}, Y., {Russeil}, D., {Sun}, M., {Toloba}, E., {Vollmer}, B., \& {Zavagno}, A. 2018, {A Virgo Environmental Survey Tracing Ionised Gas Emission (VESTIGE). III. Star formation in the stripped gas of NGC 4254}.
\newblock {\em \aap}, 615, A114.

\bibitem[{Boselli} et~al., 2022]{2022A&ARv..30....3B}
{Boselli}, A., {Fossati}, M., \& {Sun}, M. 2022, {Ram pressure stripping in high-density environments}.
\newblock {\em \aapr}, 30(1), 3.

\bibitem[{Boselli} and {Gavazzi}, 2006]{2006PASP..118..517B}
{Boselli}, A. \& {Gavazzi}, G. 2006, {Environmental Effects on Late-Type Galaxies in Nearby Clusters}.
\newblock {\em \pasp}, 118(842), 517--559.

\bibitem[{Boselli} et~al., 2014b]{2014A&A...570A..69B}
{Boselli}, A., {Voyer}, E., {Boissier}, S., {Cucciati}, O., {Consolandi}, G., {Cortese}, L., {Fumagalli}, M., {Gavazzi}, G., {Heinis}, S., {Roehlly}, Y., \& {Toloba}, E. 2014,b {The GALEX Ultraviolet Virgo Cluster Survey (GUViCS). IV. The role of the cluster environment on galaxy evolution}.
\newblock {\em \aap}, 570b, A69.

\bibitem[{Bournaud} and {Combes}, 2003]{2003A&A...401..817B}
{Bournaud}, F. \& {Combes}, F. 2003, {Formation of polar ring galaxies}.
\newblock {\em \aap}, 401, 817--833.

\bibitem[{Bournaud} et~al., 2007]{2007Sci...316.1166B}
{Bournaud}, F., {Duc}, P.-A., {Brinks}, E., {Boquien}, M., {Amram}, P., {Lisenfeld}, U., {Koribalski}, B.~S., {Walter}, F., \& {Charmandaris}, V. 2007, {Missing Mass in Collisional Debris from Galaxies}.
\newblock {\em Science}, 316(5828), 1166.

\bibitem[{Calzetti}, 2001]{2001PASP..113.1449C}
{Calzetti}, D. 2001, {The Dust Opacity of Star-forming Galaxies}.
\newblock {\em \pasp}, 113(790), 1449--1485.

\bibitem[{Cardelli} et~al., 1989]{1989ApJ...345..245C}
{Cardelli}, J.~A., {Clayton}, G.~C., \& {Mathis}, J.~S. 1989, {The Relationship between Infrared, Optical, and Ultraviolet Extinction}.
\newblock {\em \apj}, 345, 245.

\bibitem[{Casasola} et~al., 2020]{2020A&A...633A.100C}
{Casasola}, V., {Bianchi}, S., {De Vis}, P., {Magrini}, L., {Corbelli}, E., {Clark}, C.~J.~R., {Fritz}, J., {Nersesian}, A., {Viaene}, S., {Baes}, M., {Cassar{\`a}}, L.~P., {Davies}, J., {De Looze}, I., {Dobbels}, W., {Galametz}, M., {Galliano}, F., {Jones}, A.~P., {Madden}, S.~C., {Mosenkov}, A.~V., {Tr{\v{c}}ka}, A., \& {Xilouris}, E. 2020, {The ISM scaling relations in DustPedia late-type galaxies: A benchmark study for the Local Universe}.
\newblock {\em \aap}, 633, A100.

\bibitem[{Chabrier}, 2003]{2003PASP..115..763C}
{Chabrier}, G. 2003, {Galactic Stellar and Substellar Initial Mass Function}.
\newblock {\em \pasp}, 115(809), 763--795.

\bibitem[{Chien} and {Barnes}, 2010]{2010MNRAS.407...43C}
{Chien}, L.-H. \& {Barnes}, J.~E. 2010, {Dynamically driven star formation in models of NGC 7252}.
\newblock {\em \mnras}, 407(1), 43--54.

\bibitem[{Comerford} et~al., 2015]{2015ApJ...806..219C}
{Comerford}, J.~M., {Pooley}, D., {Barrows}, R.~S., {Greene}, J.~E., {Zakamska}, N.~L., {Madejski}, G.~M., \& {Cooper}, M.~C. 2015, {Merger-driven Fueling of Active Galactic Nuclei: Six Dual and Offset AGNs Discovered with Chandra and Hubble Space Telescope Observations}.
\newblock {\em \apj}, 806(2), 219.

\bibitem[{Consolandi} et~al., 2017]{2017A&A...606A..83C}
{Consolandi}, G., {Gavazzi}, G., {Fossati}, M., {Fumagalli}, M., {Boselli}, A., {Yagi}, M., \& {Yoshida}, M. 2017, {MUSE sneaks a peek at extreme ram-pressure events. III. Tomography of UGC 6697, a massive galaxy falling into Abell 1367}.
\newblock {\em \aap}, 606, A83.

\bibitem[{Cortese} et~al., 2007]{2007MNRAS.376..157C}
{Cortese}, L., {Marcillac}, D., {Richard}, J., {Bravo-Alfaro}, H., {Kneib}, J.~P., {Rieke}, G., {Covone}, G., {Egami}, E., {Rigby}, J., {Czoske}, O., \& {Davies}, J. 2007, {The strong transformation of spiral galaxies infalling into massive clusters at z \raisebox{-0.5ex}\textasciitilde 0.2}.
\newblock {\em \mnras}, 376(1), 157--172.

\bibitem[{Cowie} and {Songaila}, 1977]{1977Natur.266..501C}
{Cowie}, L.~L. \& {Songaila}, A. 1977, {Thermal evaporation of gas within galaxies by a hot intergalactic medium}.
\newblock {\em \nat}, 266, 501--503.

\bibitem[{Cramer}, 2019]{2019asrc.confE..80C}
{Cramer}, W.
\newblock {ALMA evidence for the direct ram pressure stripping of molecular gas in two cluster galaxies}.
\newblock In {\em ALMA2019: Science Results and Cross-Facility Synergies} 2019,, ~80.

\bibitem[{Cramer} et~al., 2020]{2020ApJ...901...95C}
{Cramer}, W.~J., {Kenney}, J.~D.~P., {Cortes}, J.~R., {Cortes P.~C.}, {Vlahakis}, C., {J{\'a}chym}, P., {Pompei}, E., \& {Rubio}, M. 2020, {ALMA Evidence for Ram Pressure Compression and Stripping of Molecular Gas in the Virgo Cluster Galaxy NGC 4402}.
\newblock {\em \apj}, 901(2), 95.

\bibitem[{Crowl} et~al., 2005]{2005ASPC..331..281C}
{Crowl}, H.~H., {Kenney}, J.~D.~P., {van Gorkom}, J.~H., \& {Vollmer}, B.
\newblock {Extra-planar Gas and Dust due to Ram Pressure Stripping of the Virgo Spiral NGC 4402}.
\newblock In {Braun}, R., editor, {\em Extra-Planar Gas} 2005,, volume 331 of {\em Astronomical Society of the Pacific Conference Series},  281.

\bibitem[{Deb} et~al., 2020]{2020MNRAS.494.5029D}
{Deb}, T., {Verheijen}, M. A.~W., {Gullieuszik}, M., {Poggianti}, B.~M., {van Gorkom}, J.~H., {Ramatsoku}, M., {Serra}, P., {Moretti}, A., {Vulcani}, B., {Bettoni}, D., {Jaff{\'e}}, L.~Y., {Tonnesen}, S., \& {Fritz}, J. 2020, {GASP XXV: neutral hydrogen gas in the striking jellyfish galaxy JO204}.
\newblock {\em \mnras}, 494(4), 5029--5043.

\bibitem[{Deb} et~al., 2022]{2022MNRAS.516.2683D}
{Deb}, T., {Verheijen}, M. A.~W., {Poggianti}, B.~M., {Moretti}, A., {van der Hulst}, J.~M., {Vulcani}, B., {Ramatsoku}, M., {Serra}, P., {Healy}, J., {Gullieuszik}, M., {Bacchini}, C., {Ignesti}, A., {M{\"u}ller}, A., {Zabel}, N., {Luber}, N., {Jaff{\"e}}, Y.~L., \& {Gitti}, M. 2022, {GASP XXXIX: MeerKAT hunts Jellyfish in A2626}.
\newblock {\em \mnras}, 516(2), 2683--2696.

\bibitem[{Dey} et~al., 2019]{2019AJ....157..168D}
{Dey}, A., {Schlegel}, D.~J., {Lang}, D., {Blum}, R., {Burleigh}, K., {Fan}, X., {Findlay}, J.~R., {Finkbeiner}, D., {Herrera}, D., {Juneau}, S., {Landriau}, M., {Levi}, M., {McGreer}, I., {Meisner}, A., {Myers}, A.~D., {Moustakas}, J., {Nugent}, P., {Patej}, A., {Schlafly}, E.~F., {Walker}, A.~R., {Valdes}, F., {Weaver}, B.~A., {Y{\`e}che}, C., {Zou}, H., {Zhou}, X., {Abareshi}, B., {Abbott}, T.~M.~C., {Abolfathi}, B., {Aguilera}, C., {Alam}, S., {Allen}, L., {Alvarez}, A., {Annis}, J., {Ansarinejad}, B., {Aubert}, M., {Beechert}, J., {Bell}, E.~F., {BenZvi}, S.~Y., {Beutler}, F., {Bielby}, R.~M., {Bolton}, A.~S., {Brice{\~n}o}, C., {Buckley-Geer}, E.~J., {Butler}, K., {Calamida}, A., {Carlberg}, R.~G., {Carter}, P., {Casas}, R., {Castander}, F.~J., {Choi}, Y., {Comparat}, J., {Cukanovaite}, E., {Delubac}, T., {DeVries}, K., {Dey}, S., {Dhungana}, G., {Dickinson}, M., {Ding}, Z., {Donaldson}, J.~B., {Duan}, Y., {Duckworth}, C.~J., {Eftekharzadeh}, S., {Eisenstein}, D.~J., {Etourneau}, T., {Fagrelius}, P.~A.,
  {Farihi}, J., {Fitzpatrick}, M., {Font-Ribera}, A., {Fulmer}, L., {G{\"a}nsicke}, B.~T., {Gaztanaga}, E., {George}, K., {Gerdes}, D.~W., {Gontcho}, S. G.~A., {Gorgoni}, C., {Green}, G., {Guy}, J., {Harmer}, D., {Hernandez}, M., {Honscheid}, K., {Huang}, L.~W., {James}, D.~J., {Jannuzi}, B.~T., {Jiang}, L., {Joyce}, R., {Karcher}, A., {Karkar}, S., {Kehoe}, R., {Kneib}, J.-P., {Kueter-Young}, A., {Lan}, T.-W., {Lauer}, T.~R., {Le Guillou}, L., {Le Van Suu}, A., {Lee}, J.~H., {Lesser}, M., {Perreault Levasseur}, L., {Li}, T.~S., {Mann}, J.~L., {Marshall}, R., {Mart{\'\i}nez-V{\'a}zquez}, C.~E., {Martini}, P., {du Mas des Bourboux}, H., {McManus}, S., {Meier}, T.~G., {M{\'e}nard}, B., {Metcalfe}, N., {Mu{\~n}oz-Guti{\'e}rrez}, A., {Najita}, J., {Napier}, K., {Narayan}, G., {Newman}, J.~A., {Nie}, J., {Nord}, B., {Norman}, D.~J., {Olsen}, K. A.~G., {Paat}, A., {Palanque-Delabrouille}, N., {Peng}, X., {Poppett}, C.~L., {Poremba}, M.~R., {Prakash}, A., {Rabinowitz}, D., {Raichoor}, A., {Rezaie}, M., {Robertson},
  A.~N., {Roe}, N.~A., {Ross}, A.~J., {Ross}, N.~P., {Rudnick}, G., {Safonova}, S., {Saha}, A., {S{\'a}nchez}, F.~J., {Savary}, E., {Schweiker}, H., {Scott}, A., {Seo}, H.-J., {Shan}, H., {Silva}, D.~R., {Slepian}, Z., {Soto}, C., {Sprayberry}, D., {Staten}, R., {Stillman}, C.~M., {Stupak}, R.~J., {Summers}, D.~L., {Sien Tie}, S., {Tirado}, H., {Vargas-Maga{\~n}a}, M., {Vivas}, A.~K., {Wechsler}, R.~H., {Williams}, D., {Yang}, J., {Yang}, Q., {Yapici}, T., {Zaritsky}, D., {Zenteno}, A., {Zhang}, K., {Zhang}, T., {Zhou}, R., \& {Zhou}, Z. 2019, {Overview of the DESI Legacy Imaging Surveys}.
\newblock {\em \aj}, 157(5), 168.

\bibitem[{Dorman} et~al., 1995]{1995ApJ...442..105D}
{Dorman}, B., {O'Connell}, R.~W., \& {Rood}, R.~T. 1995, {Ultraviolet Radiation from Evolved Stellar Populations. II. The Ultraviolet Upturn Phenomenon in Elliptical Galaxies}.
\newblock {\em \apj}, 442, 105.

\bibitem[{Dorman} et~al., 1993]{1993ApJ...419..596D}
{Dorman}, B., {Rood}, R.~T., \& {O'Connell}, R.~W. 1993, {Ultraviolet Radiation from Evolved Stellar Populations. I. Models}.
\newblock {\em \apj}, 419, 596.

\bibitem[{Draine} and {Salpeter}, 1979]{1979ApJ...231..438D}
{Draine}, B.~T. \& {Salpeter}, E.~E. 1979, {Destruction mechanisms for interstellar dust.}
\newblock {\em \apj}, 231, 438--455.

\bibitem[{Dressler}, 1980]{1980ApJ...236..351D}
{Dressler}, A. 1980, {Galaxy morphology in rich clusters: implications for the formation and evolution of galaxies.}
\newblock {\em \apj}, 236, 351--365.

\bibitem[{Duc} et~al., 2007]{2007IAUS..237..323D}
{Duc}, P.-A., {Bournaud}, F., \& {Boquien}, M.
\newblock {Tidal dwarf galaxies as laboratories of star formation and cosmology}.
\newblock In {Elmegreen}, B.~G. \& {Palous}, J., editors, {\em Triggered Star Formation in a Turbulent ISM} 2007,, volume 237 of {\em IAU Symposium}, pp. 323--330.

\bibitem[{Duc} and {Renaud}, 2013]{2013LNP...861..327D}
{Duc}, P.-A. \& {Renaud}, F.
\newblock {Tides in Colliding Galaxies}.
\newblock In {Souchay}, J., {Mathis}, S., \& {Tokieda}, T., editors, {\em Lecture Notes in Physics, Berlin Springer Verlag} 2013,, volume 861,  327.

\bibitem[{Ebeling} et~al., 2014]{2014ApJ...781L..40E}
{Ebeling}, H., {Stephenson}, L.~N., \& {Edge}, A.~C. 2014, {Jellyfish: Evidence of Extreme Ram-pressure Stripping in Massive Galaxy Clusters}.
\newblock {\em \apjl}, 781(2), L40.

\bibitem[{Ellison} et~al., 2010]{2010MNRAS.407.1514E}
{Ellison}, S.~L., {Patton}, D.~R., {Simard}, L., {McConnachie}, A.~W., {Baldry}, I.~K., \& {Mendel}, J.~T. 2010, {Galaxy pairs in the Sloan Digital Sky Survey - II. The effect of environment on interactions}.
\newblock {\em \mnras}, 407(3), 1514--1528.

\bibitem[{Farouki} and {Shapiro}, 1981]{1981ApJ...243...32F}
{Farouki}, R. \& {Shapiro}, S.~L. 1981, {Computer simulations of environmental influences on galaxy evolution in dense clusters. II - Rapid tidal encounters}.
\newblock {\em \apj}, 243, 32--41.

\bibitem[{Fensch} et~al., 2019]{2019A&A...628A..60F}
{Fensch}, J., {Duc}, P.-A., {Boquien}, M., {Elmegreen}, D.~M., {Elmegreen}, B.~G., {Bournaud}, F., {Brinks}, E., {de Grijs}, R., {Lelli}, F., {Renaud}, F., \& {Weilbacher}, P.~M. 2019, {Massive star cluster formation and evolution in tidal dwarf galaxies}.
\newblock {\em \aap}, 628, A60.

\bibitem[{Flaugher} et~al., 2015]{2015AJ....150..150F}
{Flaugher}, B., {Diehl}, H.~T., {Honscheid}, K., {Abbott}, T.~M.~C., {Alvarez}, O., {Angstadt}, R., {Annis}, J.~T., {Antonik}, M., {Ballester}, O., {Beaufore}, L., {Bernstein}, G.~M., {Bernstein}, R.~A., {Bigelow}, B., {Bonati}, M., {Boprie}, D., {Brooks}, D., {Buckley-Geer}, E.~J., {Campa}, J., {Cardiel-Sas}, L., {Castander}, F.~J., {Castilla}, J., {Cease}, H., {Cela-Ruiz}, J.~M., {Chappa}, S., {Chi}, E., {Cooper}, C., {da Costa}, L.~N., {Dede}, E., {Derylo}, G., {DePoy}, D.~L., {de Vicente}, J., {Doel}, P., {Drlica-Wagner}, A., {Eiting}, J., {Elliott}, A.~E., {Emes}, J., {Estrada}, J., {Fausti Neto}, A., {Finley}, D.~A., {Flores}, R., {Frieman}, J., {Gerdes}, D., {Gladders}, M.~D., {Gregory}, B., {Gutierrez}, G.~R., {Hao}, J., {Holland}, S.~E., {Holm}, S., {Huffman}, D., {Jackson}, C., {James}, D.~J., {Jonas}, M., {Karcher}, A., {Karliner}, I., {Kent}, S., {Kessler}, R., {Kozlovsky}, M., {Kron}, R.~G., {Kubik}, D., {Kuehn}, K., {Kuhlmann}, S., {Kuk}, K., {Lahav}, O., {Lathrop}, A., {Lee}, J., {Levi}, M.~E.,
  {Lewis}, P., {Li}, T.~S., {Mandrichenko}, I., {Marshall}, J.~L., {Martinez}, G., {Merritt}, K.~W., {Miquel}, R., {Mu{\~n}oz}, F., {Neilsen}, E.~H., {Nichol}, R.~C., {Nord}, B., {Ogando}, R., {Olsen}, J., {Palaio}, N., {Patton}, K., {Peoples}, J., {Plazas}, A.~A., {Rauch}, J., {Reil}, K., {Rheault}, J.~P., {Roe}, N.~A., {Rogers}, H., {Roodman}, A., {Sanchez}, E., {Scarpine}, V., {Schindler}, R.~H., {Schmidt}, R., {Schmitt}, R., {Schubnell}, M., {Schultz}, K., {Schurter}, P., {Scott}, L., {Serrano}, S., {Shaw}, T.~M., {Smith}, R.~C., {Soares-Santos}, M., {Stefanik}, A., {Stuermer}, W., {Suchyta}, E., {Sypniewski}, A., {Tarle}, G., {Thaler}, J., {Tighe}, R., {Tran}, C., {Tucker}, D., {Walker}, A.~R., {Wang}, G., {Watson}, M., {Weaverdyck}, C., {Wester}, W., {Woods}, R., {Yanny}, B., \& {DES Collaboration} 2015, {The Dark Energy Camera}.
\newblock {\em \aj}, 150(5), 150.

\bibitem[{Fumagalli} et~al., 2014]{2014MNRAS.445.4335F}
{Fumagalli}, M., {Fossati}, M., {Hau}, G. K.~T., {Gavazzi}, G., {Bower}, R., {Sun}, M., \& {Boselli}, A. 2014, {MUSE sneaks a peek at extreme ram-pressure stripping events - I. A kinematic study of the archetypal galaxy ESO137-001}.
\newblock {\em \mnras}, 445(4), 4335--4344.

\bibitem[{Gavazzi} et~al., 2005]{2005A&A...429..439G}
{Gavazzi}, G., {Boselli}, A., {van Driel}, W., \& {O'Neil}, K. 2005, {Completing H I observations of galaxies in the Virgo cluster}.
\newblock {\em \aap}, 429, 439--447.

\bibitem[{Gavazzi} et~al., 2013]{2013A&A...553A..90G}
{Gavazzi}, G., {Savorgnan}, G., {Fossati}, M., {Dotti}, M., {Fumagalli}, M., {Boselli}, A., {Guti{\'e}rrez}, L., {Hern{\'a}ndez Toledo}, H., {Giovanelli}, R., \& {Haynes}, M.~P. 2013, {H{\ensuremath{\alpha}}3: an H{\ensuremath{\alpha}} imaging survey of HI selected galaxies from ALFALFA. III. Nurture builds up the Hubble sequence in the Great Wall}.
\newblock {\em \aap}, 553, A90.

\bibitem[{George} et~al., 2025a]{2025A&A...701A..40G}
{George}, K., {Boselli}, A., {Cuillandre}, J.-C., {K{\"u}mmel}, M., {Lan{\c{c}}on}, A., {Bellhouse}, C., {Saifollahi}, T., {Mondelin}, M., {Bolzonella}, M., {Joseph}, P., {Roberts}, I.~D., {van Weeren}, R.~J., {Liu}, Q., {Sola}, E., {Urbano}, M., {Baes}, M., {Peletier}, R.~F., {Klein}, M., {Davies}, C.~T., {Zinchenko}, I.~A., {Sorce}, J.~G., {Poulain}, M., {Aghanim}, N., {Altieri}, B., {Amara}, A., {Andreon}, S., {Auricchio}, N., {Baccigalupi}, C., {Baldi}, M., {Balestra}, A., {Bardelli}, S., {Battaglia}, P., {Biviano}, A., {Bonino}, D., {Branchini}, E., {Brescia}, M., {Brinchmann}, J., {Camera}, S., {Ca{\~n}as-Herrera}, G., {Capobianco}, V., {Carbone}, C., {Carretero}, J., {Casas}, S., {Castellano}, M., {Castignani}, G., {Cavuoti}, S., {Chambers}, K.~C., {Cimatti}, A., {Colodro-Conde}, C., {Congedo}, G., {Conselice}, C.~J., {Conversi}, L., {Copin}, Y., {Courbin}, F., {Courtois}, H.~M., {Cropper}, M., {Da Silva}, A., {Degaudenzi}, H., {De Lucia}, G., {Di Giorgio}, A.~M., {Dole}, H., {Douspis}, M., {Dubath},
  F., {Dupac}, X., {Dusini}, S., {Escoffier}, S., {Farina}, M., {Faustini}, F., {Ferriol}, S., {Fotopoulou}, S., {Frailis}, M., {Franceschi}, E., {Galeotta}, S., {Gillis}, B., {Giocoli}, C., {Gracia-Carpio}, J., {Grazian}, A., {Grupp}, F., {Haugan}, S.~V.~H., {Holmes}, W., {Hook}, I.~M., {Hormuth}, F., {Hornstrup}, A., {Hudelot}, P., {Jahnke}, K., {Jhabvala}, M., {Keih{\"a}nen}, E., {Kermiche}, S., {Kiessling}, A., {Kubik}, B., {Kunz}, M., {Kurki-Suonio}, H., {Le Brun}, A.~M.~C., {Le Mignant}, D., {Ligori}, S., {Lilje}, P.~B., {Lindholm}, V., {Lloro}, I., {Mainetti}, G., {Maino}, D., {Maiorano}, E., {Mansutti}, O., {Marggraf}, O., {Markovic}, K., {Martinelli}, M., {Martinet}, N., {Marulli}, F., {Massey}, R., {Maurogordato}, S., {Medinaceli}, E., {Mei}, S., {Mellier}, Y., {Meneghetti}, M., {Merlin}, E., {Meylan}, G., {Mohr}, J.~J., {Mora}, A., {Moresco}, M., {Moscardini}, L., {Nakajima}, R., {Neissner}, C., {Nichol}, R.~C., {Niemi}, S.-M., {Nightingale}, J.~W., {Padilla}, C., {Paltani}, S., {Pasian}, F.,
  {Pedersen}, K., {Percival}, W.~J., {Pettorino}, V., {Pires}, S., {Polenta}, G., {Poncet}, M., {Popa}, L.~A., {Pozzetti}, L., {Raison}, F., {Rebolo}, R., {Renzi}, A., {Rhodes}, J., {Riccio}, G., {Romelli}, E., {Roncarelli}, M., {Rossetti}, E., {Saglia}, R., {Sakr}, Z., {Sapone}, D., {Sartoris}, B., {Schewtschenko}, J.~A., {Schirmer}, M., {Schneider}, P., {Secroun}, A., {Seidel}, G., {Seiffert}, M., {Serrano}, S., {Sirignano}, C., {Sirri}, G., {Stanco}, L., {Steinwagner}, J., {Tallada-Cresp{\'\i}}, P., {Taylor}, A.~N., {Tereno}, I., {Toft}, S., {Toledo-Moreo}, R., {Torradeflot}, F., {Tutusaus}, I., {Valentijn}, E.~A., {Valenziano}, L., {Valiviita}, J., {Vassallo}, T., {Verdoes Kleijn}, G., {Veropalumbo}, A., {Wang}, Y., {Weller}, J., {Zamorani}, G., {Zerbi}, F.~M., {Zucca}, E., {Burigana}, C., {Gabarra}, L., {Mart{\'\i}n-Fleitas}, J., \& {Scottez}, V. 2025,a {Euclid: Early Release Observations of ram-pressure stripping in the Perseus cluster: Detection of parsec-scale star formation within the low surface
  brightness stripped tails of UGC 2665 and MCG +07-07-070}.
\newblock {\em \aap}, 701a, A40.

\bibitem[{George} et~al., 2018a]{2018A&A...614A.130G}
{George}, K., {Joseph}, P., {C{\^o}t{\'e}}, P., {Ghosh}, S.~K., {Hutchings}, J.~B., {Mohan}, R., {Postma}, J., {Sankarasubramanian}, K., {Sreekumar}, P., {Stalin}, C.~S., {Subramaniam}, A., \& {Tandon}, S.~N. 2018,a {Dissecting star formation in the ``Atoms-for-Peace'' galaxy. UVIT observations of the post-merger galaxy NGC7252}.
\newblock {\em \aap}, 614a, A130.

\bibitem[{George} et~al., 2018b]{2018MNRAS.479.4126G}
{George}, K., {Poggianti}, B.~M., {Gullieuszik}, M., {Fasano}, G., {Bellhouse}, C., {Postma}, J., {Moretti}, A., {Jaff{\'e}}, Y., {Vulcani}, B., {Bettoni}, D., {Fritz}, J., {C{\^o}t{\'e}}, P., {Ghosh}, S.~K., {Hutchings}, J.~B., {Mohan}, R., {Sreekumar}, P., {Stalin}, C.~S., {Subramaniam}, A., \& {Tandon}, S.~N. 2018,b {UVIT view of ram-pressure stripping in action: star formation in the stripped gas of the GASP jellyfish galaxy JO201 in Abell 85}.
\newblock {\em \mnras}, 479b(3), 4126--4135.

\bibitem[{George} et~al., 2023]{2023MNRAS.519.2426G}
{George}, K., {Poggianti}, B.~M., {Tomi{\v{c}}i{\'c}}, N., {Postma}, J., {C{\^o}t{\'e}}, P., {Fritz}, J., {Ghosh}, S.~K., {Gullieuszik}, M., {Hutchings}, J.~B., {Moretti}, A., {Omizzolo}, A., {Radovich}, M., {Sreekumar}, P., {Subramaniam}, A., {Tandon}, S.~N., \& {Vulcani}, B. 2023, {Ultraviolet imaging observations of three jellyfish galaxies: star formation suppression in the centre and ongoing star formation in stripped tails}.
\newblock {\em \mnras}, 519(2), 2426--2437.

\bibitem[{George} et~al., 2025b]{2025A&A...700A..38G}
{George}, K., {Poggianti}, B.~M., {Vulcani}, B., {Gullieuszik}, M., {Postma}, J., {Fritz}, J., {C{\^o}t{\'e}}, P., {Jaffe}, Y.~L., {Moretti}, A., {Ignesti}, A., {Peluso}, G., {Tomi{\'c}i{\'c}}, N., {Subramaniam}, A., {Ghosh}, S.~K., \& {Tandon}, S.~N. 2025,b {Star formation at different stages of ram-pressure stripping as observed through far-ultraviolet imaging of 13 GASP galaxies}.
\newblock {\em \aap}, 700b, A38.

\bibitem[{Giunchi} et~al., 2023]{2023ApJ...949...72G}
{Giunchi}, E., {Gullieuszik}, M., {Poggianti}, B.~M., {Moretti}, A., {Werle}, A., {Scarlata}, C., {Zanella}, A., {Vulcani}, B., \& {Calzetti}, D. 2023, {HST Imaging of Star-forming Clumps in Six GASP Ram-pressure-stripped Galaxies}.
\newblock {\em \apj}, 949(2), 72.

\bibitem[{G{\'o}mez} et~al., 2003]{2003ApJ...584..210G}
{G{\'o}mez}, P.~L., {Nichol}, R.~C., {Miller}, C.~J., {Balogh}, M.~L., {Goto}, T., {Zabludoff}, A.~I., {Romer}, A.~K., {Bernardi}, M., {Sheth}, R., {Hopkins}, A.~M., {Castander}, F.~J., {Connolly}, A.~J., {Schneider}, D.~P., {Brinkmann}, J., {Lamb}, D.~Q., {SubbaRao}, M., \& {York}, D.~G. 2003, {Galaxy Star Formation as a Function of Environment in the Early Data Release of the Sloan Digital Sky Survey}.
\newblock {\em \apj}, 584(1), 210--227.

\bibitem[{Goto} et~al., 2003]{2003MNRAS.346..601G}
{Goto}, T., {Yamauchi}, C., {Fujita}, Y., {Okamura}, S., {Sekiguchi}, M., {Smail}, I., {Bernardi}, M., \& {Gomez}, P.~L. 2003, {The morphology-density relation in the Sloan Digital Sky Survey}.
\newblock {\em \mnras}, 346(2), 601--614.

\bibitem[{Gullieuszik} et~al., 2023]{2023ApJ...945...54G}
{Gullieuszik}, M., {Giunchi}, E., {Poggianti}, B.~M., {Moretti}, A., {Scarlata}, C., {Calzetti}, D., {Werle}, A., {Zanella}, A., {Radovich}, M., {Bellhouse}, C., {Bettoni}, D., {Franchetto}, A., {Fritz}, J., {Jaff{\'e}}, Y.~L., {McGee}, S.~L., {Mingozzi}, M., {Omizzolo}, A., {Tonnesen}, S., {Verheijen}, M., \& {Vulcani}, B. 2023, {UV and H{\ensuremath{\alpha}} HST Observations of Six GASP Jellyfish Galaxies}.
\newblock {\em \apj}, 945(1), 54.

\bibitem[{Gullieuszik} et~al., 2020]{2020ApJ...899...13G}
{Gullieuszik}, M., {Poggianti}, B.~M., {McGee}, S.~L., {Moretti}, A., {Vulcani}, B., {Tonnesen}, S., {Roediger}, E., {Jaff{\'e}}, Y.~L., {Fritz}, J., {Franchetto}, A., {Omizzolo}, A., {Bettoni}, D., {Radovich}, M., \& {Wolter}, A. 2020, {GASP. XXI. Star Formation Rates in the Tails of Galaxies Undergoing Ram Pressure Stripping}.
\newblock {\em \apj}, 899(1), 13.

\bibitem[{Gullieuszik} et~al., 2017]{2017ApJ...846...27G}
{Gullieuszik}, M., {Poggianti}, B.~M., {Moretti}, A., {Fritz}, J., {Jaff{\'e}}, Y.~L., {Hau}, G., {Bischko}, J.~C., {Bellhouse}, C., {Bettoni}, D., {Fasano}, G., {Vulcani}, B., {D'Onofrio}, M., \& {Biviano}, A. 2017, {GASP. IV. A Muse View of Extreme Ram-pressure-stripping in the Plane of the Sky: The Case of Jellyfish Galaxy JO204}.
\newblock {\em \apj}, 846(1), 27.

\bibitem[{Gunn} and {Gott}, 1972]{1972ApJ...176....1G}
{Gunn}, J.~E. \& {Gott}, III, J.~R. 1972, {On the Infall of Matter Into Clusters of Galaxies and Some Effects on Their Evolution}.
\newblock {\em \apj}, 176, 1.

\bibitem[{Haines} et~al., 2013]{2013ApJ...775..126H}
{Haines}, C.~P., {Pereira}, M.~J., {Smith}, G.~P., {Egami}, E., {Sanderson}, A.~J.~R., {Babul}, A., {Finoguenov}, A., {Merluzzi}, P., {Busarello}, G., {Rawle}, T.~D., \& {Okabe}, N. 2013, {LoCuSS: The Steady Decline and Slow Quenching of Star Formation in Cluster Galaxies over the Last Four Billion Years}.
\newblock {\em \apj}, 775(2), 126.

\bibitem[{Hao} et~al., 2011]{2011ApJ...741..124H}
{Hao}, C.-N., {Kennicutt}, R.~C., {Johnson}, B.~D., {Calzetti}, D., {Dale}, D.~A., \& {Moustakas}, J. 2011, {Dust-corrected Star Formation Rates of Galaxies. II. Combinations of Ultraviolet and Infrared Tracers}.
\newblock {\em \apj}, 741(2), 124.

\bibitem[{Hibbard} et~al., 1994]{1994AJ....107...67H}
{Hibbard}, J.~E., {Guhathakurta}, P., {van Gorkom}, J.~H., \& {Schweizer}, F. 1994, {Cold, Warm, and Hot Gas in the Late-Stage Merger NGC 7252}.
\newblock {\em \aj}, 107, 67.

\bibitem[{Hibbard} and {Mihos}, 1995]{1995AJ....110..140H}
{Hibbard}, J.~E. \& {Mihos}, J.~C. 1995, {Dynamical Modeling of NGC 7252 and the Return of Tidal Material}.
\newblock {\em \aj}, 110, 140.

\bibitem[{Hollenbach} and {Salpeter}, 1971]{1971ApJ...163..155H}
{Hollenbach}, D. \& {Salpeter}, E.~E. 1971, {Surface Recombination of Hydrogen Molecules}.
\newblock {\em \apj}, 163, 155.

\bibitem[{Iglesias-P{\'a}ramo} et~al., 2006]{2006ApJS..164...38I}
{Iglesias-P{\'a}ramo}, J., {Buat}, V., {Takeuchi}, T.~T., {Xu}, K., {Boissier}, S., {Boselli}, A., {Burgarella}, D., {Madore}, B.~F., {Gil de Paz}, A., {Bianchi}, L., {Barlow}, T.~A., {Byun}, Y.~I., {Donas}, J., {Forster}, K., {Friedman}, P.~G., {Heckman}, T.~M., {Jelinski}, P.~N., {Lee}, Y.~W., {Malina}, R.~F., {Martin}, D.~C., {Milliard}, B., {Morrissey}, P.~F., {Neff}, S.~G., {Rich}, R.~M., {Schiminovich}, D., {Seibert}, M., {Siegmund}, O.~H.~W., {Small}, T., {Szalay}, A.~S., {Welsh}, B.~Y., \& {Wyder}, T.~K. 2006, {Star Formation in the Nearby Universe: The Ultraviolet and Infrared Points of View}.
\newblock {\em \apjs}, 164(1), 38--51.

\bibitem[{J{\'a}chym} et~al., 2019]{2019ApJ...883..145J}
{J{\'a}chym}, P., {Kenney}, J. D.~P., {Sun}, M., {Combes}, F., {Cortese}, L., {Scott}, T.~C., {Sivanandam}, S., {Brinks}, E., {Roediger}, E., {Palou{\v{s}}}, J., \& {Fumagalli}, M. 2019, {ALMA Unveils Widespread Molecular Gas Clumps in the Ram Pressure Stripped Tail of the Norma Jellyfish Galaxy}.
\newblock {\em \apj}, 883(2), 145.

\bibitem[{J{\'a}chym} et~al., 2017]{2017ApJ...839..114J}
{J{\'a}chym}, P., {Sun}, M., {Kenney}, J. D.~P., {Cortese}, L., {Combes}, F., {Yagi}, M., {Yoshida}, M., {Palou{\v{s}}}, J., \& {Roediger}, E. 2017, {Molecular Gas Dominated 50 kpc Ram Pressure Stripped Tail of the Coma Galaxy D100}.
\newblock {\em \apj}, 839(2), 114.

\bibitem[{Kaviraj} et~al., 2007]{2007ApJS..173..619K}
{Kaviraj}, S., {Schawinski}, K., {Devriendt}, J.~E.~G., {Ferreras}, I., {Khochfar}, S., {Yoon}, S.~J., {Yi}, S.~K., {Deharveng}, J.~M., {Boselli}, A., {Barlow}, T., {Conrow}, T., {Forster}, K., {Friedman}, P.~G., {Martin}, D.~C., {Morrissey}, P., {Neff}, S., {Schiminovich}, D., {Seibert}, M., {Small}, T., {Wyder}, T., {Bianchi}, L., {Donas}, J., {Heckman}, T., {Lee}, Y.~W., {Madore}, B., {Milliard}, B., {Rich}, R.~M., \& {Szalay}, A. 2007, {UV-Optical Colors As Probes of Early-Type Galaxy Evolution}.
\newblock {\em \apjs}, 173(2), 619--642.

\bibitem[{Kennicutt} and {Evans}, 2012]{2012ARA&A..50..531K}
{Kennicutt}, R.~C. \& {Evans}, N.~J. 2012, {Star Formation in the Milky Way and Nearby Galaxies}.
\newblock {\em \araa}, 50, 531--608.

\bibitem[{Kennicutt}, 1998]{1998ARA&A..36..189K}
{Kennicutt}, Jr., R.~C. 1998, {Star Formation in Galaxies Along the Hubble Sequence}.
\newblock {\em \araa}, 36, 189--232.

\bibitem[{Konstantopoulos} et~al., 2010]{2010ApJ...723..197K}
{Konstantopoulos}, I.~S., {Gallagher}, S.~C., {Fedotov}, K., {Durrell}, P.~R., {Heiderman}, A., {Elmegreen}, D.~M., {Charlton}, J.~C., {Hibbard}, J.~E., {Tzanavaris}, P., {Chandar}, R., {Johnson}, K.~E., {Maybhate}, A., {Zabludoff}, A.~E., {Gronwall}, C., {Szathmary}, D., {Hornschemeier}, A.~E., {English}, J., {Whitmore}, B., {Mendes de Oliveira}, C., \& {Mulchaey}, J.~S. 2010, {Galaxy Evolution in a Complex Environment: A Multi-wavelength Study of HCG 7}.
\newblock {\em \apj}, 723(1), 197--217.

\bibitem[{Kovakkuni} et~al., 2023]{2023MNRAS.526.1940K}
{Kovakkuni}, N., {Lelli}, F., {Duc}, P.~A., {Boquien}, M., {Braine}, J., {Brinks}, E., {Charmandaris}, V., {Combes}, F., {Fensch}, J., {Lisenfeld}, U., {McGaugh}, S.~S., {Mihos}, J.~C., {Pawlowski}, M.~S., {Revaz}, Y., \& {Weilbacher}, P.~M. 2023, {Molecular and ionized gas in tidal dwarf galaxies: the spatially resolved star formation relation}.
\newblock {\em \mnras}, 526(2), 1940--1950.

\bibitem[{Lambas} et~al., 2012]{2012A&A...539A..45L}
{Lambas}, D.~G., {Alonso}, S., {Mesa}, V., \& {O'Mill}, A.~L. 2012, {Galaxy interactions. I. Major and minor mergers}.
\newblock {\em \aap}, 539, A45.

\bibitem[{Larson} et~al., 1980]{1980ApJ...237..692L}
{Larson}, R.~B., {Tinsley}, B.~M., \& {Caldwell}, C.~N. 1980, {The evolution of disk galaxies and the origin of S0 galaxies}.
\newblock {\em \apj}, 237, 692--707.

\bibitem[{Laudari} et~al., 2022]{2022MNRAS.509.3938L}
{Laudari}, S., {J{\'a}chym}, P., {Sun}, M., {Waldron}, W., {Chatzikos}, M., {Kenney}, J., {Luo}, R., {Nulsen}, P., {Sarazin}, C., {Combes}, F., {Edge}, T., {Voit}, M., {Donahue}, M., \& {Cortese}, L. 2022, {ESO 137-002: a large spiral undergoing edge-on ram-pressure stripping with little star formation in the tail}.
\newblock {\em \mnras}, 509(3), 3938--3956.

\bibitem[{Lee} et~al., 2020]{2020ApJ...905...31L}
{Lee}, J., {Kimm}, T., {Katz}, H., {Rosdahl}, J., {Devriendt}, J., \& {Slyz}, A. 2020, {Dual Effects of Ram Pressure on Star Formation in Multiphase Disk Galaxies with Strong Stellar Feedback}.
\newblock {\em \apj}, 905(1), 31.

\bibitem[{Longobardi} et~al., 2020a]{2020A&A...633L...7L}
{Longobardi}, A., {Boselli}, A., {Boissier}, S., {Bianchi}, S., {Andreani}, P., {Sarpa}, E., {Nanni}, A., \& {Miville-Desch{\^e}nes}, M. 2020,a {The GALEX Ultraviolet Virgo Cluster Survey (GUViCS). VIII. Diffuse dust in the Virgo intra-cluster space}.
\newblock {\em \aap}, 633a, L7.

\bibitem[{Longobardi} et~al., 2020b]{2020A&A...644A.161L}
{Longobardi}, A., {Boselli}, A., {Fossati}, M., {Villa-V{\'e}lez}, J.~A., {Bianchi}, S., {Casasola}, V., {Sarpa}, E., {Combes}, F., {Hensler}, G., {Burgarella}, D., {Schimd}, C., {Nanni}, A., {C{\^o}t{\'e}}, P., {Buat}, V., {Amram}, P., {Ferrarese}, L., {Braine}, J., {Trinchieri}, G., {Boissier}, S., {Boquien}, M., {Andreani}, P., {Gwyn}, S., \& {Cuillandre}, J.~C. 2020,b {A Virgo Environmental Survey Tracing Ionised Gas Emission (VESTIGE). VII. Bridging the cluster-ICM-galaxy evolution at small scales}.
\newblock {\em \aap}, 644b, A161.

\bibitem[{Lora} et~al., 2024]{2024ApJ...969...24L}
{Lora}, V., {Smith}, R., {Fritz}, J., {Pasquali}, A., \& {Raga}, A.~C. 2024, {Dark-matter-free Dwarf Galaxy Formation at the Tips of the Tentacles of Jellyfish Galaxies}.
\newblock {\em \apj}, 969(1), 24.

\bibitem[{Luber} et~al., 2022]{2022ApJ...927...39L}
{Luber}, N., {M{\"u}ller}, A., {van Gorkom}, J.~H., {Poggianti}, B.~M., {Vulcani}, B., {Franchetto}, A., {Bacchini}, C., {Bettoni}, D., {Deb}, T., {Fritz}, J., {Gullieuszik}, M., {Ignesti}, A., {Jaffe}, Y., {Moretti}, A., {Paladino}, R., {Ramatsoku}, M., {Serra}, P., {Smith}, R., {Tomicic}, N., {Tonnesen}, S., {Verheijen}, M., \& {Wolter}, A. 2022, {GASP XXXVII: The Most Extreme Jellyfish Galaxies Compared with Other Disk Galaxies in Clusters, an H I Study}.
\newblock {\em \apj}, 927(1), 39.

\bibitem[{Lynds} and {Toomre}, 1976]{1976ApJ...209..382L}
{Lynds}, R. \& {Toomre}, A. 1976, {On the interpretation of ring galaxies: the binary ring system II Hz 4.}
\newblock {\em \apj}, 209, 382--388.

\bibitem[{Madau} and {Dickinson}, 2014]{2014ARA&A..52..415M}
{Madau}, P. \& {Dickinson}, M. 2014, {Cosmic Star-Formation History}.
\newblock {\em \araa}, 52, 415--486.

\bibitem[{Meurer} et~al., 1999]{1999ApJ...521...64M}
{Meurer}, G.~R., {Heckman}, T.~M., \& {Calzetti}, D. 1999, {Dust Absorption and the Ultraviolet Luminosity Density at z \raisebox{-0.5ex}\textasciitilde 3 as Calibrated by Local Starburst Galaxies}.
\newblock {\em \apj}, 521(1), 64--80.

\bibitem[{Mihos}, 2004]{2004cgpc.symp..277M}
{Mihos}, J.~C.
\newblock {Interactions and Mergers of Cluster Galaxies}.
\newblock In {Mulchaey}, J.~S., {Dressler}, A., \& {Oemler}, A., editors, {\em Clusters of Galaxies: Probes of Cosmological Structure and Galaxy Evolution} 2004,,  277.

\bibitem[{Moore} et~al., 1996]{1996Natur.379..613M}
{Moore}, B., {Katz}, N., {Lake}, G., {Dressler}, A., \& {Oemler}, A. 1996, {Galaxy harassment and the evolution of clusters of galaxies}.
\newblock {\em \nat}, 379(6566), 613--616.

\bibitem[{Moretti} et~al., 2018]{2018MNRAS.480.2508M}
{Moretti}, A., {Paladino}, R., {Poggianti}, B.~M., {D'Onofrio}, M., {Bettoni}, D., {Gullieuszik}, M., {Jaff{\'e}}, Y.~L., {Vulcani}, B., {Fasano}, G., {Fritz}, J., \& {Torstensson}, K. 2018, {GASP - X. APEX observations of molecular gas in the discs and in the tails of ram-pressure stripped galaxies}.
\newblock {\em \mnras}, 480(2), 2508--2520.

\bibitem[{Moretti} et~al., 2020a]{2020ApJ...897L..30M}
{Moretti}, A., {Paladino}, R., {Poggianti}, B.~M., {Serra}, P., {Ramatsoku}, M., {Franchetto}, A., {Deb}, T., {Gullieuszik}, M., {Tomi{\v{c}}i{\'c}}, N., {Mingozzi}, M., {Vulcani}, B., {Radovich}, M., {Bettoni}, D., \& {Fritz}, J. 2020,a {The High Molecular Gas Content, and the Efficient Conversion of Neutral into Molecular Gas, in Jellyfish Galaxies}.
\newblock {\em \apjl}, 897a(2), L30.

\bibitem[{Moretti} et~al., 2020b]{2020ApJ...889....9M}
{Moretti}, A., {Paladino}, R., {Poggianti}, B.~M., {Serra}, P., {Roediger}, E., {Gullieuszik}, M., {Tomi{\v{c}}i{\'c}}, N., {Radovich}, M., {Vulcani}, B., {Jaff{\'e}}, Y.~L., {Fritz}, J., {Bettoni}, D., {Ramatsoku}, M., \& {Wolter}, A. 2020,b {GASP. XXII. The Molecular Gas Content of the JW100 Jellyfish Galaxy at z {\ensuremath{\sim}} 0.05: Does Ram Pressure Promote Molecular Gas Formation?}
\newblock {\em \apj}, 889b(1), 9.

\bibitem[{Moretti} et~al., 2023]{2023ApJ...955..153M}
{Moretti}, A., {Serra}, P., {Bacchini}, C., {Paladino}, R., {Ramatsoku}, M., {Poggianti}, B.~M., {Vulcani}, B., {Deb}, T., {Gullieuszik}, M., {Fritz}, J., \& {Wolter}, A. 2023, {The Evolution of the Cold Gas Fraction in Nearby Clusters' Ram-pressure-stripped Galaxies}.
\newblock {\em \apj}, 955(2), 153.

\bibitem[{M{\"u}ller} et~al., 2021a]{2021Galax...9..116M}
{M{\"u}ller}, A., {Ignesti}, A., {Poggianti}, B., {Moretti}, A., {Ramatsoku}, M., \& {Dettmar}, R.-J. 2021,a {Role of Magnetic Fields in Ram Pressure Stripped Galaxies}.
\newblock {\em Galaxies}, 9a(4), 116.

\bibitem[{M{\"u}ller} et~al., 2021b]{2021NatAs...5..159M}
{M{\"u}ller}, A., {Poggianti}, B.~M., {Pfrommer}, C., {Adebahr}, B., {Serra}, P., {Ignesti}, A., {Sparre}, M., {Gitti}, M., {Dettmar}, R.-J., {Vulcani}, B., \& {Moretti}, A. 2021,b {Highly ordered magnetic fields in the tail of the jellyfish galaxy JO206}.
\newblock {\em Nature Astronomy}, 5b, 159--168.

\bibitem[{Nulsen}, 1982]{1982MNRAS.198.1007N}
{Nulsen}, P.~E.~J. 1982, {Transport processes and the stripping of cluster galaxies.}
\newblock {\em \mnras}, 198, 1007--1016.

\bibitem[{O'Donnell}, 1994]{1994ApJ...422..158O}
{O'Donnell}, J.~E. 1994, {R v-dependent Optical and Near-Ultraviolet Extinction}.
\newblock {\em \apj}, 422, 158.

\bibitem[{Owen} et~al., 2006]{2006AJ....131.1974O}
{Owen}, F.~N., {Keel}, W.~C., {Wang}, Q.~D., {Ledlow}, M.~J., \& {Morrison}, G.~E. 2006, {A Deep Radio Survey of Abell 2125. III. The Cluster Core: Merging and Stripping}.
\newblock {\em \aj}, 131(4), 1974--1988.

\bibitem[{Owers} et~al., 2012]{2012ApJ...750L..23O}
{Owers}, M.~S., {Couch}, W.~J., {Nulsen}, P. E.~J., \& {Randall}, S.~W. 2012, {Shocking Tails in the Major Merger Abell 2744}.
\newblock {\em \apjl}, 750(1), L23.

\bibitem[{Park} et~al., 2008]{2008ApJ...674..784P}
{Park}, C., {Gott}, III, J.~R., \& {Choi}, Y.-Y. 2008, {Transformation of Morphology and Luminosity Classes of the SDSS Galaxies}.
\newblock {\em \apj}, 674(2), 784--796.

\bibitem[{Peng} et~al., 2010]{2010ApJ...721..193P}
{Peng}, Y.-j., {Lilly}, S.~J., {Kova{\v{c}}}, K., {Bolzonella}, M., {Pozzetti}, L., {Renzini}, A., {Zamorani}, G., {Ilbert}, O., {Knobel}, C., {Iovino}, A., {Maier}, C., {Cucciati}, O., {Tasca}, L., {Carollo}, C.~M., {Silverman}, J., {Kampczyk}, P., {de Ravel}, L., {Sanders}, D., {Scoville}, N., {Contini}, T., {Mainieri}, V., {Scodeggio}, M., {Kneib}, J.-P., {Le F{\`e}vre}, O., {Bardelli}, S., {Bongiorno}, A., {Caputi}, K., {Coppa}, G., {de la Torre}, S., {Franzetti}, P., {Garilli}, B., {Lamareille}, F., {Le Borgne}, J.-F., {Le Brun}, V., {Mignoli}, M., {Perez Montero}, E., {Pello}, R., {Ricciardelli}, E., {Tanaka}, M., {Tresse}, L., {Vergani}, D., {Welikala}, N., {Zucca}, E., {Oesch}, P., {Abbas}, U., {Barnes}, L., {Bordoloi}, R., {Bottini}, D., {Cappi}, A., {Cassata}, P., {Cimatti}, A., {Fumana}, M., {Hasinger}, G., {Koekemoer}, A., {Leauthaud}, A., {Maccagni}, D., {Marinoni}, C., {McCracken}, H., {Memeo}, P., {Meneux}, B., {Nair}, P., {Porciani}, C., {Presotto}, V., \& {Scaramella}, R. 2010, {Mass and
  Environment as Drivers of Galaxy Evolution in SDSS and zCOSMOS and the Origin of the Schechter Function}.
\newblock {\em \apj}, 721(1), 193--221.

\bibitem[{Poggianti} et~al., 2016]{2016AJ....151...78P}
{Poggianti}, B.~M., {Fasano}, G., {Omizzolo}, A., {Gullieuszik}, M., {Bettoni}, D., {Moretti}, A., {Paccagnella}, A., {Jaff{\'e}}, Y.~L., {Vulcani}, B., {Fritz}, J., {Couch}, W., \& {D'Onofrio}, M. 2016, {Jellyfish Galaxy Candidates at Low Redshift}.
\newblock {\em \aj}, 151(3), 78.

\bibitem[{Poggianti} et~al., 2019a]{2019MNRAS.482.4466P}
{Poggianti}, B.~M., {Gullieuszik}, M., {Tonnesen}, S., {Moretti}, A., {Vulcani}, B., {Radovich}, M., {Jaff{\'e}}, Y., {Fritz}, J., {Bettoni}, D., {Franchetto}, A., {Fasano}, G., {Bellhouse}, C., \& {Omizzolo}, A. 2019,a {GASP XIII. Star formation in gas outside galaxies}.
\newblock {\em \mnras}, 482a(4), 4466--4502.

\bibitem[{Poggianti} et~al., 2019b]{2019ApJ...887..155P}
{Poggianti}, B.~M., {Ignesti}, A., {Gitti}, M., {Wolter}, A., {Brighenti}, F., {Biviano}, A., {George}, K., {Vulcani}, B., {Gullieuszik}, M., {Moretti}, A., {Paladino}, R., {Bettoni}, D., {Franchetto}, A., {Jaff{\'e}}, Y.~L., {Radovich}, M., {Roediger}, E., {Tomi{\v{c}}i{\'c}}, N., {Tonnesen}, S., {Bellhouse}, C., {Fritz}, J., \& {Omizzolo}, A. 2019,b {GASP XXIII: A Jellyfish Galaxy as an Astrophysical Laboratory of the Baryonic Cycle}.
\newblock {\em \apj}, 887b(2), 155.

\bibitem[{Poggianti} et~al., 2017a]{2017Natur.548..304P}
{Poggianti}, B.~M., {Jaff{\'e}}, Y.~L., {Moretti}, A., {Gullieuszik}, M., {Radovich}, M., {Tonnesen}, S., {Fritz}, J., {Bettoni}, D., {Vulcani}, B., {Fasano}, G., {Bellhouse}, C., {Hau}, G., \& {Omizzolo}, A. 2017,a {Ram-pressure feeding of supermassive black holes}.
\newblock {\em \nat}, 548a(7667), 304--309.

\bibitem[{Poggianti} et~al., 2017b]{2017ApJ...844...48P}
{Poggianti}, B.~M., {Moretti}, A., {Gullieuszik}, M., {Fritz}, J., {Jaff{\'e}}, Y., {Bettoni}, D., {Fasano}, G., {Bellhouse}, C., {Hau}, G., {Vulcani}, B., {Biviano}, A., {Omizzolo}, A., {Paccagnella}, A., {D'Onofrio}, M., {Cava}, A., {Sheen}, Y.~K., {Couch}, W., \& {Owers}, M. 2017,b {GASP. I. Gas Stripping Phenomena in Galaxies with MUSE}.
\newblock {\em \apj}, 844b(1), 48.

\bibitem[{Postma} and {Leahy}, 2017]{2017PASP..129k5002P}
{Postma}, J.~E. \& {Leahy}, D. 2017, {CCDLAB: A Graphical User Interface FITS Image Data Reducer, Viewer, and Canadian UVIT Data Pipeline}.
\newblock {\em \pasp}, 129(981), 115002.

\bibitem[{Postma} and {Leahy}, 2021]{2021JApA...42...30P}
{Postma}, J.~E. \& {Leahy}, D. 2021, {UVIT data reduction pipeline: A CCDLAB and UVIT tutorial}.
\newblock {\em Journal of Astrophysics and Astronomy}, 42(2), 30.

\bibitem[{Rakhi} et~al., 2023]{2023MNRAS.522.1196R}
{Rakhi}, R., {Santhosh}, G., {Joseph}, P., {George}, K., {Subramanian}, S., {Kavila}, I., {Postma}, J., {Duc}, P.-A., {C{\^o}t{\'e}}, P., {Cortese}, L., {Ghosh}, S.~K., {Subramaniam}, A., {Tandon}, S., {Hutchings}, J., {Wesley}, P.~S., {Bharadwaj}, A., \& {Niroula}, N. 2023, {UVIT view of NGC 5291: Ongoing star formation in tidal dwarf galaxies at 0.35 kpc resolution}.
\newblock {\em \mnras}, 522(1), 1196--1207.

\bibitem[{Ramatsoku} et~al., 2020]{2020A&A...640A..22R}
{Ramatsoku}, M., {Serra}, P., {Poggianti}, B.~M., {Moretti}, A., {Gullieuszik}, M., {Bettoni}, D., {Deb}, T., {Franchetto}, A., {van Gorkom}, J.~H., {Jaff{\'e}}, Y., {Tonnesen}, S., {Verheijen}, M.~A.~W., {Vulcani}, B., {Andati}, L.~A.~L., {de Blok}, E., {J{\'o}zsa}, G.~I.~G., {Kamphuis}, P., {Kleiner}, D., {Maccagni}, F.~M., {Makhathini}, S., {Moln{\'a}r}, D.~C., {Ramaila}, A.~J.~T., {Smirnov}, O., \& {Thorat}, K. 2020, {GASP. XXVI. HI gas in jellyfish galaxies: The case of JO201 and JO206}.
\newblock {\em \aap}, 640, A22.

\bibitem[{Ramatsoku} et~al., 2019]{2019MNRAS.487.4580R}
{Ramatsoku}, M., {Serra}, P., {Poggianti}, B.~M., {Moretti}, A., {Gullieuszik}, M., {Bettoni}, D., {Deb}, T., {Fritz}, J., {van Gorkom}, J.~H., {Jaff{\'e}}, Y.~L., {Tonnesen}, S., {Verheijen}, M.~A.~W., {Vulcani}, B., {Hugo}, B., {J{\'o}zsa}, G.~I.~G., {Maccagni}, F.~M., {Makhathini}, S., {Ramaila}, A., {Smirnov}, O., \& {Thorat}, K. 2019, {GASP - XVII. H I imaging of the jellyfish galaxy JO206: gas stripping and enhanced star formation}.
\newblock {\em \mnras}, 487(4), 4580--4591.

\bibitem[{Rawle} et~al., 2014]{2014MNRAS.442..196R}
{Rawle}, T.~D., {Altieri}, B., {Egami}, E., {P{\'e}rez-Gonz{\'a}lez}, P.~G., {Richard}, J., {Santos}, J.~S., {Valtchanov}, I., {Walth}, G., {Bouy}, H., {Haines}, C.~P., \& {Okabe}, N. 2014, {Star formation in the massive cluster merger Abell 2744}.
\newblock {\em \mnras}, 442(1), 196--206.

\bibitem[{Roberts} et~al., 2021a]{2021A&A...650A.111R}
{Roberts}, I.~D., {van Weeren}, R.~J., {McGee}, S.~L., {Botteon}, A., {Drabent}, A., {Ignesti}, A., {Rottgering}, H.~J.~A., {Shimwell}, T.~W., \& {Tasse}, C. 2021,a {LoTSS jellyfish galaxies. I. Radio tails in low redshift clusters}.
\newblock {\em \aap}, 650a, A111.

\bibitem[{Roberts} et~al., 2021b]{2021A&A...652A.153R}
{Roberts}, I.~D., {van Weeren}, R.~J., {McGee}, S.~L., {Botteon}, A., {Ignesti}, A., \& {Rottgering}, H.~J.~A. 2021,b {LoTSS jellyfish galaxies. II. Ram pressure stripping in groups versus clusters}.
\newblock {\em \aap}, 652b, A153.

\bibitem[{Robotham} et~al., 2018]{2018MNRAS.476.3137R}
{Robotham}, A.~S.~G., {Davies}, L.~J.~M., {Driver}, S.~P., {Koushan}, S., {Taranu}, D.~S., {Casura}, S., \& {Liske}, J. 2018, {ProFound: Source Extraction and Application to Modern Survey Data}.
\newblock {\em \mnras}, 476(3), 3137--3159.

\bibitem[{Safranek-Shrader} et~al., 2017]{2017MNRAS.465..885S}
{Safranek-Shrader}, C., {Krumholz}, M.~R., {Kim}, C.-G., {Ostriker}, E.~C., {Klein}, R.~I., {Li}, S., {McKee}, C.~F., \& {Stone}, J.~M. 2017, {Chemistry and radiative shielding in star-forming galactic discs}.
\newblock {\em \mnras}, 465(1), 885--905.

\bibitem[{Salpeter}, 1955]{1955ApJ...121..161S}
{Salpeter}, E.~E. 1955, {The Luminosity Function and Stellar Evolution.}
\newblock {\em \apj}, 121, 161.

\bibitem[{Santhosh} et~al., 2025]{2025PASA...42...73S}
{Santhosh}, G., {Rajalakshmi}, R., {George}, K., {Subramanian}, S., \& {Indulekha}, K. 2025, {Star formation in interacting galaxy systems: UVIT imaging of NGC 7252 and NGC 5291}.
\newblock {\em \pasa}, 42, e073.

\bibitem[{Schawinski} et~al., 2007]{2007ApJS..173..512S}
{Schawinski}, K., {Kaviraj}, S., {Khochfar}, S., {Yoon}, S.~J., {Yi}, S.~K., {Deharveng}, J.~M., {Boselli}, A., {Barlow}, T., {Conrow}, T., {Forster}, K., {Friedman}, P.~G., {Martin}, D.~C., {Morrissey}, P., {Neff}, S., {Schiminovich}, D., {Seibert}, M., {Small}, T., {Wyder}, T., {Bianchi}, L., {Donas}, J., {Heckman}, T., {Lee}, Y.~W., {Madore}, B., {Milliard}, B., {Rich}, R.~M., \& {Szalay}, A. 2007, {The Effect of Environment on the Ultraviolet Color-Magnitude Relation of Early-Type Galaxies}.
\newblock {\em \apjs}, 173(2), 512--523.

\bibitem[{Schlafly} and {Finkbeiner}, 2011]{2011ApJ...737..103S}
{Schlafly}, E.~F. \& {Finkbeiner}, D.~P. 2011, {Measuring Reddening with Sloan Digital Sky Survey Stellar Spectra and Recalibrating SFD}.
\newblock {\em \apj}, 737(2), 103.

\bibitem[{Schweizer} et~al., 1987]{1987ApJ...320..454S}
{Schweizer}, F., {Ford}, Jr., W.~K., {Jedrzejewski}, R., \& {Giovanelli}, R. 1987, {The Structure and Evolution of Hoag's Object}.
\newblock {\em \apj}, 320, 454.

\bibitem[{Sivanandam} et~al., 2014]{2014ApJ...796...89S}
{Sivanandam}, S., {Rieke}, M.~J., \& {Rieke}, G.~H. 2014, {Tracing Ram-pressure Stripping with Warm Molecular Hydrogen Emission}.
\newblock {\em \apj}, 796(2), 89.

\bibitem[{Skibba} et~al., 2009]{2009MNRAS.399..966S}
{Skibba}, R.~A., {Bamford}, S.~P., {Nichol}, R.~C., {Lintott}, C.~J., {Andreescu}, D., {Edmondson}, E.~M., {Murray}, P., {Raddick}, M.~J., {Schawinski}, K., {Slosar}, A., {Szalay}, A.~S., {Thomas}, D., \& {Vandenberg}, J. 2009, {Galaxy Zoo: disentangling the environmental dependence of morphology and colour}.
\newblock {\em \mnras}, 399(2), 966--982.

\bibitem[{Skryabina} et~al., 2024]{2024MNRAS.532..883S}
{Skryabina}, M.~N., {Adams}, K.~R., \& {Mosenkov}, A.~V. 2024, {Tidal features and disc thicknesses of edge-on galaxies in the SDSS Stripe 82}.
\newblock {\em \mnras}, 532(1), 883--902.

\bibitem[{Smith} et~al., 2010]{2010MNRAS.408.1417S}
{Smith}, R.~J., {Lucey}, J.~R., {Hammer}, D., {Hornschemeier}, A.~E., {Carter}, D., {Hudson}, M.~J., {Marzke}, R.~O., {Mouhcine}, M., {Eftekharzadeh}, S., {James}, P., {Khosroshahi}, H., {Kourkchi}, E., \& {Karick}, A. 2010, {Ultraviolet tails and trails in cluster galaxies: a sample of candidate gaseous stripping events in Coma}.
\newblock {\em \mnras}, 408(3), 1417--1432.

\bibitem[{Sol Alonso} et~al., 2006]{2006MNRAS.367.1029S}
{Sol Alonso}, M., {Lambas}, D.~G., {Tissera}, P., \& {Coldwell}, G. 2006, {Effects of galaxy interactions in different environments}.
\newblock {\em \mnras}, 367(3), 1029--1038.

\bibitem[{Solanes} et~al., 2002]{2002AJ....124.2440S}
{Solanes}, J.~M., {Sanchis}, T., {Salvador-Sol{\'e}}, E., {Giovanelli}, R., \& {Haynes}, M.~P. 2002, {The Three-dimensional Structure of the Virgo Cluster Region from Tully-Fisher and H I Data}.
\newblock {\em \aj}, 124(5), 2440--2452.

\bibitem[{Subramaniam} et~al., 2016]{2016SPIE.9905E..1FS}
{Subramaniam}, A., {Tandon}, S.~N., {Hutchings}, J., {Ghosh}, S.~K., {George}, K., {Girish}, V., {Kamath}, P.~U., {Kathiravan}, S., {Kumar}, A., {Lancelot}, J.~P., {Mahesh}, P.~K., {Mohan}, R., {Murthy}, J., {Nagabhushana}, S., {Pati}, A.~K., {Postma}, J., {Rao}, N.~K., {Sankarasubramanian}, K., {Sreekumar}, P., {Sriram}, S., {Stalin}, C.~S., {Sutaria}, F., {Sreedhar}, Y.~H., {Barve}, I.~V., {Mondal}, C., \& {Sahu}, S.
\newblock {In-orbit performance of UVIT on ASTROSAT}.
\newblock In {den Herder}, J.-W.~A., {Takahashi}, T., \& {Bautz}, M., editors, {\em Space Telescopes and Instrumentation 2016: Ultraviolet to Gamma Ray} 2016,, volume 9905 of {\em Society of Photo-Optical Instrumentation Engineers (SPIE) Conference Series},  99051F.

\bibitem[{Sun} et~al., 2006]{2006ApJ...637L..81S}
{Sun}, M., {Jones}, C., {Forman}, W., {Nulsen}, P.~E.~J., {Donahue}, M., \& {Voit}, G.~M. 2006, {A 70 Kiloparsec X-Ray Tail in the Cluster A3627}.
\newblock {\em \apjl}, 637(2), L81--L84.

\bibitem[{Tandon} et~al., 2020]{2020AJ....159..158T}
{Tandon}, S.~N., {Postma}, J., {Joseph}, P., {Devaraj}, A., {Subramaniam}, A., {Barve}, I.~V., {George}, K., {Ghosh}, S.~K., {Girish}, V., {Hutchings}, J.~B., {Kamath}, P.~U., {Kathiravan}, S., {Kumar}, A., {Lancelot}, J.~P., {Leahy}, D., {Mahesh}, P.~K., {Mohan}, R., {Nagabhushana}, S., {Pati}, A.~K., {Rao}, N.~K., {Sankarasubramanian}, K., {Sriram}, S., \& {Stalin}, C.~S. 2020, {Additional Calibration of the Ultraviolet Imaging Telescope on Board AstroSat}.
\newblock {\em \aj}, 159(4), 158.

\bibitem[{Tandon} et~al., 2017]{2017AJ....154..128T}
{Tandon}, S.~N., {Subramaniam}, A., {Girish}, V., {Postma}, J., {Sankarasubramanian}, K., {Sriram}, S., {Stalin}, C.~S., {Mondal}, C., {Sahu}, S., {Joseph}, P., {Hutchings}, J., {Ghosh}, S.~K., {Barve}, I.~V., {George}, K., {Kamath}, P.~U., {Kathiravan}, S., {Kumar}, A., {Lancelot}, J.~P., {Leahy}, D., {Mahesh}, P.~K., {Mohan}, R., {Nagabhushana}, S., {Pati}, A.~K., {Kameswara Rao}, N., {Sreedhar}, Y.~H., \& {Sreekumar}, P. 2017, {In-orbit Calibrations of the Ultraviolet Imaging Telescope}.
\newblock {\em \aj}, 154(3), 128.

\bibitem[{Tomi{\v{c}}i{\'c}} et~al., 2024]{2024ApJ...976...90T}
{Tomi{\v{c}}i{\'c}}, N., {Werle}, A., {Vulcani}, B., {Ignesti}, A., {Moretti}, A., {Wolter}, A., {George}, K., {Poggianti}, B.~M., \& {Gullieuszik}, M. 2024, {Spatially Resolved Comparison of SFRs from UV and H{\ensuremath{\alpha}} in GASP Gas-stripped Galaxies}.
\newblock {\em \apj}, 976(1), 90.

\bibitem[{Tonnesen} et~al., 2007]{2007ApJ...671.1434T}
{Tonnesen}, S., {Bryan}, G.~L., \& {van Gorkom}, J.~H. 2007, {Environmentally Driven Evolution of Simulated Cluster Galaxies}.
\newblock {\em \apj}, 671(2), 1434--1445.

\bibitem[{Toomre} and {Toomre}, 1972]{1972ApJ...178..623T}
{Toomre}, A. \& {Toomre}, J. 1972, {Galactic Bridges and Tails}.
\newblock {\em \apj}, 178, 623--666.

\bibitem[{Verdugo} et~al., 2015]{2015A&A...582A...6V}
{Verdugo}, C., {Combes}, F., {Dasyra}, K., {Salom{\'e}}, P., \& {Braine}, J. 2015, {Ram pressure stripping in the Virgo Cluster}.
\newblock {\em \aap}, 582, A6.

\bibitem[{Vogelsberger} et~al., 2019]{2019MNRAS.487.4870V}
{Vogelsberger}, M., {McKinnon}, R., {O'Neil}, S., {Marinacci}, F., {Torrey}, P., \& {Kannan}, R. 2019, {Dust in and around galaxies: dust in cluster environments and its impact on gas cooling}.
\newblock {\em \mnras}, 487(4), 4870--4883.

\bibitem[{Vulcani} et~al., 2020]{2020ApJ...892..146V}
{Vulcani}, B., {Fritz}, J., {Poggianti}, B.~M., {Bettoni}, D., {Franchetto}, A., {Moretti}, A., {Gullieuszik}, M., {Jaff{\'e}}, Y., {Biviano}, A., {Radovich}, M., \& {Mingozzi}, M. 2020, {GASP XXIV. The History of Abruptly Quenched Galaxies in Clusters}.
\newblock {\em \apj}, 892(2), 146.

\bibitem[{Vulcani} et~al., 2018]{2018ApJ...866L..25V}
{Vulcani}, B., {Poggianti}, B.~M., {Gullieuszik}, M., {Moretti}, A., {Tonnesen}, S., {Jaff{\'e}}, Y.~L., {Fritz}, J., {Fasano}, G., \& {Bettoni}, D. 2018, {Enhanced Star Formation in Both Disks and Ram-pressure-stripped Tails of GASP Jellyfish Galaxies}.
\newblock {\em \apjl}, 866(2), L25.

\bibitem[{Wang} et~al., 2025]{2025MNRAS.538..327W}
{Wang}, S., {Wang}, J., {Lee-Waddell}, K., {Yang}, D., {Lin}, X., \& {Staveley-Smith}, L. 2025, {FEASTS: the fate of gas and star formation in interacting galaxies}.
\newblock {\em \mnras}, 538(1), 327--350.

\bibitem[{Werle} et~al., 2024]{2024A&A...682A.162W}
{Werle}, A., {Giunchi}, E., {Poggianti}, B., {Gullieuszik}, M., {Moretti}, A., {Zanella}, A., {Tonnesen}, S., {Fritz}, J., {Vulcani}, B., {Bacchini}, C., {Akerman}, N., {Kulier}, A., {Tomicic}, N., {Smith}, R., \& {Wolter}, A. 2024, {The history of star-forming regions in the tails of six GASP jellyfish galaxies observed with the Hubble Space Telescope}.
\newblock {\em \aap}, 682, A162.

\bibitem[{Whitmore} et~al., 1993]{1993ApJ...407..489W}
{Whitmore}, B.~C., {Gilmore}, D.~M., \& {Jones}, C. 1993, {What Determines the Morphological Fractions in Clusters of Galaxies?}
\newblock {\em \apj}, 407, 489.

\bibitem[{Wilkins} et~al., 2012]{2012MNRAS.424.1522W}
{Wilkins}, S.~M., {Gonzalez-Perez}, V., {Lacey}, C.~G., \& {Baugh}, C.~M. 2012, {Predictions for the intrinsic UV continuum properties of star-forming galaxies and the implications for inferring dust extinction}.
\newblock {\em \mnras}, 424(2), 1522--1529.

\bibitem[{Zabel} et~al., 2019]{2019MNRAS.483.2251Z}
{Zabel}, N., {Davis}, T.~A., {Smith}, M. W.~L., {Maddox}, N., {Bendo}, G.~J., {Peletier}, R., {Iodice}, E., {Venhola}, A., {Baes}, M., {Davies}, J.~I., {de Looze}, I., {Gomez}, H., {Grossi}, M., {Kenney}, J. D.~P., {Serra}, P., {van de Voort}, F., {Vlahakis}, C., \& {Young}, L.~M. 2019, {The ALMA Fornax Cluster Survey I: stirring and stripping of the molecular gas in cluster galaxies}.
\newblock {\em \mnras}, 483(2), 2251--2268.

\end{thebibliography}
\bibliographystyle{paslike}

\end{document}